\documentclass[preprint]{aastex}  
\usepackage{color}

\newcommand{\noprint}[1]{}
\newcommand{\figsetstart}{{\bf Fig. Set} }
\newcommand{\figsetend}{}
\newcommand{\figsetgrpstart}{}
\newcommand{\figsetgrpend}{}
\newcommand{\figsetnum}[1]{{\bf #1.}}
\newcommand{\figsettitle}[1]{ {\bf #1} }
\newcommand{\figsetgrpnum}[1]{\noprint{#1}}
\newcommand{\figsetgrptitle}[1]{\noprint{#1}}
\newcommand{\figsetplot}[1]{\noprint{#1}}
\newcommand{\figsetgrpnote}[1]{\noprint{#1}}

\newcommand{\swift}{{\it Swift}}
\newcommand{\rosat}{{\it ROSAT}}
\newcommand{\xmm}{{\it XMM}}
\newcommand{\cxo}{{\it Chandra}}
\newcommand{\FeX}{$[$\ion{Fe}{10}$]$}
\newcommand{\total}{62}  
\newcommand{\totalswift}{52} 
\newcommand{\totalSSS}{26} 
\newcommand{\totallimitSSS}{38} 
\newcommand{\totalswiftSSS}{16} 
\newcommand{\totallimitswiftSSS}{19} 
\newcommand{\newturnoffperiod}{10} 
\newcommand{\newturnoffperiodlimit}{10} 

\newcommand{\btxt}[1]{{#1}}

\shorttitle{\swift II}
\shortauthors{Schwarz et al.}

\begin{document}

\title{\swift\ X-Ray Observations of Classical Novae. II. 
The Super Soft Source sample}

\author{Greg J. Schwarz}
\affil{American Astronomical Society, 2000 Florida Ave., NW, Suite 400, 
DC 20009-1231, Greg.Schwarz@aas.org}

\author{Jan-Uwe Ness}
\affil{XMM-Newton Science Operations Centre, ESAC, Apartado 78, 28691 
Villanueva de la Canada, Madrid, Spain}

\author{J.P. Osborne, K.L. Page, P.A. Evans, A.P. Beardmore}
\affil{Department of Physics and Astronomy, University of Leicester, 
LE1 7RH, UK}

\author{Frederick M. Walter}
\affil{Department of Physics and Astronomy, Stony Brook University, 
Stony Brook, NY, 11794-3800}

\author{L. Andrew Helton}
\affil{SOFIA Science Center, USRA, NASA Ames Research Center,
M.S. N211-3, Moffett Field, CA 94035}

\author{Charles E. Woodward}
\affil{Department of Astronomy, School of Physics and Astronomy, 
116 Church Street S.E., University of Minnesota, Minneapolis, MN 55455}

\author{Mike Bode}
\affil{Astrophysics Research Institute, Liverpool John Moores University,
Birkenhead, CH41 1LD, UK}

\author{Sumner Starrfield}
\affil{School of Earth and Space Exploration, Arizona State University,
P.O. Box 871404, Tempe, AZ 85287-1404}

\author{Jeremy J. Drake}
\affil{Smithsonian Astrophysical Observatory, 60 Garden St., MS 3, 
Cambridge, MA 02138}


\begin{abstract}
The \swift\ GRB satellite is an excellent facility for studying novae. Its 
rapid response time and sensitive X-ray detector provides an unparalleled 
opportunity to investigate the previously poorly sampled evolution of novae 
in the X-ray regime. 
This paper presents \swift\ observations of \totalswift\
Galactic/Magellanic Cloud novae.  We included the XRT (0.3-10 keV) 
X-ray instrument count rates and the UVOT (1700-8000\AA) filter photometry.  
Also included in the analysis are the publicly available pointed observations 
of 10 additional novae the X-ray archives.  This is the largest X-ray 
sample of Galactic/Magellanic Cloud novae yet assembled and consists of 
\totalSSS\ novae with super soft X-ray emission, 19 from \swift\ 
observations.  The data set shows that the faster novae have an early 
hard X-ray phase that is usually missing in slower novae.  The 
Super Soft X-ray phase occurs earlier and does not last as long in fast 
novae compared to slower novae.  All the \swift\ novae with sufficient 
observations show that novae are highly variable with rapid variability 
and different periodicities.  In the majority of cases, nuclear burning 
ceases less than 3 years after the outburst begins.  Previous relationships, 
such as the nuclear burning duration vs. t$_2$ or the expansion velocity
of the eject and nuclear burning duration 
vs. the orbital period, are shown to be poorly correlated with the full 
sample indicating that additional factors beyond the white dwarf mass and 
binary separation play important roles in the evolution of a nova outburst.
Finally, we confirm two optical phenomena that are correlated with strong, 
soft X-ray emission which can be used to further increase the efficiency of 
X-ray campaigns.
\end{abstract}

\keywords{novae, cataclysmic variables --- X-rays: stars --- ultraviolet: stars}

\section{INTRODUCTION}

Novae occur in binary systems in which a Roche lobe filling secondary is
losing hydrogen-rich material through the inner Lagrangian point onto a
white dwarf (WD) primary.  Mass transfer can also occur in long period 
systems if the secondary has a significant wind, {\it e.g.} the giant 
secondary in RS Oph or V407 Cyg.  Core material is mixed into the accreted 
material and is violently ejected into space when the pressure at the 
WD-accretion interface becomes great enough to initiate a thermonuclear 
runaway (TNR).  Novae eject, into the interstellar medium (ISM), a 
mixture of material accreted from the companion star, highly processed 
material from the underlying WD, and products of nucleosynthesis occurring 
during the TNR.  As a result of the TNR, up to 10$^{-4}$ M$_{\odot}$ of 
material can be ejected from the WD enriched in C, N, O, Ne, Mg, Al 
and other species 
\btxt{\citep{2006NuPhA.777..550J}}
at $v \sim 10^2 - 10^4$ km s$^{-1}$.  Any remaining 
hydrogen still bound to the WD continues to burn in hydrostatic 
equilibrium until it is consumed or ejected via a wind.

Initially, the radiative output of a nova occurs in the optical but as the
photosphere of the WD recedes, the spectral energy distribution
shifts to higher energies \citep{1978ARA&A..16..171G}.  The rate of the 
optical decline defines a nova's primary characteristics
\citep[{\it e.g.},][and references therein]{Warner2008}, namely the time 
to decline 2 magnitudes from visual maximum, t$_2$.  The decline 
rate depends on the amount of mass ejected, its velocity, composition,
and if it runs into circumbinary material.  The bolometric luminosity 
during the outburst is high, near or exceeding the Eddington limit
(for the fastest novae), and thus additional 
material is ejected via a strong stellar wind
\citep{1998MNRAS.300..931S,2001MNRAS.320..103S}.  In some novae the 
collision between this fast wind and the initial exploded mass or any
pre-existing circumbinary material can produce X-ray emission from shocks.  
The emission from this early X-ray phase is hard, has a low luminosity, 
of order 10$^{33-35}$ erg s$^{-1}$, and declines relatively rapidly 
\citep{1998ApJ...499..395B,2001MNRAS.326L..13O}.  As fuel continues to burn, 
mass loss causes the photosphere of the WD to shrink 
\citep{1985ApJ...294..263M}.  
\btxt{The effective temperature increases, 
peaking in the soft X-rays, at (2-8)$\times$10$^5$ K 
\citep{1996ApJ...456..788K,1996ApJ...463L..21S,2010ApJ...717..363R}}.  
Once the ejecta have cleared sufficiently, and if the line of sight extinction 
is not severe, some novae exhibit characteristics similar to the
Super Soft X-ray binary sources 
\citep[SSSs:][]{1997ARA&A..35...69K} with strong and soft, 
E$_{peak}$ $<$ 1 keV, X-ray emission.  This point in novae evolution is 
called the SSS phase.  At low spectral resolution, the UV/X-ray
spectral energy distributions (SED) resembles
blackbodies, but higher resolution \cxo\ or \xmm\ grating observations
reveal a significantly more complex picture.  The spectra frequently have
P-Cygni profiles or emission lines superimposed on a line blanketed
atmosphere.  Models sophisticated enough to interpret the high resolution
data are only now becoming available \citep{2010AN....331..175V}.  Once 
nuclear burning ends, the X-ray light curve rapidly declines as the WD cools 
marking the end of the SSS phase and the outburst.  At some point mass
transfer resumes and eventually another eruption occurs.  These are
called classical novae (CNe) until a second outburst is observed then they
become recurrent novae (RNe).  
\btxt{Detailed reviews of nova evolution are 
presented by \citet{Starrfield08} and \citet{2010AN....331..160B}.
\citet{2010AN....331..169H} discuss the theoretical implications of X-ray 
observations of novae while \citet{2010ApJS..187..275S} discusses the 
current understanding of the RN class.}

An important, but not the sole
driver of the nova phenomenon is the mass of the WD. 
Explosions on larger mass WDs expel less mass but at higher velocities. 
They have larger luminosities, are in outburst for less time, and (should)
have shorter recurrence times than novae on lower mass WDs.  
High mass ($>$ 1.25 M$_{\odot}$) WDs reach TNR ignition more rapidly
than low mass WDs and thus do not have the chance to accrete as much
material.  They also reach higher peak temperatures during the TNR leading
to a more energetic explosion.  However, other factors are believed to 
play important roles leading to a nova event.  These include
the composition of the WD, either CO or ONe, 
the initial temperature of the WD during accretion, the mass accretion 
rate \citep{2005ApJ...623..398Y}, the composition of the accreted 
material \citep{2000AIPC..522..379S}, and the mixing history
of the core/envelope. All impact  
how much mass can be transfered to the WD before 
an outburst begins.  Models show that different combinations of these 
characteristics can reproduce a wide range of nova outbursts 
\citep{2005ApJ...623..398Y,WoodStar11}. 
Unfortunately very few of these parameters
have been observationally verified in any nova.

The X-ray regime is a crucial component for the study of novae providing 
insight into TNR burning processes, WD mass and composition, accretion and 
mixing mechanisms, dust grain formation and destruction, and mass loss 
processes.  In addition, high mass novae such as RNe are potential
SN Ia progenitors via the single degenerate scenario 
\citep[e.g.][]{2008A&A...484L...9W,2010ASPC..429..173W,2010Ap&SS.329..287M}.
To make progress understanding the physics of these important astrophysical
phenomena, observations of a large number of novae are required 
to sample all the contributing 
factors. Prior to the launch of \swift\ the general X-ray temporal
evolution of novae was far from complete 
as only a few novae had been observed at more than one epoch in X-rays.  

\swift\ is an excellent facility for studying novae as it has a superb
soft X-ray response with its XRT instrument \citep{2005SSRv..120..165B}.
\citet{2007ApJ...663..505N} show how the XRT favorably compares with 
currently available X-ray instruments.  
\swift\ also has a co-aligned UV/optical instrument, UVOT
\citep[see][for details]{2005SSRv..120...95R}, which provides either 6 
filter photometry or low resolution grism spectroscopy.  The other 
\swift\ instrument is a $\gamma$-ray detector, BAT. 
\btxt{However, novae are generally not strong $\gamma$-ray 
sources \citep{2005NuPhA.758..721H}.}
The decay of $^{22}$Na
(half-life 2.6 yrs) generates a 1275 keV emission line but only $>$ 
1.25 M$_{\odot}$ WDs are predicted to produce sufficient 
$^{22}$Na during the TNR.  This line 
has not yet been definitively detected by satellites \citep{Hernanz08}
but there is a recent claim by \citet{2010ApJ...723L..84S}
that their models with Compton decay of $^{22}$Na can account for
the hard X-ray flux in V2491 Cyg provided an exceptionally large amount
of $^{22}$Na, 3$\times$10$^{-5}$M$_{\odot}$, was synthesized.
Another $\gamma$-ray emission mechanism is electron-positron 
annihilation very early in the outburst but this is expected to be
detectable only in nearby novae.  
The symbiotic RN RS Oph, at 1.6 kpc, was clearly detected 
in the lowest energy channels of the \swift/BAT \citep{2008A&A...485..223S},
but that emission is consistent with that from high temperature
shocks as the outburst ejecta plow into the pre-existing red giant
wind.  Recently the symbiotic RN V407 Cyg was detected in the GeV band 
by {\it Fermi-LAT} \citep{2010Sci...329..817A} only a few days after
visual maximum.  \citet{2010Sci...329..817A} show that the $\gamma$-ray
emission can be explained by either Compton scattering of infrared photons 
in the red giant wind or $\pi^0$ decay from proton-proton collisions.
\citet{2011arXiv1101.6013L} predict that $\pi^0$ $\gamma$-rays
will be created in the high circumbinary densities of very long orbital 
periods systems such as V407 Cyg, with a period of $\sim$ 43 years 
\citep{1990MNRAS.242..653M,2011arXiv1109.5397S}.

\swift\ has a rapid response ToO procedure and flexible scheduling 
which is critical in obtaining well sampled X-ray light curves of 
transient events.  Initial \swift\ results of 11 novae were 
presented in \citet{2007ApJ...663..505N}.
Since that time significantly more data have been obtained by the 
\swift\ Nova-CV group\footnote{The current members of the group and 
observation strategy are provided at \url{http://www.swift.ac.uk/nova-cv/}.}
which has devised an observing strategy to efficiently utilize the
satellite's unique capabilities and maximize the science return by 
observing interesting and bright novae with low extinction
recently discovered in the Milky Way and Magellanic Clouds.  In five 
years \swift\ has performed multiple visits for \totalswift\ classical and 
recurrent Galactic/Magellanic Cloud novae totaling well over 2 Ms of 
exposure time.  

Here we present a summary of all the Galactic/Magellanic Cloud \swift\
nova observations from launch (2004 November 20) to 2010 July 31 using
the XRT (0.3-10 keV) X-ray instrument (count rates and hardness ratios) 
and the available UVOT (1700-8000\AA) filter photometry. \swift\ 
observations of novae in the M31 group are reported in 
\citet{2010A&A...523A..89H}, \citet{2010AN....331..187P} and 
references within.  We combine the \swift\ Galactic/Magellanic Cloud 
data with archival pointed observations of CNe and RNe from \rosat, 
\xmm, \cxo, {\it BeppoSax}, {\it RXTE}, and {\it ASCA} to produce
the most comprehensive X-ray sample of local nova.  
The sample includes \totalSSS\ systems that were observed during the SSS phase.

In Section 2, we summarize the properties of the \total\ novae in the
X-ray sample.  The averaged \swift\ XRT count rates and UVOT magnitudes
for each observational session are also provided.  Studies of high 
frequency phenomena in individual objects are either left for future
work or have previously been published \citep[V458 Vul, V2941 Cyg, V598 Pup, 
RS Oph, V407 Cyg, and V723 Cas in][ respectively]{2009AJ....137.4160N,
2010MNRAS.401..121P,2009A&A...507..923P,2011ApJ...727..124O,
2011A&A...527A..98S,2008AJ....135.1328N}.  
Sections 3 and 4 detail the observations and results 
during the hard and SSS phases, respectively.  A discussion 
follows in Section 5 articulating trends between
the SSS duration and t$_2$, expansion velocity of the ejecta, and
orbital period plus the role of SSS emission in 
dust-forming novae.  Also included is a discussion on the origin of the 
different variability observed in the X-ray and UV light curves of the 
\swift\ sources.  Optical characteristics indicative of SSS emission
in CN and RN are also presented.  The last section, Section 6, 
provides a summary of this work.

\section{THE X-RAY DATA SET}

\subsection{Characteristics}

Table \ref{chartable} presents the primary characteristics of all the 
Galactic/Magellanic Cloud novae with pointed X-ray observations prior
to July 31st, 2010.  In addition to the \swift\ data, the sample includes 
all the publicly available pointed observations from the \rosat, \xmm, 
\cxo, {\it BeppoSax}, {\it RXTE}, and {\it ASCA} archives.
The columns give the nova name, visual magnitude at maximum, Julian date of
visual maximum, time to decline two magnitudes from visual maximum,
the Full-Width at Half-Maximum, FWHM, of H$\alpha$ or H$\beta$ taken near
visual maximum, E(B-V) and averaged Galactic hydrogen column density, N$_H$, 
along the line of sight, proposed orbital period, estimated distance, 
whether the nova was observed to form dust, and if the nova is a 
known RN.  The numbers in the parentheses are the literature 
references given in the table notes.  The names of novae with \swift\ 
observations are shown in {\it bold}.

\begin{deluxetable}{lllrrrrrrll}
\tablecaption{Observable Characteristics of 
Galactic/Magellanic Cloud novae with X-ray observations\label{chartable}}
\rotate
\tablewidth{700pt}
\tabletypesize{\scriptsize}
\tablehead{
\colhead{Name\tablenotemark{a}} & \colhead{V$_{max}$\tablenotemark{b}} & 
\colhead{Date\tablenotemark{c}} & \colhead{t$_2$\tablenotemark{d}} & 
\colhead{FWHM\tablenotemark{e}} & \colhead{E(B-V)} & 
\colhead{N$_H$\tablenotemark{f}} & \colhead{Period} & 
\colhead{D} & \colhead{Dust?\tablenotemark{g}} & \colhead{RN?} \\ 
\colhead{} & \colhead{(mag)} & \colhead{(JD)} & \colhead{(d)} & 
\colhead{(km s$^{-1}$)} & \colhead{(mag)} & \colhead{(cm$^{-2}$)} &
\colhead{(d)} & \colhead{(kpc)} & \colhead{} & \colhead{}
} 
\startdata
CI Aql & 8.83 (1) & 2451665.5 (1) & 32 (2) & 2300 (3) & 0.8$\pm0.2$ (4) & 1.2e+22 & 0.62 (4) & 6.25$\pm5$ (4) & N & Y \\
{\bf CSS081007}\tablenotemark{h} & \nodata & 2454596.5\tablenotemark{i} & \nodata & \nodata & 0.146 & 1.1e+21 & 1.77 (5) & 4.45$\pm1.95$ (6) & \nodata & \nodata \\
GQ Mus & 7.2 (7) & 2445352.5 (7) & 18 (7) & 1000 (8) & 0.45 (9) & 3.8e+21  & 0.059375 (10) & 4.8$\pm1$ (9) & N (7) & \nodata \\
IM Nor & 7.84 (11) & 2452289 (2) & 50 (2) & 1150 (12) & 0.8$\pm0.2$ (4) & 8e+21 & 0.102 (13) & 4.25$\pm3.4$ (4) & N & Y \\
{\bf KT Eri} & 5.42 (14) & 2455150.17 (14) & 6.6 (14) & 3000 (15) & 0.08 (15) & 5.5e+20 & \nodata & 6.5 (15) & N & M \\
{\bf LMC 1995} & 10.7 (16) & 2449778.5 (16) & 15$\pm2$ (17) & \nodata & 0.15 (203) & 7.8e+20  & \nodata & 50 & \nodata & \nodata \\
LMC 2000 & 11.45 (18) & 2451737.5 (18) & 9$\pm2$ (19) & 1700 (20) & 0.15 (203) & 7.8e+20  & \nodata & 50 & \nodata & \nodata \\
{\bf LMC 2005} & 11.5 (21) & 2453700.5 (21) & 63 (22) & 900 (23) & 0.15 (203) & 1e+21 & \nodata & 50  & M (24) & \nodata \\
{\bf LMC 2009a} & 10.6 (25) & 2454867.5 (25) & 4$\pm1$  & 3900 (25) & 0.15 (203)  & 5.7e+20 & 1.19 (26) & 50 & N & Y \\
{\bf SMC 2005} & 10.4 (27) & 2453588.5 (27) & \nodata & 3200 (28) & \nodata & 5e+20  & \nodata & 61 & \nodata & \nodata \\
{\bf QY Mus} & 8.1 (29) & 2454739.90 (29) & 60:  & \nodata & 0.71 (30) & 4.2e+21  & \nodata & \nodata & M & \nodata \\
{\bf RS Oph} & 4.5 (31) & 2453779.44 (14) & 7.9 (14) & 3930 (31) & 0.73 (32) & 2.25e+21 & 456 (33) & 1.6$\pm0.3$ (33) & N (34) & Y \\
{\bf U Sco} & 8.05 (35) & 2455224.94 (35) & 1.2 (36) & 7600 (37) & 0.2$\pm0.1$ (4) & 1.2e+21 & 1.23056 (36) & 12$\pm2$ (4) & N & Y \\
{\bf V1047 Cen} & 8.5 (38) & 2453614.5 (39) & 6 (40) & 840 (38) & \nodata & 1.4e+22  & \nodata & \nodata & \nodata & \nodata \\
{\bf V1065 Cen} & 8.2 (41) & 2454123.5 (41) & 11 (42) & 2700 (43) & 0.5$\pm0.1$ (42) & 3.75e+21 & \nodata & 9.05$\pm2.8$ (42) & Y (42) & \nodata \\
V1187 Sco & 7.4 (44) & 2453220.5 (44) & 7: (45) & 3000 (44) & 1.56 (44) & 8.0e+21 & \nodata & 4.9$\pm0.5$ (44) & N & \nodata \\
{\bf V1188 Sco} & 8.7 (46) & 2453577.5 (46) & 7 (40) & 1730 (47) & \nodata & 5.0e+21  & \nodata & 7.5 (39) & \nodata & \nodata \\
{\bf V1213 Cen} & 8.53 (48) & 2454959.5 (48) & 11$\pm2$ (49) & 2300 (50) & 2.07 (30) & 1.0e+22 & \nodata & \nodata & \nodata & \nodata \\
{\bf V1280 Sco} & 3.79 (51) & 2454147.65 (14) & 21 (52) & 640 (53) & 0.36 (54) & 1.6e+21  & \nodata & 1.6$\pm0.4$ (54) & Y (54) & \nodata \\
{\bf V1281 Sco} & 8.8 (55) & 2454152.21 (55) & 15:& 1800 (56) & 0.7 (57) & 3.2e+21 & \nodata & \nodata & N & \nodata \\
{\bf V1309 Sco} & 7.1 (58) & 2454714.5 (58) & 23$\pm2$ (59) & 670 (60) & 1.2 (30) & 4.0e+21 & \nodata & \nodata & \nodata & \nodata \\
{\bf V1494 Aql} & 3.8 (61) & 2451515.5 (61) & 6.6$\pm0.5$ (61) & 1200 (62) & 0.6 (63) & 3.6e+21  & 0.13467 (64) & 1.6$\pm0.1$ (63) & N & \nodata \\
{\bf V1663 Aql} & 10.5 (65) & 2453531.5 (65) & 17 (66) & 1900 (67) & 2: (68) & 1.6e+22  & \nodata & 8.9$\pm3.6$ (69) & N & \nodata \\
V1974 Cyg & 4.3 (70) & 2448654.5 (70) & 17 (71) & 2000 (19) & 0.36$\pm0.04$ (71) & 2.7e+21  & 0.081263 (70) & 1.8$\pm0.1$ (72) & N & \nodata \\
{\bf V2361 Cyg} & 9.3 (73) & 2453412.5 (73) & 6 (40) & 3200 (74) & 1.2: (75) & 7.0e+21 & \nodata & \nodata & Y (40) & \nodata \\
{\bf V2362 Cyg} & 7.8 (76) & 2453831.5 (76) & 9 (77) & 1850 (78) & 0.575$\pm0.015$ (79) & 4.4e+21  & 0.06577 (80) & 7.75$\pm3$ (77) & Y (81) & \nodata \\
{\bf V2467 Cyg} & 6.7 (82) & 2454176.27 (82) & 7 (83) & 950 (82) & 1.5 (84) & 1.4e+22  & 0.159 (85) & 3.1$\pm0.5$ (86) & M (87) & \nodata \\
{\bf V2468 Cyg} & 7.4 (88) & 2454534.2 (88) & 10: & 1000 (88) & 0.77 (89) & 1.0e+22  & 0.242 (90) & \nodata & N & \nodata \\
{\bf V2491 Cyg} & 7.54 (91) & 2454567.86 (91) & 4.6 (92) & 4860 (93) & 0.43 (94) & 4.7e+21  & 0.09580: (95) & 10.5 (96) & N & M \\
V2487 Oph & 9.5 (97) & 2450979.5 (97) & 6.3 (98) & 10000 (98) & 0.38$\pm0.08$ (98) & 2.0e+21 & \nodata & 27.5$\pm3$ (99) & N (100) & Y (101) \\
{\bf V2540 Oph} & 8.5 (102) & 2452295.5 (102) & \nodata & \nodata & \nodata & 2.3e+21 & 0.284781 (103) & 5.2$\pm0.8$ (103) & N & \nodata \\
V2575 Oph & 11.1 (104) & 2453778.8 (104) & 20: & 560 (104) & 1.4 (105) & 3.3e+21 & \nodata & \nodata & N (105) & \nodata \\
{\bf V2576 Oph} & 9.2 (106) & 2453832.5 (106) & 8: & 1470 (106) & 0.25 (107) & 2.6e+21  & \nodata & \nodata & N & \nodata \\
{\bf V2615 Oph} & 8.52 (108) & 2454187.5 (108) & 26.5 (108) & 800 (109) & 0.9 (108) & 3.1e+21  & \nodata & 3.7$\pm0.2$ (108) & Y (110) & \nodata \\
{\bf V2670 Oph} & 9.9 (111) & 2454613.11 (111) & 15: & 600 (112) & 1.3: (113) & 2.9e+21  & \nodata & \nodata & N (114) & \nodata \\
{\bf V2671 Oph} & 11.1 (115) & 2454617.5 (115) & 8: & 1210 (116) & 2.0 (117) & 3.3e+21  & \nodata & \nodata & M (117) & \nodata \\
{\bf V2672 Oph} & 10.0 (118) & 2455060.02 (118) & 2.3 (119) & 8000 (118) & 1.6$\pm0.1$ (119) & 4.0e+21  & \nodata & 19$\pm2$ (119) & \nodata & M \\
V351 Pup & 6.5 (120) & 2448617.5 (120) & 16 (121) & \nodata & 0.72$\pm0.1$ (122) & 6.2e+21 & 0.1182 (123) & 2.7$\pm0.7$ (122) & N & \nodata \\
{\bf V382 Nor} & 8.9 (124) & 2453447.5 (124) & 12 (40) & 1850 (23) & \nodata & 1.7e+22 & \nodata & \nodata & \nodata & \nodata \\
V382 Vel & 2.85 (125) & 2451320.5 (125) & 4.5 (126) & 2400 (126) & 0.05: (126) & 3.4e+21  & 0.146126 (127) & 1.68$\pm0.3$ (126) & N & \nodata \\
{\bf V407 Cyg} & 6.8 (128) & 2455266.314 (128) & 5.9 (129) & 2760 (129) & 0.5$\pm0.05$ (130) & 8.8e+21 & 15595 (131) & 2.7 (131) & \nodata & Y \\
{\bf V458 Vul} & 8.24 (132) & 2454322.39 (132) & 7 (133) & 1750 (134) & 0.6 (135) & 3.6e+21 & 0.06812255 (136) & 8.5$\pm1.8$ (133) & N (135) & \nodata \\
{\bf V459 Vul} & 7.57 (137) & 2454461.5 (137) & 18 (138) & 910 (139) & 1.0 (140) & 5.5e+21  & \nodata & 3.65$\pm1.35$ (138) & Y (140) & \nodata \\
V4633 Sgr & 7.8 (141) & 2450895.5 (141) & 19$\pm3$ (142) & 1700 (143) & 0.21 (142) & 1.4e+21  & 0.125576 (144) & 8.9$\pm2.5$ (142) & N & \nodata \\
{\bf V4643 Sgr} & 8.07 (145) & 2451965.867 (145) & 4.8 (146) & 4700 (147) & 1.67 (148) & 1.4e+22 & \nodata & 3 (148) & N & \nodata \\
{\bf V4743 Sgr} & 5.0 (149) & 2452537.5 (149) & 9 (150) & 2400 (149) & 0.25 (151) & 1.2e+21 & 0.281 (152) & 3.9$\pm0.3$ (151) & N & \nodata \\
{\bf V4745 Sgr} & 7.41 (153) & 2452747.5 (153) & 8.6 (154) & 1600 (155) & 0.1 (154) & 9.0e+20  & 0.20782 (156) & 14$\pm5$ (154) & \nodata & \nodata \\
{\bf V476 Sct} & 10.3 (157) & 2453643.5 (157) & 15 (158) & \nodata & 1.9 (158) & 1.2e+22  & \nodata & 4$\pm1$ (158) & M (159) & \nodata \\
{\bf V477 Sct} & 9.8 (160) & 2453655.5 (160) & 3 (160) & 2900 (161) & 1.2: (162) & 4e+21  & \nodata & \nodata & M (163) & \nodata \\
{\bf V5114 Sgr} & 8.38 (164) & 2453081.5 (164) & 11 (165) & 2000 (23) & \nodata & 1.5e+21  & \nodata & 7.7$\pm0.7$ (165) & N (166) & \nodata \\
{\bf V5115 Sgr} & 7.7 (167) & 2453459.5 (167) & 7 (40) & 1300 (168) & 0.53 (169) & 2.3e+21  & \nodata & \nodata & N (169) & \nodata \\
{\bf V5116 Sgr} & 8.15 (170) & 2453556.91 (170) & 6.5 (171) & 970 (172) & 0.25 (173) & 1.5e+21 & 0.1238 (171) & 11$\pm3$ (173) & N (174) & \nodata \\
{\bf V5558 Sgr} & 6.53 (175) & 2454291.5 (175) & 125 (176) & 1000 (177) & 0.80 (178) & 1.6e+22  & \nodata & 1.3$\pm0.3$ (176) & N (179) & \nodata \\
{\bf V5579 Sgr} & 5.56 (180) & 2454579.62 (180) & 7: & 1500 (23) & 1.2 (181) & 3.3e+21 & \nodata & \nodata & Y (181) & \nodata \\
{\bf V5583 Sgr} & 7.43 (182) & 2455051.07 (182) & 5: & 2300 (182) & 0.39 (30) & 2.0e+21 & \nodata & 10.5 & \nodata & \nodata \\
{\bf V574 Pup} & 6.93 (183) & 2453332.22 (183) & 13 (184) & 2800 (184) & 0.5$\pm0.1$  & 6.2e+21 & \nodata & 6.5$\pm1$  & M (185) & \nodata \\
{\bf V597 Pup} & 7.0 (186) & 2454418.75 (186) & 3: & 1800 (187) & 0.3 (188) & 5.0e+21  & 0.11119 (189) & \nodata & N (188) & \nodata \\
{\bf V598 Pup} & 3.46 (14) & 2454257.79 (14) & 9$\pm1$ (190) & \nodata & 0.16 (190) & 1.4e+21 & \nodata & 2.95$\pm0.8$ (190) & \nodata & \nodata \\
{\bf V679 Car} & 7.55 (191) & 2454797.77 (191) & 20: & \nodata & \nodata & 1.3e+22  & \nodata & \nodata & \nodata & \nodata \\
{\bf V723 Cas} & 7.1 (192) & 2450069.0 (192) & 263 (2) & 600 (193) & 0.5 (194) & 2.35e+21  & 0.69 (195) & 3.86$\pm0.23$ (196) & N & \nodata \\
V838 Her & 5 (197) & 2448340.5 (197) & 2 (198) & \nodata & 0.5$\pm0.1$ (198) & 2.6e+21  & 0.2975 (199) & 3$\pm1$ (198) & Y (200) & \nodata \\
{\bf XMMSL1 J06}\tablenotemark{j} & 12 (201) & 2453643.5 (202) & 8$\pm2$ (202) & \nodata & 0.15 (203) & 8.7e+20 & \nodata & 50 & \nodata & \nodata \\
\enddata
\tablenotetext{a}{Novae with \swift\ observations are presented in {\it bold}.}
\tablenotetext{b}{Visual maximum.}
\tablenotetext{c}{Date of visual maximum.}
\tablenotetext{d}{As measured from the visual light curve. A ":" indicates
an uncertain value due to an estimate from the AAVSO light curve.}
\tablenotetext{e}{Of Balmer lines measured at or near visual maximum.}
\tablenotetext{f}{Average Galactic N$_H$ within 0.5$\arcdeg$ of the nova 
position as given in the HEASARC N$_H$ tool.}
\tablenotetext{g}{Dust forming nova? (Y)es, (N)o, or (M)abye.
Novae with "N" but no dust reference were sufficiently observed
but no dust was specifically reported in any of the references.
}
\tablenotetext{h}{Full nova name is CSS081007030559+054715.}
\tablenotetext{i}{An averaged date based on available photometry.}
\tablenotetext{j}{Full nova name is XMMSL1 J060636.2-694933.} \\
\tablecomments{Numbers in parenthesis are the reference codes.} 
\tablerefs{
1 = \citet{2000IAUC.7411....3H};
2 = \citet{2010AJ....140...34S};
3 = \citet{2000IAUC.7409....1T};
4 = \citet{2010ApJS..187..275S};
5 = \citet{2010AN....331..156B};
6 = \citet{2008ATel.1847....1S};
7 = \citet{1984MNRAS.211..421W};
8 = \citet{1983IAUC.3766....1C};
9 = \citet{1984A&A...137..307K};
10 = \citet{1989ApJ...339L..41D};
11 = \citet{2002IAUC.7791....2K};
12 = \citet{2002IAUC.7799....3D};
13 = \citet{2003MNRAS.343..313W};
14 = \citet{2010ApJ...724..480H};
15 = \citet{2009ATel.2327....1R};
16 = \citet{1995IAUC.6143....2L};
17 = \citet{2004IBVS.5582....1L};
18 = \citet{2000IAUC.7453....1L};
19 = \citet{2003A&A...405..703G};
20 = \citet{2000IAUC.7457....1D};
21 = \citet{2005IAUC.8635....1L};
22 = \citet{2007JAVSO..35..359L};
23 = Average of our SMARTS spectra;
24 = Evidence from our SMARTS IR lightcurve;
25 = \citet{2009IAUC.9019....1L};
26 = \citet{2009ATel.2001....1B};
27 = \citet{2005IAUC.8582....2L};
28 = \citet{2005IAUC.8582....3M};
29 = \citet{2008IAUC.8990....2L};
30 = \citet{1998ApJ...500..525S};
31 = \citet{2007ApJ...665L..63B};
32 = \citet{1985IAUC.4067....2S};
33 = \citet{2008ApJ...673.1067N};
34 = \citet{2007ApJ...671L.157E};
35 = \citet{2010ATel.2419....1S};
36 = \citet{2001A&A...378..132E};
37 = \citet{2010ATel.2411....1A};
38 = \citet{2005IAUC.8596....1L};
39 = \citet{2007ApJ...663..505N};
40 = \citet{2007ApJ...662..552H};
41 = \citet{2007IAUC.8800....1L};
42 = \citet{2010AJ....140.1347H};
43 = \citet{2007IAUC.8800....2W};
44 = \citet{2006ApJ...638..987L};
45 = \citet{2004AAS...20515003S};
46 = \citet{2005IAUC.8574....1P};
47 = \citet{2005IAUC.8576....2N};
48 = \citet{2009IAUC.9043....1P};
49 = From AAVSO lightcurve;
50 = \citet{2009IAUC.9043....2P};
51 = \citet{2007IAUC.8807....1Y};
52 = \citet{2008MNRAS.391.1874D};
53 = \citet{2007CBET..852....1M};
54 = \citet{2008A&A...487..223C};
55 = \citet{2007IAUC.8810....1Y};
56 = \citet{2007IAUC.8812....2N};
57 = \citet{2007IAUC.8846....2R};
58 = \citet{2008IAUC.8972....1N};
59 = From AAVSO light curve;
60 = \citet{2008IAUC.8972....2N};
61 = \citet{2000A&A...355L...9K};
62 = \citet{1999IAUC.7324....1F};
63 = \citet{2003A&A...404..997I};
64 = \citet{2001IAUC.7665....2B};
65 = \citet{2005IAUC.8540....1P};
66 = \citet{2006JBAA..116..320B};
67 = Average of our SMARTS spectra;
68 = \citet{2005IAUC.8640....2P};
69 = \citet{2007ApJ...669.1150L};
70 = \citet{1994ApJ...431L..47D};
71 = \citet{1996AJ....111..869A};
72 = \citet{1997A&A...318..908C};
73 = \citet{2005IAUC.8483....1N};
74 = \citet{2005IAUC.8484....1N};
75 = \citet{2005IAUC.8524....2R};
76 = \citet{2006CBET..466....2W};
77 = \citet{2008A&A...479L..51K};
78 = \citet{2006ATel..792....1C};
79 = \citet{2006IAUC.8702....2S};
80 = \citet{2009ATel.2137....1B};
81 = \citet{2008AJ....136.1815L};
82 = \citet{2007IAUC.8821....1N};
83 = \citet{2009AAS...21349125L};
84 = \citet{2007IAUC.8848....1M};
85 = \citet{2008ATel.1723....1S};
86 = \citet{2009AN....330...77P};
87 = \citet{WoodStar11};
88 = \citet{2008IAUC.8927....2N};
89 = \citet{2008IAUC.8936....2R};
90 = \citet{2009ATel.2157....1S};
91 = \citet{2008IAUC.8934....1N};
92 = \citet{2008ATel.1485....1T};
93 = \citet{2008ATel.1475....1T};
94 = \citet{2008IAUC.8938....2R};
95 = \citet{2008ATel.1514....1B} but Darnley et al. (2011, submitted) do not find convincing evidence of this period in their data.;
96 = \citet{2008CBET.1379....1H};
97 = \citet{1998IAUC.6941....1N};
98 = \citet{2000ApJ...541..791L};
99 = \citet{2000ApJ...541..791L};
100 = \citet{1998IAUC.7049....1R};
101 = \citet{2009AJ....138.1230P};
102 = \citet{2002IAUC.7809....1R};
103 = \citet{2005PASA...22..298A};
104 = \citet{2006IAUC.8671....1P};
105 = \citet{2006IAUC.8710....2R};
106 = \citet{2006IAUC.8700....1W};
107 = \citet{2006IAUC.8730....5L};
108 = \citet{2008MNRAS.387..344M};
109 = \citet{2007IAUC.8824....1N};
110 = \citet{2007IAUC.8846....2R};
111 = \citet{2008IAUC.8948....2A};
112 = \citet{2008IAUC.8948....3N};
113 = \citet{2008IAUC.8956....1R};
114 = \citet{2008IAUC.8998....3S};
115 = \citet{2008IAUC.8950....1N};
116 = \citet{2008CBET.1448....1H};
117 = \citet{2008IAUC.8957....1R};
118 = \citet{2009IAUC.9064....1N};
119 = \citet{2010MNRAS.tmp.1484M};
120 = \citet{1992IAUC.5422....1C};
121 = \citet{1996ApJ...466..410O};
122 = \citet{1996MNRAS.279..280S};
123 = \citet{2001MNRAS.328..159W};
124 = \citet{2005IAUC.8497....1L};
125 = \citet{1999IAUC.7176....1L};
126 = \citet{2002A&A...390..155D};
127 = \citet{2006AJ....131.2628B};
128 = \citet{2010CBET.2199....1H};
129 = \citet{2011MNRAS.410L..52M};
130 = \citet{2011A&A...527A..98S};
131 = \citet{1990MNRAS.242..653M};
132 = \citet{2007CBET.1029....1M};
133 = \citet{2008Ap&SS.315...79P};
134 = \citet{2007IAUC.8862....2B};
135 = \citet{2007IAUC.8883....1L};
136 = \citet{2010MNRAS.407L..21R};
137 = \citet{2007CBET.1183....1M};
138 = \citet{2010NewA...15..170P};
139 = \citet{2007CBET.1181....1N};
140 = \citet{2008IAUC.8936....3R};
141 = \citet{1998IAUC.6847....2N};
142 = \citet{2001MNRAS.328.1169L};
143 = \citet{1998IAUC.6848....1D};
144 = \citet{2008MNRAS.387..289L};
145 = \citet{2001IAUC.7590....2K};
146 = \citet{2001IBVS.5138....1B};
147 = \citet{2001IAUC.7589....2A};
148 = \citet{2008AstL...34..249B};
149 = \citet{2002IAUC.7975....2K};
150 = \citet{2003MNRAS.344..521M};
151 = \citet{2007AAS...210.0402V};
152 = \citet{2003IAUC.8176....3W};
153 = \citet{2003IAUC.8126....2L};
154 = \citet{2005A&A...429..599C};
155 = \citet{2003IAUC.8132....2K};
156 = \citet{2006MNRAS.371..459D};
157 = \citet{2005IAUC.8607....1S};
158 = \citet{2006MNRAS.369.1755M};
159 = \citet{2005IAUC.8638....1P};
160 = \citet{2006A&A...452..567M};
161 = \citet{2005IAUC.8617....1P};
162 = \citet{2005IAUC.8644....1M};
163 = \citet{2005IAUC.8644....1M};
164 = \citet{2004IAUC.8306....1N};
165 = \citet{2006A&A...459..875E};
166 = \citet{2004IAUC.8368....3L};
167 = \citet{2005IAUC.8502....1S};
168 = \citet{2005IAUC.8501....2K};
169 = \citet{2005IAUC.8523....4R};
170 = \citet{2005IAUC.8559....2G};
171 = \citet{2008A&A...478..815D};
172 = \citet{2005IAUC.8559....1L};
173 = \citet{2008ApJ...675L..93S};
174 = \citet{2005IAUC.8579....3R};
175 = \citet{2007CBET.1010....1M};
176 = \citet{2008NewA...13..557P};
177 = \citet{2007CBET.1006....1I};
178 = \citet{2007IAUC.8884....2R};
179 = \citet{2009AAS...21442806R};
180 = \citet{2008CBET.1352....1M};
181 = \citet{2008IAUC.8948....1R};
182 = \citet{2009IAUC.9061....1N};
183 = \citet{2004IAUC.8445....3S};
184 = \citet{2005IBVS.5638....1S};
185 = \citet{Heltonthesis};
186 = \citet{2007IAUC.8895....1P};
187 = \citet{2007IAUC.8896....2N};
188 = \citet{2008IAUC.8911....2N};
189 = \citet{2009MNRAS.397..979W};
190 = \citet{2008A&A...482L...1R};
191 = \citet{2008IAUC.8999....1W};
192 = \citet{1996A&A...315..166M};
193 = \citet{1996IAUC.6365....2I};
194 = \citet{2008AJ....135.1328N};
195 = \citet{2007AstBu..62..125G};
196 = \citet{2009AJ....138.1090L};
197 = \citet{1991IAUC.5222....1S};
198 = \citet{1996MNRAS.282..563V};
199 = \citet{1994ApJ...420..830S};
200 = \citet{1992ApJ...384L..41W};
201 = \citet{2009A&A...506.1309R};
202 = \citet{2009A&A...506.1309R};
203 = Standard LMC value.
}
\end{deluxetable}

Although P-Cygni absorption profiles provide the best values for
the early velocities of the ejecta, they are not nearly as well reported 
in the literature as FWHMs of Balmer lines near maximum.  Since nearly 
every nova has a FWHM citation as part of the spectroscopic confirmation of 
the initial visual detection, they are used as the expansion velocity
proxy.  Expansion velocities provide another way to classify a nova since 
more massive WDs eject less mass and at a greater velocity than low mass WDs.  
This characteristic can be preferable to t$_2$ since the rate of decline 
can be difficult to determine for novae with secondary maxima, dust 
formation, or that have poorly sampled early light curves.  Both FWHM 
and t$_2$ are used as simple proxies for the WD mass.
\footnote{While many other parameters also affect these observables, such as 
the accretion rate, these parameters are generally not known for specific
novae and thus their contributions to the secular evolution can not be
determined.}
The N$_H$ values were 
obtained from the HEASARC N$_H$ 
tool\footnote{http://heasarc.gsfc.nasa.gov/cgi-bin/Tools/w3nh/w3nh.pl}
using the averaged LAB \citep{2005A&A...440..775K} and DL 
\citep{1990ARA&A..28..215D} maps within a 0.5$\arcdeg$ area around each nova.

\begin{figure*}[htbp]
\plotone{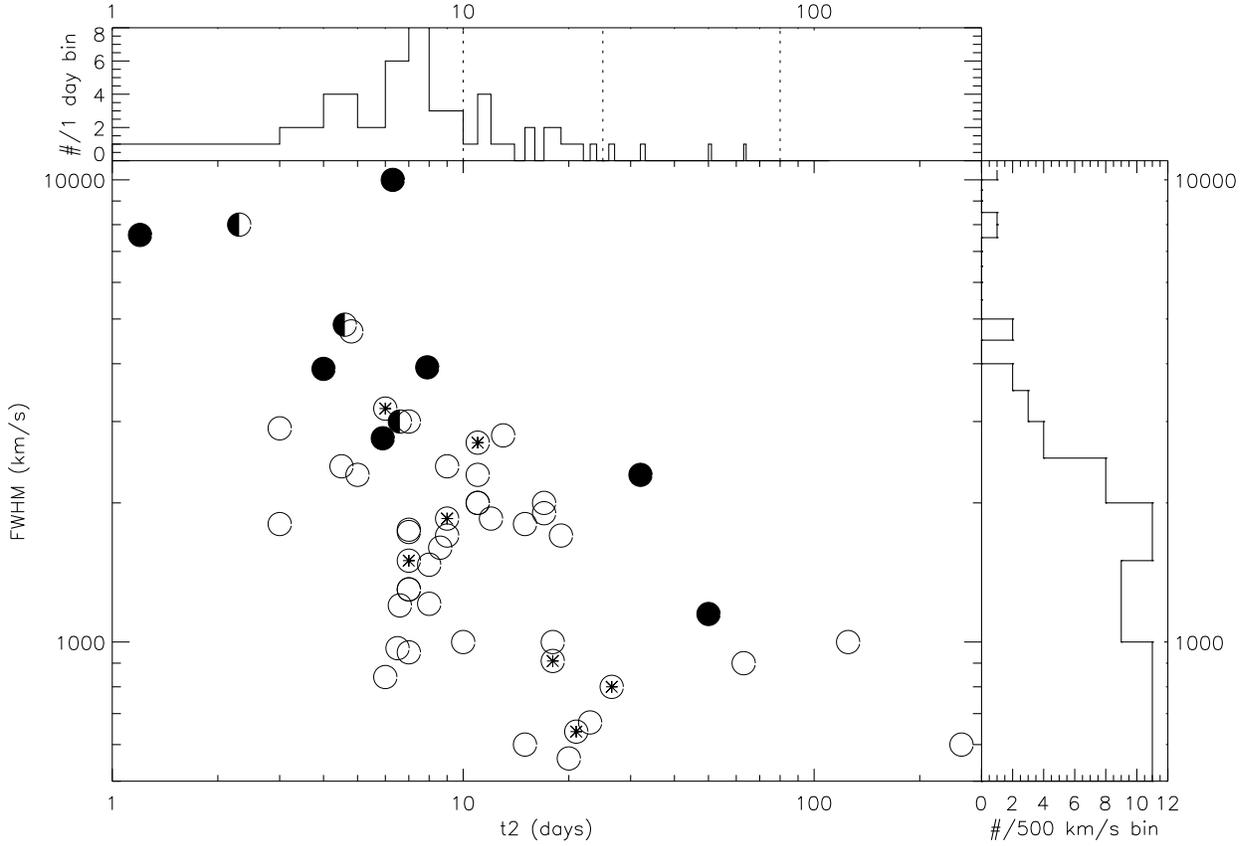}
\caption{The t$_2$ vs. FWHM near maximum for the novae in the sample.  Filled
circles are known RNe.  Half filled circles are suspected
RNe based on their characteristics.  Circles with asterisks inside
indicate dusty novae.  The distribution histograms for t$_2$ and the FWHM 
are also shown in the secondary graphs.  The dotted lines in the t$_2$ 
histogram show the boundaries between the "very fast", "fast", "moderately 
fast", and "slow" light curve classifications \citep{Warner2008}.  The 
majority of our novae belong to the "fast" or "very fast" classifications.
\label{t2vsfwhm}}
\end{figure*}

Figure \ref{t2vsfwhm} shows the t$_2$ and FWHM distribution for all the novae 
with both values.  The 7 filled circles are known RNe while the 3
half filled circles are suspected RNe.  Dusty novae have an asterisk
inside their circle symbols.  As expected, the RNe tend 
toward large FWHM and fast t$_2$ times.  In this sample the dusty novae 
are scattered throughout the FWHM-t$_2$ phase space showing no particular 
preference for any type of nova.  Figure \ref{t2vsfwhm} also shows
that there is a wide dispersion between FWHM and t$_2$, e.g. novae
with t$_2$ of 10 days have FWHM values between 1000 and 3000 km s$^{-1}$.

The top panel of Figure \ref{t2vsfwhm} shows the distribution of t$_2$ in 
one day bins.  Using the light curve classifications of \citet{Warner2008},
the sample is heavily weighted toward very fast (t$_2$ $<$ 10 days) and 
fast (11 $>$ t$_2$ $<$ 25 days) novae.  These are intrinsically more luminous,
with a larger rise from quiescence to maximum light.  The peak is 
at 8 days and with a median t$_2$ of 9 days.  There are only 5 novae 
in the entire sample with t$_2$ times greater than 50 days, 
IM Nor, LMC 2005, QY Mus, V723 Cas, and V5558 Sgr. The far
right panel in Figure \ref{t2vsfwhm} gives the distribution for 
FWHM in 500 km s$^{-1}$ bins.  The majority of the novae in the sample
have low expansion velocities with the peak in the 1500-2000 km s$^{-1}$ bin. 
The median FWHM is 1800 km s$^{-1}$.  There are only 5 novae with 
FWHM $\geq$ 4000 km s$^{-1}$ in the sample and all but V4643 Sgr 
are RNe or suspected RNe.

The X-ray sample is biased toward fast novae for multiple reasons. 
The bulk of the observations are from \swift, and \swift\ has only been 
operational for 5 years.  Fast systems, like the RN RS Oph will rise 
and fall on time-scales of months (see Section \ref{var}) while slow 
novae, such as V1280 Sco, have not yet had sufficient time to evolve 
into soft X-ray sources (and may not) thus are therefore underrepresented. 
Slow novae also require more observing time 
to be monitored over their lifetime, particularly if the same coverage of
the X-ray evolution is desired.  Allocations of \swift\ observing time
over multiple cycles are difficult to justify and execute unless a compelling
scientific rationale is forthcoming, such as unusual or significant 
spectral variations (see Section \ref{fex}), count rate oscillations, 
abundance pattern changes, etc.
Slow and old novae (many tens of months post-outburst)
are generally sampled once a year in part due to their slow evolution.
However, the main reason the sample depicted in Fig. \ref{t2vsfwhm}
favors fast novae is due to the strong 
selection effect toward outbursts on high mass WDs.  While high mass WDs, 
{\it e.g.} $\geq$ 1.2 M$_{\odot}$, are relatively rare in the field,
the time-scale between outbursts is significantly shorter than for low-mass 
WDs, meaning they dominate any observational sample
\citep{1994ApJ...425..797L}.  Finally, high mass WDs give rise to more
luminous outbursts and the \swift\ Nova-CV group has a V $<$ 8 magnitude
selection criterion which leads to preferentially selecting brighter sources.

\begin{figure*}[htbp]
\plotone{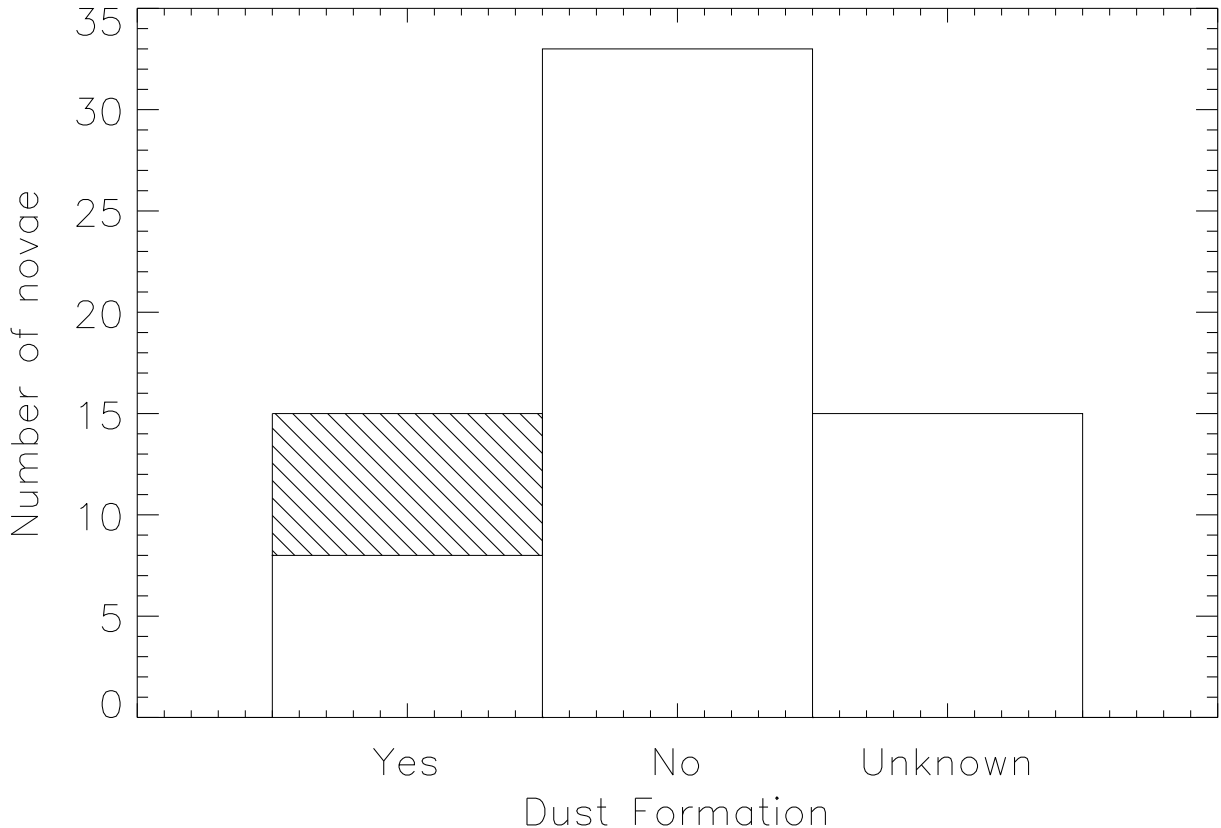}
\caption{The distribution of dusty novae in the sample.  The cross-hashed
region is for novae that showed strong dust characteristics; however, the
presence of dust in these systems has not been spectrophotometrically 
corroborated at IR wavelengths. The majority of novae in our X-ray
selected sample (Table \ref{chartable}) did not form dust.
\label{dusthistogram}}
\end{figure*}

Figure \ref{dusthistogram} shows the distribution of our sample 
(Table \ref{chartable}) with respect to dust formation frequency.  
Only $\sim$ 16\% of the novae in the sample had clear indications
in the literature of dust formation in the ejecta.  The dust formation 
frequency increases to 31\% when including the 7 novae where dust 
likely formed based on characteristics of the visual light curve but not 
yet confirmed by a measured SED excess in the thermal- and mid-infrared, 
{\it e.g.} the "maybe"s in column 10 of Table \ref{chartable}.
This is consistent with the expectations of the general population
of dusty novae which ranges from 18\% to $\gtrsim 40$\%.  The lower limit
is set by \citet{2010AJ....140...34S} who find that 93 well sampled 
American Association of Variable Star Observers (AAVSO) novae have
the large dip in their visual light curves indicative of strong 
dust formation \citep[see][]{1998PASP..110....3G}.  The upper limit
is from a recent \textit{Spitzer} survey of IR bright novae that finds
many novae have weak dust emission signatures with little or no dip in the 
visual light curve especially at late epochs (many 100s of days post-outburst) 
when emission from the dust envelope is a few $\mu$Jy
\citep{WoodStar11,Heltonthesis}

In order to obtain the best X-ray and UV data, it is desirable to target
novae with low extinction along the line of sight.  However, determining
the extinction early in the outburst is challenging.  N$_H$ maps are crude
since they sample large regions of the sky.  The region size used to 
derive the N$_H$ values in Table \ref{chartable} was 0.5$^{\arcdeg}$.
Typically just a handful of sight lines are available in regions of this
relatively small size.  The problem is exacerbated in inhomogeneous areas 
like the Galactic plane where most novae are found.  The 
extinction maps of \citet{1998ApJ...500..525S} can be used to obtain E(B-V) 
since their spatial resolution is significantly higher. However, the
\citet{1998ApJ...500..525S} maps suffer from large errors in the Galactic
plane, $|b| <$ 5$^{\arcdeg}$.  Maps also give the total Galactic line of
sight with no information versus distance and thus provide only an upper 
limit.  E(B-V) can also be determined from
indirect methods but these require either high resolution spectroscopy 
to measure ISM absorption lines 
\citep[{\it e.g.} \ion{Na}{1}~D $\lambda$5890\AA;][]{munari97}, 
the line strengths of optical and near-IR spectroscopy of \ion{O}{1} lines
\citep{rudy89}, or extensive B and V photometry during
the early outburst \citep[{\it e.g.} intrinsic (B-V) at V$_{max}$ or
t$_2$;][]{1987A&AS...70..125V}.  Finally, 
E(B-V) estimates can be affected by other factors occurring during 
the outburst such as dust formation or intrinsic absorption from the ejecta
while the expanding material is still dense.

\begin{figure*}[htbp]
\plotone{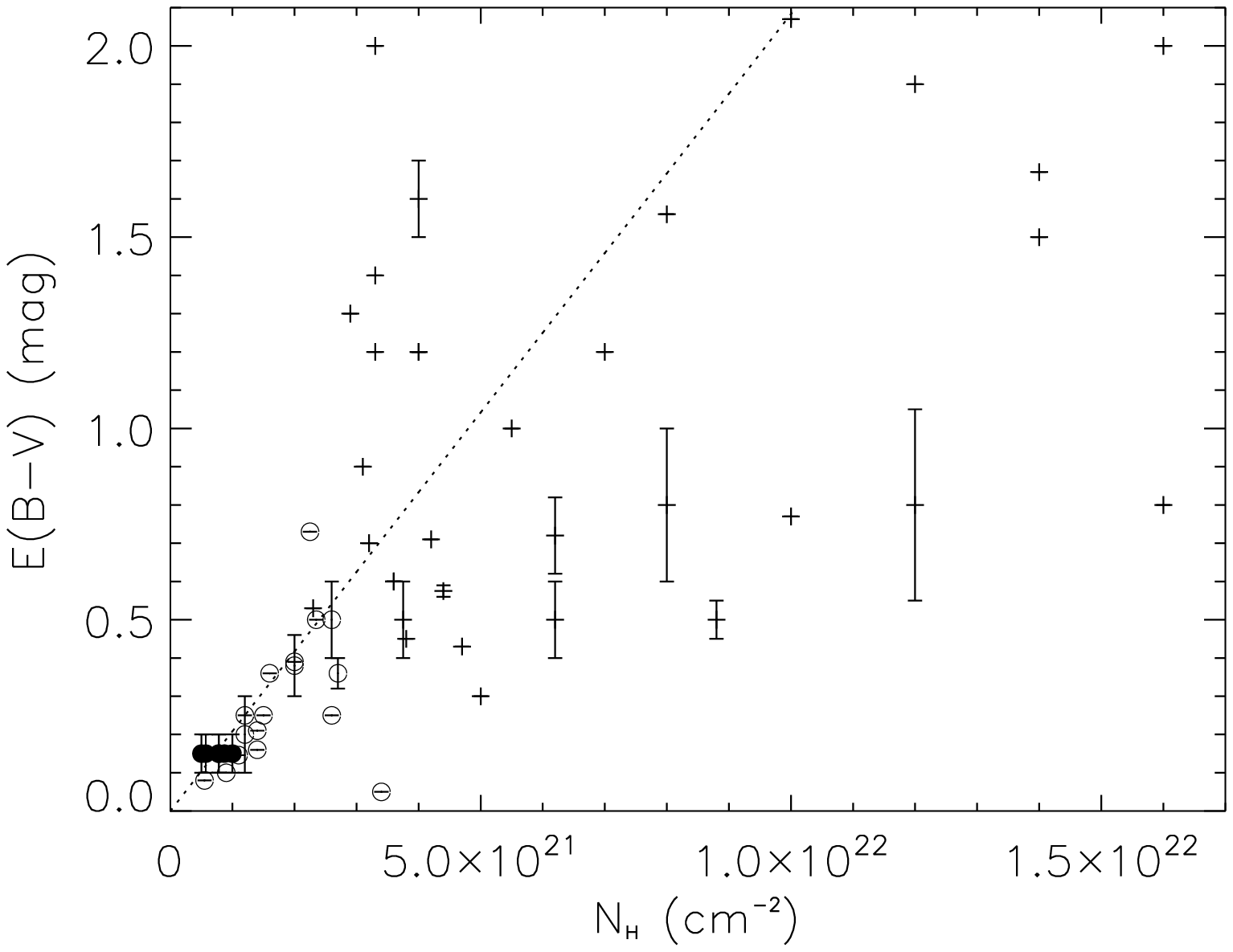}
\caption{Local N$_H$ value versus the estimated E($B-V$). The values
are from Table~\ref{chartable}. The dotted line shows the E(B-V) vs. N$_H$ 
relationship of \citet{1978ApJ...224..132B}. Circles are $|b| \geq$ 
5$^{\arcdeg}$ novae, the $|b| <$ 5$^{\arcdeg}$ novae are shown as pluses,
and filled circles are Magellanic novae.  Errors bars are given when
available.
\label{ebmvnh}}
\end{figure*}

It is therefore desirable to check 
that the general relationship between N$_H$ and E(B-V) holds for novae.  
Figure \ref{ebmvnh} shows N$_H$ versus E(B-V) for the novae in this 
paper with the dotted line showing the average Milky Way extinction law, 
E(B-V) = N$_H$/4.8$\times$10$^{21}$ \citep{1978ApJ...224..132B}, as the 
dotted line.  The circles represent novae with Galactic latitudes, 
$|b| \geq$ 5$^{\arcdeg}$ while pluses are novae found within the disk,
$|b| <$ 5$^{\arcdeg}$.  Filled circles are Magellanic novae.  Errors are 
present on all sources when available in the literature.  There is good 
agreement with the relationship for novae with E(B-V) $\leq$0.6 and 
N$_H$ $\leq$ 2.9$\times$10$^{21}$ with a correlation coefficient of 0.85. 
These are primarily novae found outside of the galactic disk and thus
fit the relationship well.  Novae with these low extinction values and
column densities are ideal for soft X-ray detection.  The relationship 
breaks down at larger values with a lower correlation coefficient of 0.64 
for the entire sample as it is dominated by novae embedded within
the Galactic disk.  Novae with E(B-V) values greater than 1.5 generally 
make poor SSS candidates due to the large extinction.

The maximum magnitude vs. rate of decline relationship of
\citet{1995ApJ...452..704D} provides an estimate of the distances
for the Galactic novae in Table \ref{chartable}.  The distance estimate
range extends from the relatively nearby V1280 Sco ($\sim$ 1 kpc) 
to the other side of the Galaxy for V2576 Oph ($\sim$ 28 kpc).
The median Galactic distance from this relationship is 5.5 kpc
for this sample.

\subsection{X-ray evolution\label{xrayevol}}

All the available \swift\ XRT and UVOT data of novae in the public archive
up to 2010 July 31 are presented in Table \ref{fullswift}.  The data were
primarily obtained from pointed observations but a few serendipitous 
observations are also included.  The full data 
set is available in the electronic edition with only V1281 Sco shown as an
example here.  The columns provide the \swift\ observation identification,
exposure time, day of the observation from visual maximum (see Table 
\ref{chartable}), XRT total (0.3-10 keV) count rate, the Hard (1-10 keV)
to Soft (0.3-1 keV) hardness ratio, HR1, the Soft and Hard band count rates,
the (Hard-Soft)/(Hard+Soft) hardness ratio, HR2, and the uvw2 
($\lambda_c$ = 1928\AA), uvm2 (2246\AA), uvw1 (2600\AA), $u$ (3465\AA), 
$b$ (4392\AA), and $v$ (5468\AA) UVOT filter magnitudes if available.  
The UVOT magnitudes do not include the systematic photometric calibration 
errors from \citet[][Table 6]{2008MNRAS.383..627P}.

There is one row in the table per observation ID, however this is not a 
fixed unit of time; most observation IDs are less than 0.13 days 
duration and the median exposure time is 1.76 ks. For this exposure time, 
our 3 sigma detection limit is 0.0037 counts s$^{-1}$ (0.3-10 keV, corrected 
for typical PSF coverage). This corresponds to an unabsorbed flux limit in 
the same band, assuming absorption by N$_H$ = 3$\times$ 10$^{20}$ of 
1.5$\times$ 10$^{-13}$ and 2.0$\times$ 10$^{-13}$ erg cm$^{-2}$ s$^{-1}$ for
a 5 keV optically thin thermal spectrum and a 50 eV blackbody spectrum, 
respectively.

To create a self consistent dataset for Table~\ref{fullswift} we used the 
software described by \citet{2009MNRAS.397.1177E,2007A&A...469..379E}.
This extracts source and background event lists from the data (using an 
annular source region where necessary to eliminate pile up), and then bins 
these data to form the light curve, applying corrections for pile up, bad 
pixels and the finite size of the source region as necessary. 
Since novae tend to be soft, we chose the energy bands for the
hardness ratio to be 0.3--1 keV and 1--10 keV. There is also evidence
that, for very soft sources, pile up occurs at lower count rates than for
hard sources; we thus set the threshold at which pile up is considered a
risk to be 0.3 (80) count s$^{-1}$ in PC (WT) mode \citep[the defaults from]
[are 0.6 and 150 respectively]{2007A&A...469..379E}.

We chose to group the data in one bin per \swift\ observation. In the
current version of the online software (for this binning mode), background
subtraction is only carried out using Gaussian statistics, and does not
produce upper limits if this results in a non-detection. We thus took
the `detailed' light curves produced by the web tools, which include the
number of measured counts in each bin, the exposure time, and the
correction factor (accounting for pile up etc.). Following the approach
of \citet{2007A&A...469..379E} for other binning methods, where any bins had
fewer than 15 detected source counts, we used the Bayesian method of
\citet{1991ApJ...374..344K} to determine whether the source was
detected at the 3-sigma level. If this was not the case, a 3-$\sigma$
upper limit was produced using this Bayesian method, otherwise a data
point with standard 1-$\sigma$ uncertainty was produced using the 
\citet{1991ApJ...374..344K} approach.

The hardness ratios were always calculated using Gaussian statistics, 
unless one band had zero detected source photons: in this case no ratio
could be produced.  The hardness ratios were defined as HR1 = H/S and 
HR2 = (H-S)/(H+S) where H = 1.0-10 keV and S = 0.3-1.0 keV.

\begin{deluxetable}{cccccccccccccc}
\tablecaption{\swift\ XRT/UVOT data for novae in the archive\label{fullswift}}
\tablewidth{0pt}
\tabletypesize{\scriptsize}
\rotate
\tablecolumns{14}
\tablehead{
\colhead{ObsID} & \colhead{Exp} & \colhead{Day\tablenotemark{a}} & 
\colhead{CR\tablenotemark{b}} & 
\colhead{HR1\tablenotemark{c}} & \colhead{Soft} &
\colhead{Hard} & \colhead{HR2\tablenotemark{c}} &
\colhead{uvw2} & \colhead{uvm2} & \colhead{uvw1} & \colhead{u} & 
\colhead{b} & \colhead{v} \\ 
\colhead{} & \colhead{(ksec)} & \colhead{(d)} & \colhead{(ct s$^{-1}$)} &
\colhead{} & \colhead{(ct s$^{-1}$)} & \colhead{(ct s$^{-1}$)} &
\colhead{} & \colhead{(mag)} & \colhead{(mag)} & \colhead{(mag)} & 
\colhead{(mag)} & \colhead{(mag)} & \colhead{(mag)}
}
\startdata
\cutinhead{V1281\,Sco}
00030891001 & 3.87&   2.95&$<$ 0.0030                       & \nodata           & \nodata&\nodata&\nodata&\nodata&\nodata &\nodata&\nodata&\nodata&\nodata \\
00037164001 & 5.24& 338.66&    0.1634$^{+0.0079}_{-0.0079}$ & 0.0090$\pm$0.0074 & 0.1619& 0.0015 &-0.98  &\nodata&\nodata &\nodata&\nodata&\nodata&\nodata \\
00037164002 & 3.45& 344.07&    0.2429$^{+0.0120}_{-0.0120}$ & 0.0062$\pm$0.0057 & 0.2414& 0.0015 &-0.99  &19.50  &19.64   &18.20  &\nodata&\nodata&\nodata \\
00037164003 & 4.24& 351.05&    0.6376$^{+0.0282}_{-0.0282}$ & 0.0047$\pm$0.0053 & 0.6346& 0.0030 &-0.99  &20.32  &\nodata &\nodata&\nodata&\nodata&\nodata \\
00037164005 & 1.66& 361.10&    0.2727$^{+0.0185}_{-0.0185}$ & 0.0012$\pm$0.0081 & 0.2723& 0.0003 &-1.00  &20.43  &\nodata &\nodata&\nodata&\nodata&\nodata \\
00037164006 & 2.89& 366.41&    0.2284$^{+0.0129}_{-0.0129}$ & 0.0097$\pm$0.0089 & 0.2262& 0.0022 &-0.98  &20.42  &\nodata &\nodata&\nodata&\nodata&\nodata \\
00037164007 & 2.02& 432.69&    0.0853$^{+0.0079}_{-0.0079}$ & 0.0002$\pm$0.0063 & 0.0853& 0.0000 &-1.00  &20.22  &\nodata &19.11  &\nodata&\nodata&\nodata \\
00090248001 & 4.68& 819.99&$<$ 0.0013                       & \nodata           & \nodata&\nodata&\nodata&20.32  &$>$20.56&20.07  &\nodata&\nodata&19.30 \\
\enddata
\tablenotetext{a}{Days after visual maximum, see Table \ref{chartable}.}
\tablenotetext{b}{corrected for PSF losses and bad columns.
The 3 sigma upper limits are given when there is no detection 
better than 3 sigma.}
\tablenotetext{c}{Hardness ratios defined as HR1=H/S and HR2=(H-S)/(H+S) with 
Hard(H)=1-10\,keV and Soft(S)=0.3-1\,keV}
\tablecomments{Table \ref{fullswift} is published in its entirety in the 
electronic edition of the {\it Astrophysical Journal}.  A portion is shown 
here for guidance regarding its form and content.}
\end{deluxetable}

\begin{figure*}[htbp]
\includegraphics[angle=90,scale=0.60]{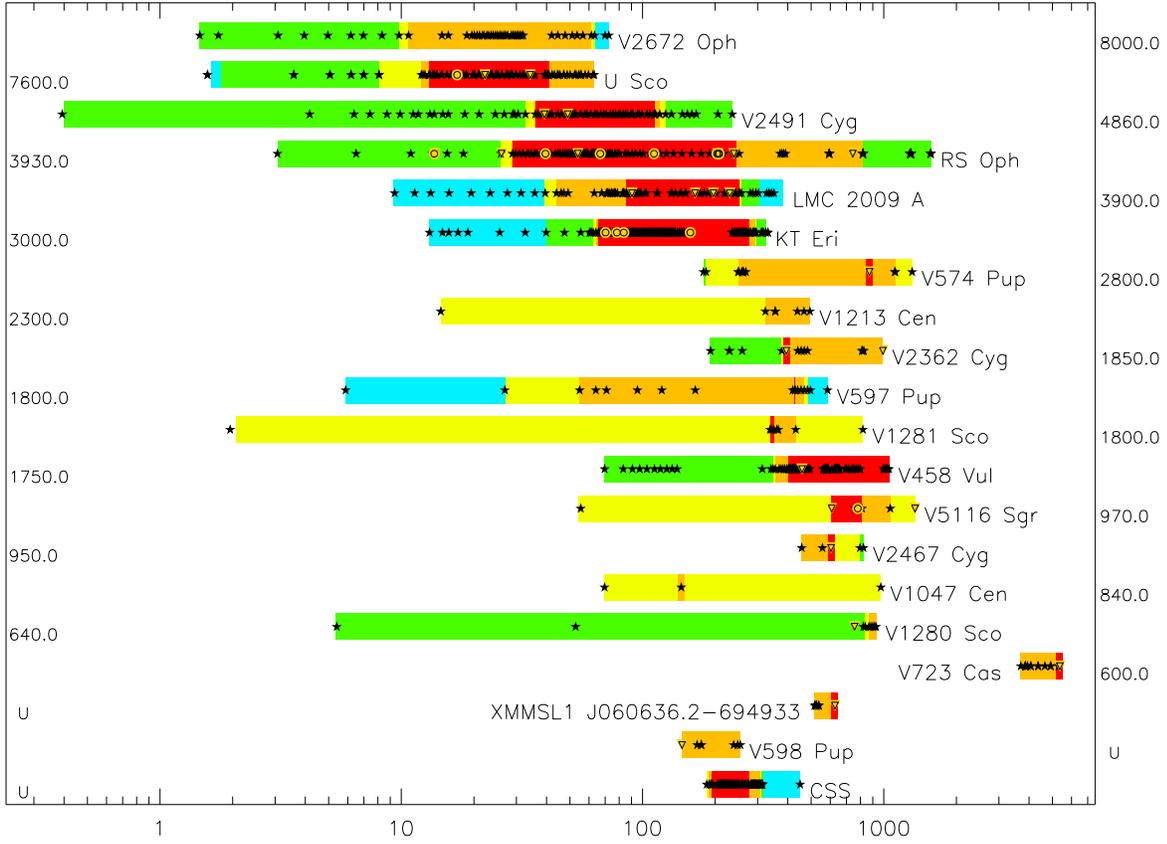}
\caption{The X-ray epochs for \swift\ sources with the best SSS
phase coverage.
The novae are arranged by increasing optical emission line FWHM with 
the FWHM values shown either left or right of the source.  "U" is used 
for novae with unknown FWHM velocities.  Refer to Table \ref{colordescriptions}
for a summary of the color coding.  \label{sssgood}}
\end{figure*}

\begin{figure*}[htbp]
\includegraphics[angle=90,scale=0.60]{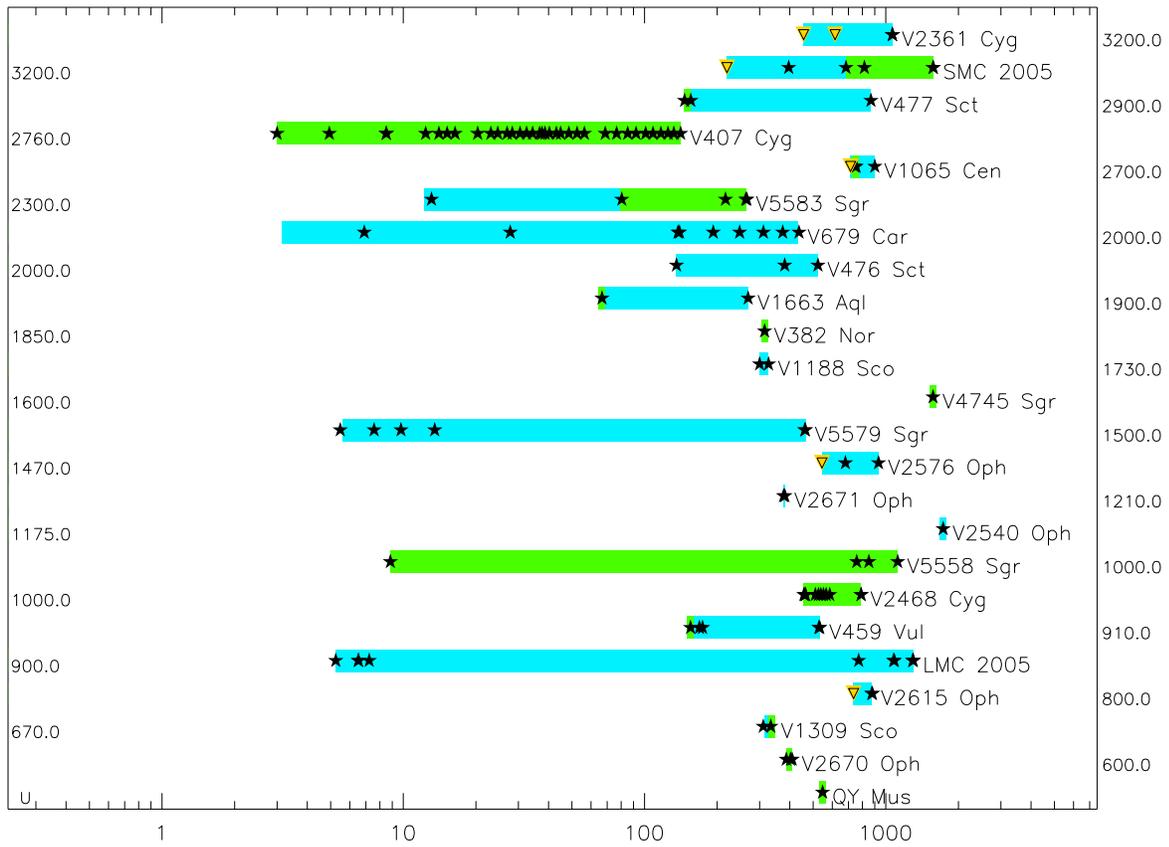}
\caption{Same as Figure \ref{sssgood} except for the \swift\ sources without
significant SSS detections.  \label{sssbad}}
\end{figure*}

\begin{figure*}[htbp]
\includegraphics[angle=90,scale=0.60]{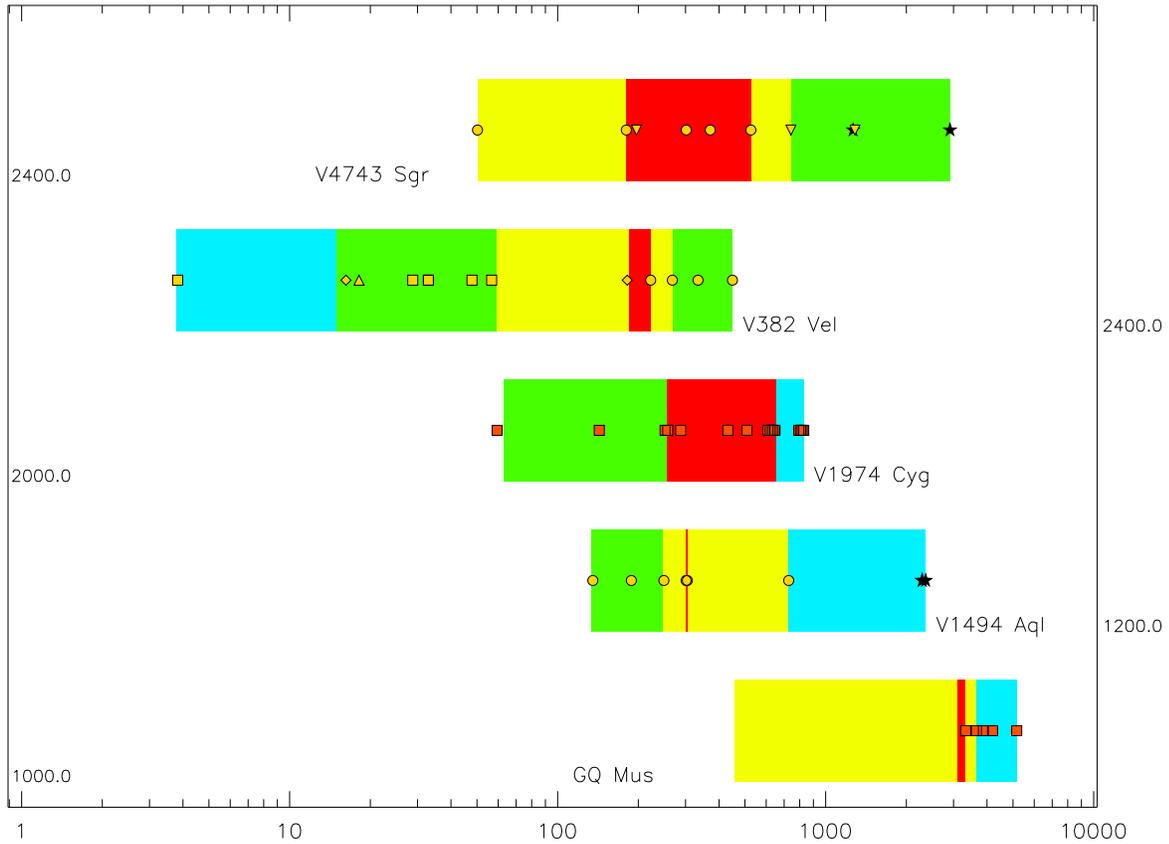}
\caption{Same as Figure \ref{sssgood} but for pre-\swift\ 
SSS novae.\label{sssother}}
\end{figure*}

Figures \ref{sssgood} and \ref{sssbad} show the X-ray observations of the 
\swift\ novae as a function of time since visual maximum.  The novae are 
organized from bottom to top in increasing FWHM values (Table 
\ref{chartable}), with the FWHM alternating on the left and right sides
of the figures.  Novae with unknown FWHM are labeled "U" and placed
at the bottom.  Figure \ref{sssgood} shows the novae with confirmed SSS 
emission while Figure \ref{sssbad} shows the
novae with no current SSS detections.  Note that some novae in 
Figure \ref{sssbad}, particularly the slowly evolving ones V5558 Sgr 
and V2468 Cyg,
may eventually evolve to the SSS phase.  Figure \ref{sssother} is the same 
but for well observed SSS novae observed prior to the launch of \swift;
these novae typically have much poorer observational coverage.
The black stars are the individual \swift\ observations.  
The figures also contain supplemental observations obtained with other
X-ray facilities, \cxo, \xmm, {\it ASCA}, {\it RXTE}, 
{\it BeppoSax}, and \rosat\ which are shown as circles, downward pointing 
triangles, upward pointing triangles, yellow squares, diamonds, and red 
squares, respectively.  The colors associated with each bar give the 
type of emission observed based on the hardness ratio.  Red bars indicate
time intervals when the HR2 of an individual source was $\lesssim -0.3$ 
and the uncertainty in the relative error was $<$ 5\%; the photons in this
case are primarily soft and these regions are associated with the SSS phase.
Orange bars designate observations with the same hardness ratio but 
larger errors.  Yellow
shows regions between observations where hard/soft change occurred.
The orange and yellow regions represent the maximum limits of the soft 
phase since the transition occurred at some point during these times.  
Section \ref{sssphase} describes the SSS phase in greater detail.  
Green regions show times when the overall detected spectrum was 
hard, HR2 $>$ -0.3 and section \ref{xrayevol} discusses this phase.  
Finally, blue represents 
times of non-detections.  Table \ref{colordescriptions} also gives the 
color descriptions for Figures \ref{sssgood} - \ref{sssother}.

\begin{deluxetable}{lll}
\tablecaption{Figures \ref{sssgood} - \ref{sssother} detection
definitions and descriptions and symbol legend\label{colordescriptions}}
\tablewidth{0pt}
\tabletypesize{\scriptsize}
\tablehead{
\colhead{Color} & \colhead{HR2\tablenotemark{a} and error} & 
\colhead{X-ray emission} 
}
\startdata
Blue   & \nodata & Undetected \\
Green  & $>$-0.3 & Hard \\
Yellow & \nodata & Transition between Green and Orange/Red classification. \\
Orange & $\lesssim$-0.3 and $>$ 5\% error & Soft but with large uncertainty, highly variable during initial rise. \\
Red    & $\lesssim$-0.3 and $<$ 5\% error & Soft X-rays \\
\hline
\multicolumn{3}{c}{Symbol legend} \\
\hline
\swift & stars & \\
\cxo & circles & \\
\xmm & downward pointing triangles & \\
{\it ASCA} & downward pointing triangles & \\
{\it RXTE} & yellow squares & \\
{\it BeppoSax} & diamonds & \\
{\it ROSAT} & red squares & \\
\enddata
\tablenotetext{a}{HR2=(H-S)/(H+S) with
Hard(H)=1-10\,keV and Soft(S)=0.3-1\,keV}
\end{deluxetable}

Several trends are evident in Figures \ref{sssgood} - \ref{sssother}.
As the FWHM decreases, the novae in the sample become SSS later and remain 
in the SSS phase longer.  This behavior is consistent with 
larger expansion velocity novae originating on higher mass WDs
\citep{2009cfdd.confE.199S}.  In addition the early, hard detections 
are generally only observed in the high FWHM novae.  The trends evident in 
Figures \ref{sssgood} and \ref{sssother} allow for a straightforward 
interpretation of Figure \ref{sssbad} - fast novae (loci at the top 
of the panel) are infrequently observed in the SSS phase as 
early X-ray observations of these systems is often absent.  The slower 
novae at the bottom of the figure have not been followed with 
sufficient temporal coverage late in their evolution or they 
have not yet reached the SSS phase or have ceased nuclear burning before
their ejecta clears sufficiently to observe SSS emission.

A note of caution about using Figures \ref{sssgood} and \ref{sssbad} to
determine nuclear burning time scales is appropriate.  These figures 
are based only on the strength and error of HR2 as provided in 
Table \ref{fullswift} which is based on a fixed hardness threshold for
all novae in the table.
Novae that have significant intrinsic hard emission such as V407 Cyg
may not be classified as SSSs by this definition even though they 
have soft X-ray light curves typical of nuclear burning and cessation
on the WD (see Section \ref{v407cygSSS}). 
High extinction will have a similar effect.  The red regions
generally also overestimate the duration of the SSS since that phase is
also defined by a tremendous rise in the soft X-ray count rate.
Sections \ref{S:ton} and \ref{S:toff} provide the determination of nuclear 
burning time scales for the X-ray nova sample.

For completeness, Table \ref{othertable} gives a summary of all the 
publicly available, pointed \xmm\ and \cxo\ nova observations.
The columns are the nova name, the observational 
identifier, the exposure time, Julian date and day after visual 
maximum of the observation, and a short comment on the result of 
the observation.  The instrument set up is also given in the 2nd 
column for the \cxo\ observations.  In some cases this data set 
provides important information on the SSS status of some sources
due to a lack of or weak \swift\ detections. 
An example would be the \xmm\ observations
of V574 Pup which confirms that there was a strong SSS during the 
interval between 2005 and 2007 when there were no \swift\ data 
\citep{Heltonthesis}.

\begin{deluxetable}{llrcrl}
\tablecaption{Pointed Chandra and XMM observations of recent novae 
\label{othertable}}
\tablewidth{0pt}
\tablehead{
\colhead{Name} & \colhead{Obs ID\tablenotemark{a}} & 
\colhead{Exp (ks)} & \colhead{JD} & \colhead{Days\tablenotemark{b}} & 
\colhead{Result\tablenotemark{c}}
}
\startdata
CI Aql    & 2465 (ACIS-S)     &  2.2 & 2452062 &  396 & Faint (1) \\
          & 2492 (ACIS-S)     & 19.9 & 2452123 &  457 & Faint (1) \\
          & 0652760201        & 26.9 & 2455577 & 3912 & NPA \\
CSS081007\tablenotemark{d} & 9970 (HRC-S/LETG) & 35.2 & 2454818 &  222:& SSS (2) \\ 
IM Nor    & 3434 (ACIS-S)     &  5.6 & 2452317 &   28 & Not detected (3) \\ 
          & 2672 (ACIS-S)     &  4.9 & 2452425 &  136 & Faint and hard (3) \\
KT Eri    & 12097 (HRC-S/LETG)& 15.2 & 2455219 &   69 & SSS (4) \\
          & 12100 (HRC-S/LETG)&  5.1 & 2455227 &   77 & SSS \\
          & 12101 (HRC-S/LETG)&  5.1 & 2455233 &   83 & SSS \\
	  & 12203 (HRC-S/LETG)&  5.1 & 2455307 &  157 & SSS \\
Nova LMC 2000 & 0127720201    & 16.3 & 2451751 &   14 & Faint and hard (5) \\
              & 0127720301    & 10.0 & 2451785 &   48 & Hard (5) \\
	      & 0127720401    & 10.5 & 2451998 &  291 & Not detected (5) \\
Nova LMC 2009a& 0610000301    & 37.7 & 2454957 &   90 & SSS \\
              & 0610000501    & 58.1 & 2455032 &  165 & SSS \\
              & 0604590301    & 31.9 & 2455063 &  196 & SSS \\ 
              & 0604590401     & 51.2 & 2455097 &  230 & SSS \\ 
Nova SMC 2005 & 0311590601    & 11.6 & 2453807 &  219 & Not detected \\
U Sco     & 12102 (HRC-S/LETG)& 23.2 & 2455241 &   17 & SSS (6) \\
          & 0650300201        & 63.8 & 2455247 &   23 & SSS (7) \\
          & 0561580301        & 62.8 & 2455259 &   35 & SSS (7) \\
V1065 Cen & 0555690301        &  9.4 & 2454837 &  714 & Not detected \\
V1187 Sco & 4532 (ACIS-S)     &  5.2 & 2453305 &   96 & $<$ 2keV + NVII line \\
          & 4533 (HRC-S/LETG) & 26.1 & 2453401 &  181 & Not detected \\
          & 0404431101        &  4.7 & 2454161 &  941 & Not detected \\
          & 0404430301        &  9.4 & 2454161 &  941 & Not detected \\
          & 0555691001        &  7.1 & 2454904 & 1684 & Not detected \\ 
V1280 Sco & 0555690601        &  4.5 & 2454903 &  756 & 1-keV emission \\
V2361 Cyg & 0405600101        & 11.1 & 2453868 &  456 & Not detected \\
          & 0405600401        & 14.9 & 2454028 &  616 & Not detected \\
V2362 Cyg & 0506050101        &  9.9 & 2454225 &  394 & thermal plasma (8) \\
          & 0550190501        & 27.9 & 2454821 &  990 & Very weak \\
V2467 Cyg & 0555690501        &  7.0 & 2454780 &  605 & SSS \\
V2487 Oph & 0085580401        &  8.3 & 2451965 &  986 & thermal plasma (9)\\
          & 0085581401        &  8.1 & 2452157 & 1178 & thermal plasma (9)\\
          & 0085581701        &  7.6 & 2452331 & 1352 & thermal plasma (9)\\
          & 0085582001        &  8.5 & 2452541 & 1562 & thermal plasma + Fe K$\alpha$ line (9)\\
V2491 Cyg & 0552270501        & 39.3 & 2454606 &   39 & SSS (10) \\
          & 0552270601        & 30.0 & 2454616 &   49 & SSS (11) \\
V2575 Oph & 0506050201        & 14.9 & 2454347 &  569 & Not detected \\ 
V2576 Oph & 0506050301        & 11.5 & 2454376 &  544 & Not detected \\
V2615 Oph & 0555690401        &  9.7 & 2454922 &  735 & Not detected \\
V351 Pup  & 0304010101        & 51.8 & 2453525 & 4908 & Faint \\
V458 Vul  & 0555691401        & 11.7 & 2454780 &  459 & weak 1-keV emission \\
V4633 Sgr & 0085580301        & 10.2 & 2451828 &  933 & weak (12) \\
          & 0085581201        &  7.3 & 2451977 & 1082 & weak (12)\\
	  & 0085581301        & 11.6 & 2452159 & 1264 & weak (12)\\
V4643 Sgr & 0148090101        & 11.9 & 2452716 &  750 & Not detected \\
          & 0148090501        & 11.0 & 2452894 &  928 & Not detected \\
V5114 Sgr & 0404430401        &  7.9 & 2454167 & 1086 & Not detected \\
          & 0404431201        &  3.6 & 2454167 & 1086 & Not detected \\
V5115 Sgr & 0405600301        &  9.2 & 2454005 &  566 & weak SSS \\
          & 0550190201        & 14.9 & 2454925 & 1486 & weak detection \\ 
V5116 Sgr & 0405600201        & 12.9 & 2454164 &  608 & SSS (13) \\
          & 7462 (HRC-S/LETG) & 35.2 & 2454336 &  780 & SSS (14) \\
          & 0550190101        & 26.6 & 2454893 & 1337 & Not detected \\ 
V574 Pup  & 0404430201        & 16.6 & 2454203 &  872 & SSS  \\
V598 Pup  & 0510010901        &  5.5 & 2454402 &  146 & SSS (15) \\
XMMSL1 J060636\tablenotemark{d} & 0510010501   &  8.9 & 2454270 &  627 & SSS (16) \\
\enddata
\tablenotetext{a}{\cxo\ observations have a four digit IDs and are followed
by the instrument configuration. \xmm\ observations have 10 digit IDs.}
\tablenotetext{b}{Days since visual maximum, see Table \ref{chartable}.}
\tablenotetext{c}{The number in parenthesis is the code to the published data.
NPA stands for "Not Publicly Available" and indicates proprietary observations
at the time of this publication.}
\tablenotetext{d}{Full novae names are CSS081007030559+054715 and
XMMSL1 J060636.2-694933.} \\
\tablecomments{
(1) \citet{2002A&A...387..944G};
(2) \citet{2009ATel.1910....1N};
(3) \citet{2005ApJ...620..938O};
(4) \citet{2010ATel.2418....1N};
(5) \citet{2003A&A...405..703G};
(6) \citet{2010ATel.2451....1O};
(7) \citet{2010ATel.2469....1N};
(8) \citet{2007ATel.1226....1H};
(9) \citet{2007ASPC..372..519F}; 
(10) \citet{2008ATel.1561....1N};
(11) \citet{2008ATel.1573....1N};
(12) \citet{2007ApJ...664..467H};
(13) \citet{2008ApJ...675L..93S};
(14) \citet{2007ATel.1202....1N};
(15) \citet{2008A&A...482L...1R};
(16) \citet{2009A&A...506.1309R}.
}
\end{deluxetable}

\section{THE EARLY HARD X-RAY PHASE}

Some novae have hard X-ray emission, e.g. $>$ 1 keV, early 
in the outburst.  These novae tend to be fast or recurrent novae.  This 
initial hard emission is thought to arise from shock heated gas 
inside the ejecta or from collisions with external material, {\it e.g} the 
wind of the red giant secondary in RS Oph \citep{2006ApJ...652..629B,
2006Natur.442..276S,2007ApJ...665..654V, 2009ApJ...691..418D}.  Early hard 
X-ray emission observed in the very fast nova V838 Her has been attributed 
to intra-ejecta shocks from a secularly increasing ejection velocity 
\citep{1992Natur.356..222L,1994MNRAS.271..155O}.  Much later in the outburst
when nuclear burning has ceased hard X-rays can again dominate.  These hard
X-rays come from line emission from the ejected shell and/or emission 
from the accretion disk \citep{2002AIPC..637..345K}, or in the case of
RS Oph, the re-emergence of the declining shocked wind emission once the 
SSS emission has faded \citep{2008ASPC..401..269B}.

Every nova with a FWHM $\ge$ 3000 km s$^{-1}$ and observations within 100 
days after visual maximum in the \swift\ sample exhibited hard X-rays.  
This detection rate is partially due to the fact that many of these novae 
were high interest targets, {\it e.g.} very bright at visual maximum (KT Eri),
extreme ejection velocity (V2672 Oph), RN (RS Oph, V407 Cyg and
U Sco), detected prior to outbursts as an X-ray source (V2491 Cyg), etc.; 
thus their early X-ray evolution was well documented.  In addition, 
a higher cadence of observations during the early phases greatly 
increased the probability of discovery.  

The evidence of initial hard X-ray emission for slow novae is sparse
as few were well sampled early in their outbursts.  Only V458 Vul
\citep{2009AJ....137.4160N,2009PASJ...61S..69T} had early 
observations which showed a hard component with a duration of hundreds
of days from its first observation $\sim$ 70 days after visual maximum.
The lack of significant evidence of hard emission in the early outburst 
of slow novae is consistent with
shocks, either within the ejecta or with a pre-outburst ambient medium, 
being the primary source of early hard X-ray emission in the
faster novae.  Slower novae have lower ejection speeds
and thus should either have less or delayed shock emission
\citep[see equ. 3 in][]{2006ApJ...652..629B}.  Hard X-rays
were also detected late in the outburst of novae with extreme and multiple 
ejection events.
The best example of this is V2362 Cyg 
which at the time of its unusually bright secondary maximum had
already doubled the width of its emission lines and was detected 
as a hard X-ray source \citep{2008AJ....136.1815L}.  
Similarly the slow nova V5558 Sgr was also 
a late hard source.  Its early light curve was marked by numerous 
secondary maxima similar to V723 Cas \citep{2008NewA...13..557P}.  

Another interesting case is the slow nova V1280 Sco which was detected 
multiple times between days 834 - 939 after outburst as an X-ray source.  
\citet{2009ATel.2063....1N} found that the X-ray count rate was 
relatively low and the SED was best fit with multiple thermal plasma 
models consistent with line emission.  They attributed the lines 
from shock heating of the ejecta but this is difficult to reconcile 
with how rapidly shock emission declines.  \citet{2010ApJ...724..480H} showed
that V1280 Sco had two bright secondary peaks after maximum. Thus, it is
possible that this nova experienced additional ejection events later in
the outburst that contributed the necessary energy to power shocks.
Contemporary optical spectra from our Small and Moderate Aperture
Research Telescope System (SMARTS) nova monitoring program
show that the photosphere of V1280 Sco remains optically thick with 
P-Cygni profiles still present more than 4 years after outburst. 
Alternatively, the line emission may be from circumstellar gas 
photoionized by the initial X-ray pulse of the explosion.  Given the 
relative proximity of V1280 Sco, ranging from 0.63$\pm$0.10 kpc
\citep{2010ApJ...724..480H} to 1.6 kpc \citep{2008A&A...487..223C}, any X-ray 
emission lines would be much brighter than most novae in our sample which has
a larger median distance of 5.5 kpc.  Unfortunately, V1280 Sco was X-ray 
faint making it impossible to determine the source of its X-ray emission.


\section{THE SSS PHASE\label{sssphase}}

\subsection{Rise to X-ray Maximum and the ``Turn-on'' Time\label{S:ton}}

The unprecedented temporal coverage of the early outburst in X-rays with
\swift\ has fully revealed a new phenomenon during the rise to X-ray maximum.
Prior to \swift, V1974 Cyg had the best sampled X-ray light curve
\citep[see Fig. 1 in ][]{1996ApJ...456..788K}.  The 18 \rosat\ 
observations showed a slow and monotonic rise to maximum. This light 
curve evolution was expected as the obscuration from the ejecta clears 
and the effective temperature of the WD photosphere increases
\citep{1985ApJ...294..263M}.  However \cxo\ observations of
V1494 Aql \citep{2003ApJ...584..448D} and V4743 Sgr
\citep{2003ApJ...594L.127N} hinted that this transition was not as smooth
as previously observed, with short term "bursts", periodic oscillations,
and sudden declines.  

\begin{figure*}[htbp]
\plotone{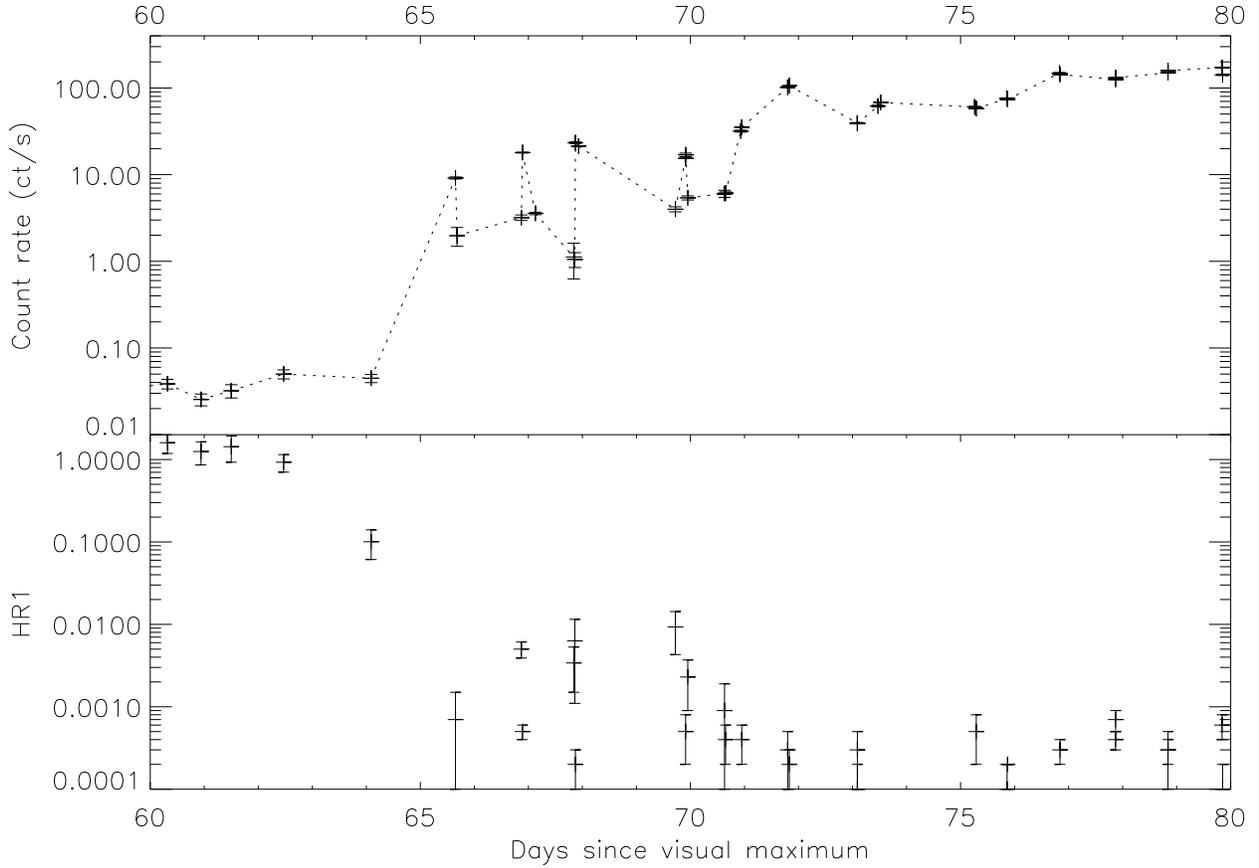}
\caption{The early X-ray light curve of KT Eri in days since visual maximum.
The top panel shows the count rate and the lower panel gives the hardness
ratio, HR1.  Dotted lines are added to the top panel to emphasize the
variability.  Prior to day 65 KT Eri was faint and hard.
Between days 65 and 75 the source transitioned to the bright SSS phase
with large amplitude oscillations in the count rate and some corresponding
changes in HR1.  After day 76 the both the count rate and
hardness ratio significantly stabilized but still showed variability
(see Section \ref{var}).
\label{kteriearlylc}}
\end{figure*}

With daily and sometimes hourly \swift\ coverage,
the rise to X-ray maximum is unequivocally highly chaotic with large 
changes in the count rate evident in all well observed \swift\ 
novae to date.  Figure \ref{kteriearlylc} illustrates this 
phenomenon in KT Eri from the data available in Table \ref{fullswift}.
During the initial rise to X-ray maximum, it exhibited large
oscillations.  The numerous large declines are even more dramatic when the
observational data sets are not grouped by observation
ID number as in Table \ref{fullswift} 
but broken into small increments (Walter et al. in prep).
At 76 days after visual maximum the variability became much smaller 
and the count rate stabilized around $\sim$150 ct s$^{-1}$. 
In addition to KT Eri \citep{2010ATel.2392....1B},
RS Oph \citep{2011ApJ...727..124O}, U Sco \citep{2010ATel.2430....1S},
nova LMC 2009a \citep{2009ATel.2025....1B,Bode2011},
V2672 Oph \citep{2009ATel.2173....1S}, 
V2491 Cyg \citep{2010MNRAS.401..121P,2011arXiv1103.4543N}, 
and V458 Vul \citep{2009AJ....137.4160N} 
all showed this large amplitude variability.  The first three novae 
are known RNe while the next two and KT Eri are suspected to be
RNe based on their observational characteristics.  
The fact that the less energetic V458 Vul 
also exhibited this phenomenon indicates that it is not just associated
with very fast or recurrent novae. 
See Section \ref{var} for further discussion of nova variability.

\begin{figure*}[htbp]
\plotone{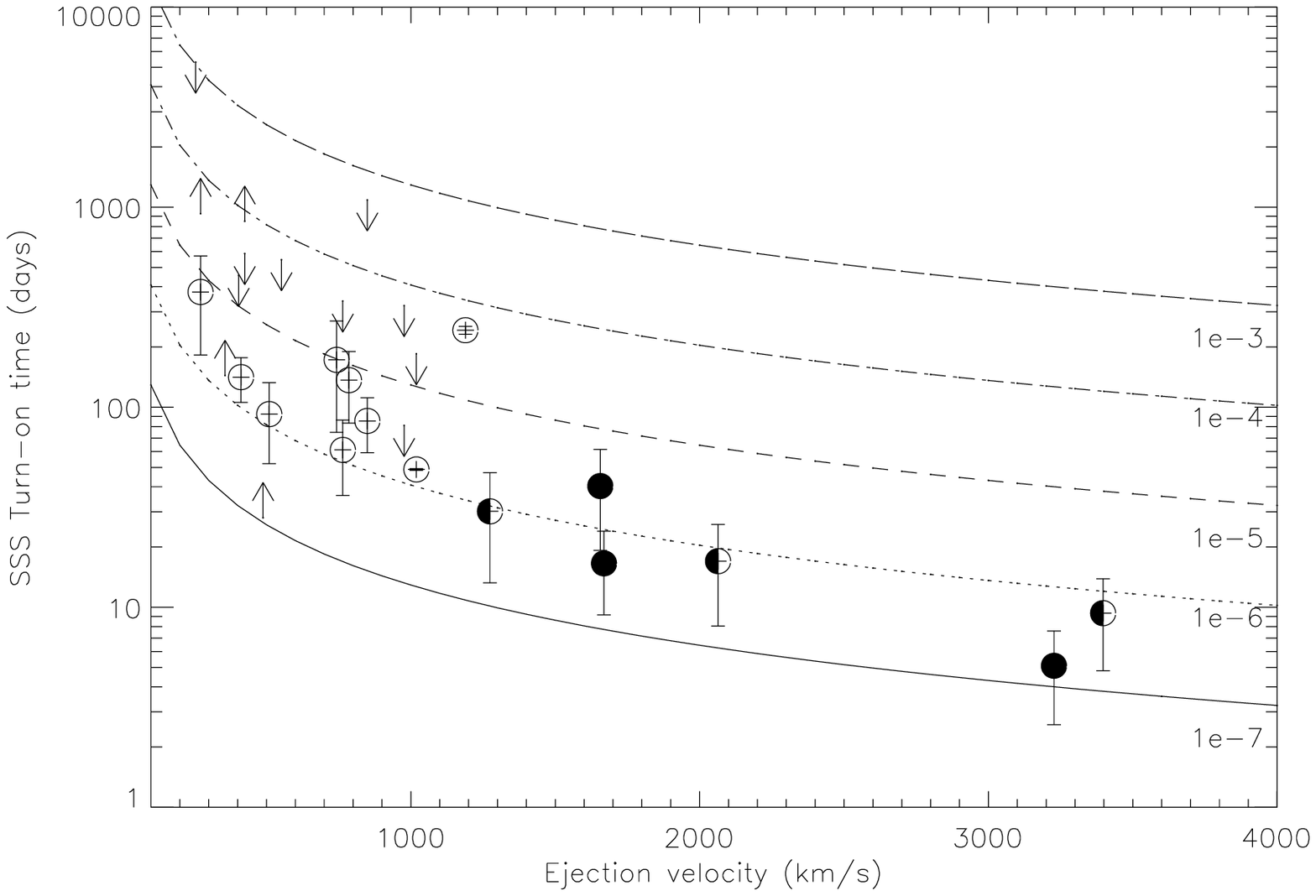}
\caption{SSS turn-on time of novae (Table \ref{timescales})
as a function of the ejection velocity (estimated from the FWHMs
in Table \ref{chartable}).  
Filled circles are known RNe.  Half filled circles
are suspected RNe based on their characteristics.
From the top to the bottom the lines
show the relationship from Shore (2008; Eqn. 9.2) for ejected masses
of 1$\times$10$^{-3}$, 1$\times$10$^{-4}$, 1$\times$10$^{-5}$,
1$\times$10$^{-6}$, and 1$\times$10$^{-7}$ M$_{\odot}$, respectively.
The downward and upward arrows are estimated upper and lower limits.
\label{velturnon}}
\end{figure*}

The emergence of the SSS, referred to as ``turn-on'' time or t$_{on}$ 
hereafter, provides information on the mass of the ejected shell.  
The turn-on times for the novae in this sample are given in 
Table \ref{timescales}.  The t$_{on}$ time is defined as the time 
after visual maximum when the HR2 $<$ -0.8 and there is significant
increase in the soft count rate.  Similarly the ``turn-off'' 
time (t$_{off}$) is defined as the time after t$_{on}$ when the
hardness ratio becomes harder, HR2 $>$ -0.8, and the soft count rate
declines rapidly as nuclear burning ends.  Note that these
definitions should not be confused with the SSS phases as shown in 
Figures \ref{sssgood} - \ref{sssother} as t$_{on}$ and t$_{off}$ also
include the change in the soft count rate.  SSS emission can only be 
observed when the ejecta column density declines 
to the point where the source can be observed. With the expansion
velocity and turn-on time, upper limits on the ejected mass can be
established.  \citet{Shore08} gives the relationship (see Equation 9.2),
\begin{equation}
M_{eject} \sim 6 \times 10^{-7} \phi N_{H}(22) v_{exp}(1000)^2 t_{on}^2 M_{\odot} \\
\end{equation}
where $\phi$ is the filling factor, N$_H$(22) is the column density 
in units of 10$^{22}$ cm$^{-2}$, v$_{exp}$(1000) is the expansion velocity in 
units of 1000\ km\ s$^{-1}$, t$_{on}$ is the soft X-ray turn-on time
in days and assumes spherical geometry. In this study, $\phi$ = 0.1 and a
column density of 10$^{22}$ cm$^{-2}$ is used as the minimum N$_H$ for
the ejected shell to become transparent to soft X-rays.
The expansion velocity is determined from v$_{exp}$ = FWHM/2.355
\citep{2010PASP..122..898M} where FWHM is the width of the Balmer
lines near visual maximum as given in Table \ref{chartable}.
Using the t$_{on}$ times from Table \ref{timescales}, 
Figure \ref{velturnon} shows the estimated ejected masses as a function 
of ejection velocity.  
\btxt{Note that the velocities derived from these FWHMs are lower limits
as the X-ray opacity in the ejecta depends on faster material.  This has
the effect of shifting all the points in Figure \ref{velturnon} to the 
right.}
Accordingly the fastest novae, at the bottom 
right, U Sco and V2672 Oph, must have ejected much less than 10$^{-5}$ 
M$_{\odot}$ otherwise they would not have been observed as SSS sources 
so early after outburst.  This inference is consistent with independent 
ejected mass estimates 
\citep[{\it e.g.} U Sco,][]{2000AJ....119.1359A,2010AJ....140.1860D,
2010ApJ...720L.195D}.
Conversely, novae in the upper left corner must eject a significant 
amount of material.  Large mass ejection events are also inferred from 
the optical spectra of novae like V1280 Sco which still showed P-Cygni 
lines 3 years after outburst \citep{2010PASJ...62L...5S} and a year 
later in our recent SMARTS spectroscopy.  

Note that external extinction from the ISM is not taken into account 
in Figure \ref{velturnon} nor is the evolution of the 
effective temperature of the WD photosphere. Novae with large extinction
may never be observed in the SSS phase while a slow increase in the 
WD temperature after the ejecta has sufficiently cleared will delay
the onset of t$_{on}$ resulting in an overestimate of the ejected mass
derived from Equ. 1.   Both factors along with deviations
from the underlying assumptions such as different filling factors
and non-spherical symmetry, can lead to different mass values given
in Figure \ref{velturnon}.  These limitations explain why two novae
with the same ejection velocities, V2468 Cyg and V5558 Sgr at 
425 km s$^{-1}$, can have divergent mass estimates due to different
turn-on times.

\begin{deluxetable}{lcc}
\tablecaption{SSS X-ray time scales\label{timescales}}
\tablewidth{0pt}
\tablehead{
\colhead{Name} & \colhead{turn-on} & \colhead{turn-off} \\
\colhead{} & \colhead{(d)} & \colhead{(d)}
}
\startdata
CI Aql & \nodata & $<$396 \\
CSS 081007 & 185$\pm$68 & 314$\pm$68 \\
GQ Mus & \nodata & 3484.5$\pm$159.5 \\
IM Nor & $>$28 & $<$136 \\
KT Eri & 71$\pm$1 & 280$\pm$10 \\
LMC 1995 & $<$1087 & 2545$\pm$426 \\
LMC 2000 & $>$48 & $<$291 \\
LMC 2009a & 95$\pm$5 & 270$\pm$10 \\
RS Oph & 35$\pm$5 & 70$\pm$2 \\
U Sco & 23$\pm$1 & 34$\pm$1 \\
V1047 Cen & $>$144 & $<$972 \\
V1065 Cen & \nodata & $<$744\tablenotemark{a} \\
V1187 Sco & \nodata & $<$181\tablenotemark{a} \\
V1213 Cen & $<$322 & $>$494 \\
V1280 Sco & $>$928 & \nodata \\
V1281 Sco & $<$339 & 627$\pm$194 \\
V1494 Aql & 217.5$\pm$30.5 & 515.5$\pm$211.5 \\
V1974 Cyg & 201$\pm$54 & 561.5$\pm$50.5 \\
V2361 Cyg & \nodata & $<$456 \\
V2362 Cyg & \nodata & $<$990 \\ 
V2467 Cyg & $<$456 & 702$\pm$97 \\
V2468 Cyg & $<$586  & \nodata \\
V2487 Oph & \nodata & $<$986 \\
V2491 Cyg & 40$\pm$2 & 44$\pm$1 \\
V2672 Oph & 22$\pm$2 & 28$\pm$2 \\
V351 Pup & \nodata & $<$490 \\
V382 Vel & $<$185 & 245.5$\pm$22.5 \\
V407 Cyg & 15$\pm$5 & 30$\pm$5 \\
V458 Vul & 406$\pm$4 & $>$1051 \\
V4633 Sgr & \nodata & $<$934 \\
V4743 Sgr & 115$\pm$65 & 634$\pm$108 \\
V5114 Sgr & $<$1086 & \nodata \\
V5115 Sgr & $<$546 & 882$\pm$336 \\
V5116 Sgr & 332.75$\pm$275.25 & 938$\pm$126 \\
V5558 Sgr & $>$850\tablenotemark{b} & \nodata \\
V5583 Sgr & $<$81 & 149$\pm$68 \\
V574 Pup & 571$\pm$302 & 1192.5$\pm$82.5 \\
V597 Pup & 143$\pm$23 & 455$\pm$15 \\
V598 Pup & \nodata & $<$127 \\
V723 Cas & $<$3698 & $>$5308 \\
V838 Her & \nodata & $<$365 \\
XMMSL1 J060636 & \nodata & $<$291  \\
\enddata
\tablecomments{t$_{on}$ and t$_{off}$ bracket the time after visual maximum
when the hardness ratio HR2 is softer than -0.8.}
\tablenotetext{a}{Evolution of $[$\ion{Fe}{7}$]$ (6087\AA) and lack of 
$[$\ion{Fe}{10}$]$ (6375\AA) in our SMARTS optical
spectra are consistent with this upper limit from the X-ray non-detection.
See Section \ref{fex}.}
\tablenotetext{b}{Optical spectra are slowly becoming more ionized which
is consistent with slowly increasing SSS emission observed with \swift.}
\end{deluxetable}

\subsection{Turn-off time\label{S:toff}}

\begin{figure*}[htbp]
\plotone{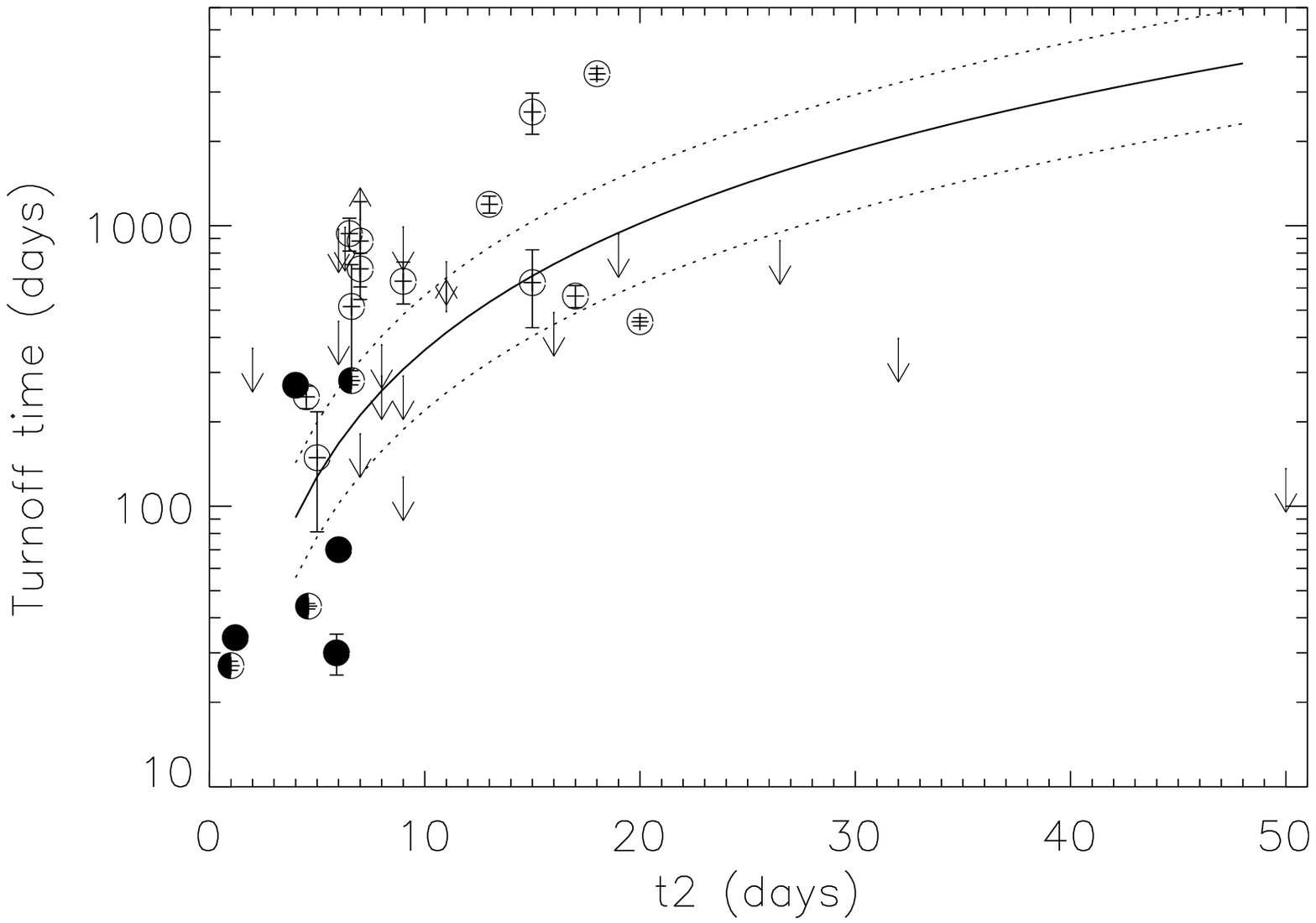}
\caption{SSS turn-off time as a function of t$_2$ time with the
\citet{2010ApJ...709..680H} relationship (solid line) and its
associated uncertainty (dotted lines) overplotted.  Upper and lower
limits are also shown. 
Filled circles are known RNe.  Half filled circles 
are suspected RNe based on their characteristics.
\label{t2turnoff}}
\end{figure*}

Table \ref{timescales} shows t$_{off}$ times or upper/lower limits
for the novae in our sample.  If optical light curve decline times,
{\it e.g.} t$_2$, are used as 
simple proxies for WD masses then 
there should be a relationship between 
t$_2$ and duration of the SSS phase.  In Figure \ref{t2turnoff} the
turn-off time, t$_{off}$, is shown versus t$_2$.
Overplotted as the solid line is the turn-off versus decline relationship
of \citet[][Equ. 31]{2010ApJ...709..680H} where t$_3$ was converted 
to t$_2$ using Equ. 7 in \citet{2007ApJ...662..552H}.
The combined uncertainties of both equations is represented by 
the two dotted lines.  \citet{2010ApJ...709..680H} find that the time when
nuclear burning ends is $\propto$ t$_{break}^{1.5}$ (Equ. 26), where
t$_{break}$ is the time of the steepening of their model free-free 
optical-IR light curves.  This relationship is derived using a series of 
steady state models with a decreasing envelope mass to fit the observed 
multiwavelength light curves.  The X-ray and UV light curves are fit with 
blackbodies while the optical and IR curves use optically thin, free-free 
emission.  The parameters of the model are the WD mass, composition of
the WD envelope, and its mass prior to outburst.  While the general trend
is similar, the observed data do not fit the \citet{2010ApJ...709..680H} 
relationship, especially when the sample is expanded to include the novae 
with only upper or lower limits.

The relationship derived by \citet{2010ApJ...709..680H} utilizes the
t$_2$ derived from the $y$ band light curve instead of the $V$ band as in 
this paper.  The $y$ band is used by \citet{2010ApJ...709..680H} since it 
generally samples the continuum where as the $V$ band can have a contribution
in the red wing from strong H$\alpha$ line emission. However, the difference
in filters can not explain the poor agreement between the data and the
relationship in Fig. \ref{t2turnoff} since there are similar numbers of
novae that fall above the line as below.  If a contribution from H$\alpha$
in $V$ was significant then the disagreement would not be symmetric.

\begin{figure*}[htbp]
\plotone{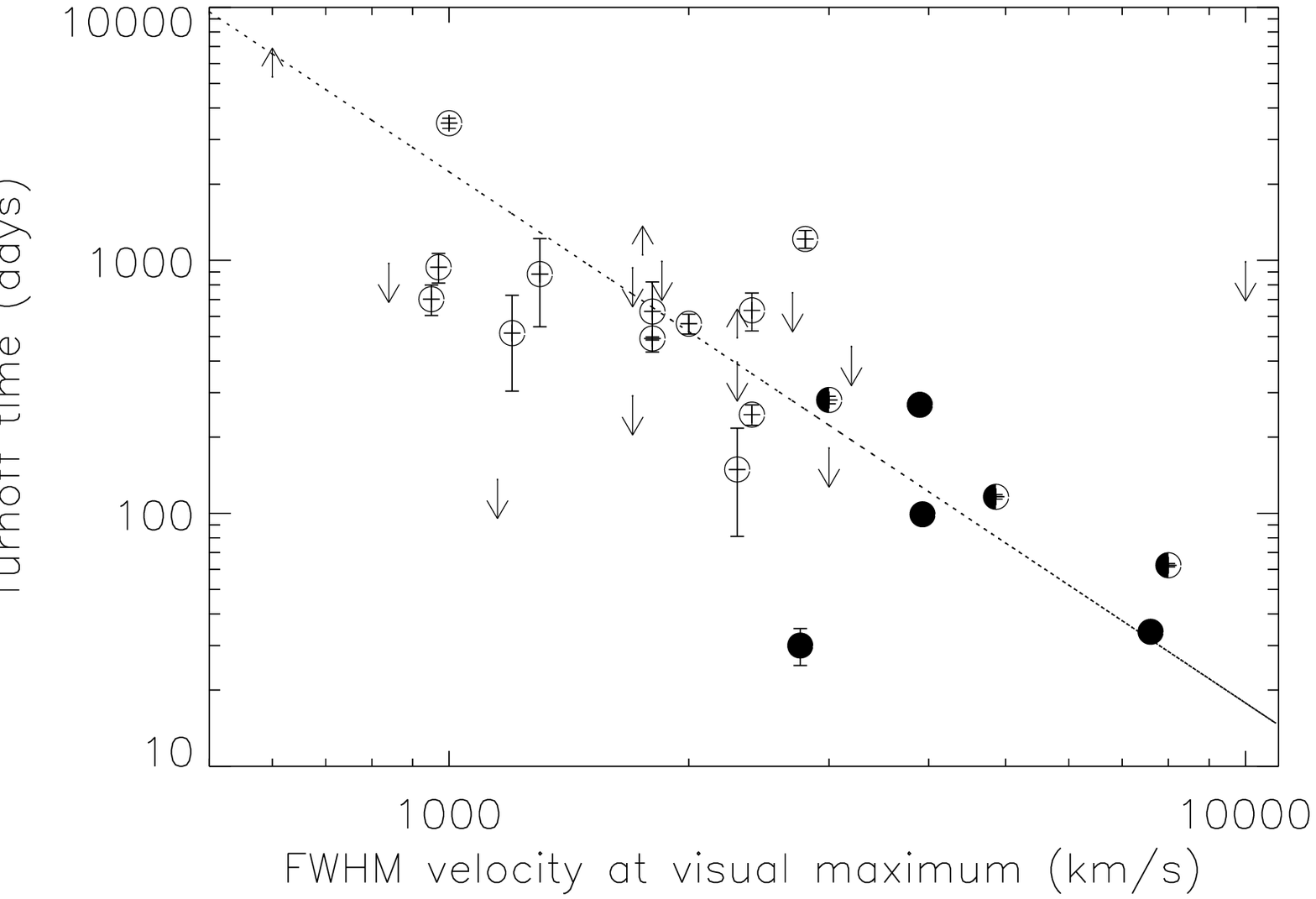}
\caption{SSS turn-off time as a function of the FWHM of H$\alpha$ or
H$\beta$ near visual maximum.  The relationship of \citet{2003A&A...405..703G}
is shown as the dotted line.  Upper and lower limits are also shown.
Filled circles are known RNe.  Half filled circles
are suspected RNe based on their characteristics.
\label{velturnoff}}
\end{figure*}

Similarly, Figure \ref{velturnoff} shows the relationship between the 
FWHM and turn-off time with the dotted line depicting the 
\citet{2003A&A...405..703G} turn-off vs. velocity relation.  This relationship
was derived from all the SSS nova data available at the time which was
only 4 well constrained SSS novae and 4 novae with turnoff limits.  With 
the significantly larger sample currently available it is clear that 
there is not a tight fit to the relationship.
This discrepancy is particularly acute for the slower novae in our sample 
which have turned off much sooner than expected.  These Figures illustrate 
that the gross behavior of novae is still poorly understood
and confirm that the observational characteristics of an individual
nova is governed by more than just the WD mass.

Accurate determinations of the duration of nuclear burning
can also provide an independent ejected mass estimate.  Recently,
\citet{2010ApJ...712L.143S} found that the ejected mass is only
dependent on the total radiated energy, E$_{rad}$, and does not
require knowledge about the geometry and structure of the shell
as with other methods. E$_{rad}$ is not a trivial value to determine
as it depends on the bolometric luminosity of the source and the
duration of the outburst. \swift\ observations can potentially
determine the bolometric flux both when the bulk of the emission is
in a narrow wavelength region such as the early, optically thick
phases in the UV and optical or later in the soft X-ray during the
SSS phase.  Estimates of the luminosity during both phases requires
an accurate determination of the extinction and the distance.
Perhaps the best example to use the \citet{2010ApJ...712L.143S}
technique on is RS Oph.  With a bolometric luminosity of
3$\times$10$^{4}$ L$_{\odot}$ from TMAP atmosphere models
\citep{2011ApJ...727..124O} and a t$_{off}$ of 70 days the estimated
ejected mass $\sim$ 2$\times$10$^{-6}$ M$_{\odot}$.  This is consistent
with the low mass estimates from the radio
\citep[(4$\pm$2)$\times$10$^{-7}$ M$_{\odot}$;][]{2009MNRAS.395.1533E}
and hydro-dynamical models of the X-ray behavior
\citep[1.1$\times$10$^{-6}$ M$_{\odot}$;][]{1992MNRAS.255..683O} and
\citep[$\sim$5$\times$10$^{-6}$ M$_{\odot}$;][]{2009ApJ...691..418D}

\subsubsection{SSS phase durations}

\begin{figure*}[htbp]
\plottwo{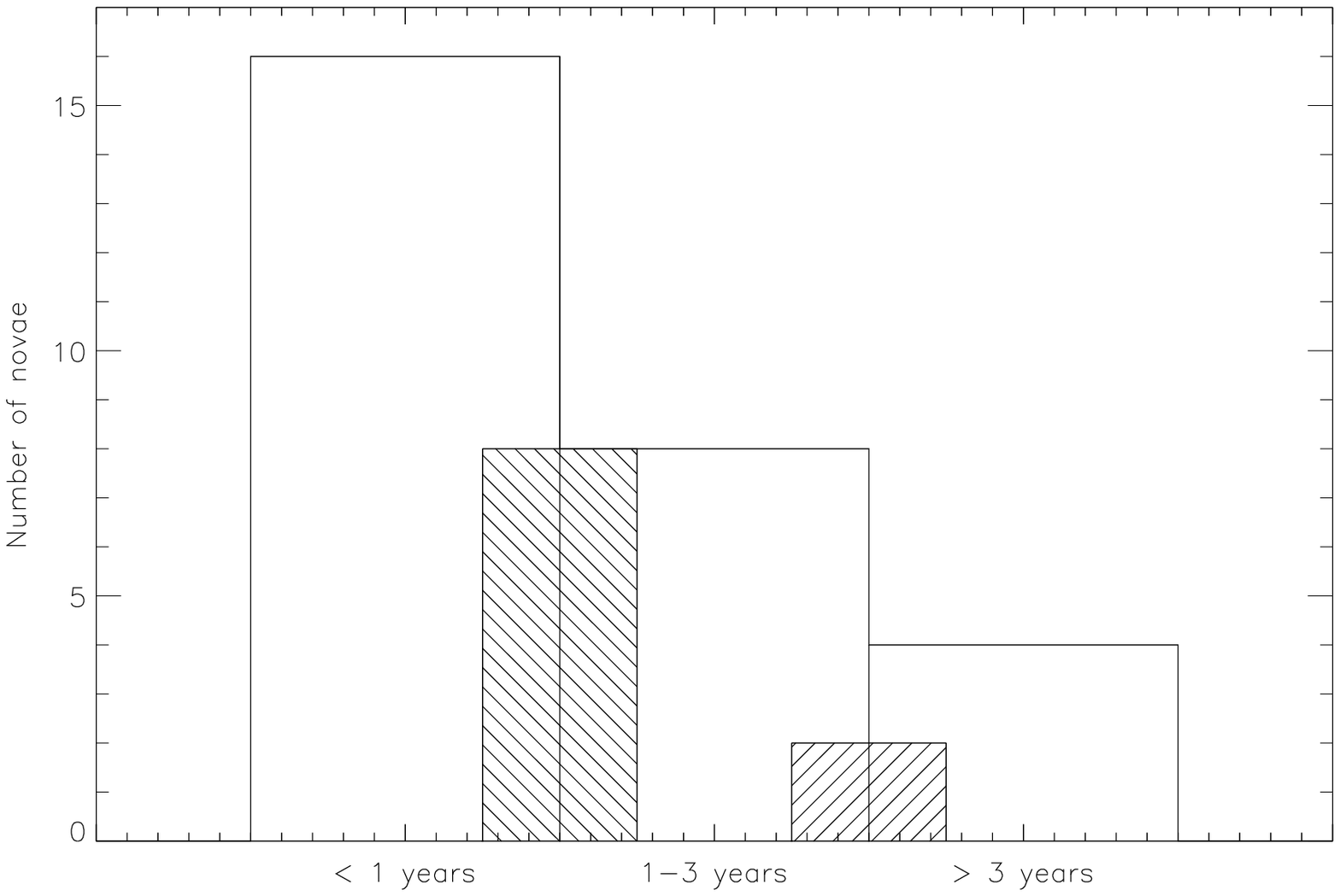}{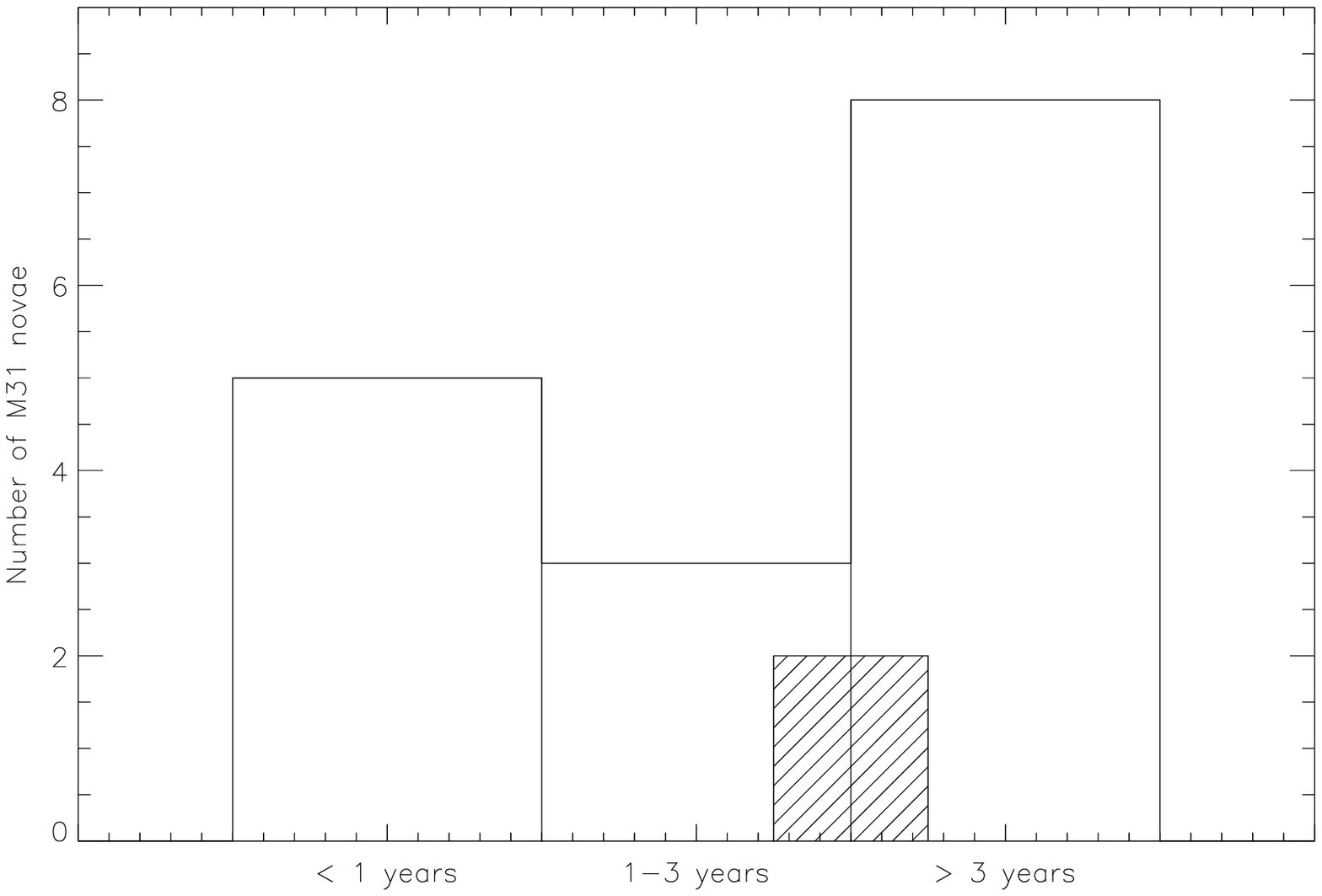}
\caption{Distribution of the durations of well established SSS novae 
in the Galaxy/Magellanic Clouds (left figure) and M31 (right figure)
from \citet{2010A&A...523A..89H}.
The three duration bins are less than one year, between 1 and 3 years, 
and greater than three years.  The hashed areas include the novae with 
only limits on their turnoff time that precludes placing them in a 
specific bin.
\label{histogram}}
\end{figure*}

Figure \ref{histogram}a shows the distribution of the duration of the SSS 
phase for this sample of novae.  Since there are still relatively few 
novae with well established turn-off times a coarse histogram with only three 
bins is used.  The bins have durations of less than one year, between 
one and three years, and greater than three years.  Due to large 
uncertainties in their exact turn-off times, ten of the sample novae 
cannot be placed within a single bin and thus are shown as the smaller 
cross-hatched columns between the bins in which they might belong.
Of the \totallimitSSS\ novae with detected SSS emission or with strong 
limits on the duration of the SSS phase, 89\%, have turned off in under 
3 years.  There are only four novae, GQ Mus, LMC 1995, V574 Pup, and 
V723 Cas, with detected SSS emission beyond 3 years.  V458 Vul and 
V1213 Cen were still SSSs at their last observations and could also 
exceed 3 years.  A similar rapid turn-off was inferred from a search of
the \rosat\ archive of novae with SSS detections.  \citet{2001A&A...373..542O} 
found only 3 SSS novae among the 39 Galactic and Magellanic cloud novae 
in the \rosat\ archive observed at least once within 10 years after 
visual maximum.  The median age of the 19 novae with documented turn-off 
times from this sample is 1.4 years.


The situation is different for nova surveys of M31 
\citep{2007A&A...465..375P,2010A&A...523A..89H} where SSSs identified as 
classical novae 5-10 years after outburst are fairly common, {\it e.g.}
1995-05b, 1995-11c, and 1999-10a.  Figure \ref{histogram}b shows the same
bins as before but with the 18 M31 novae detected as SSSs given in Table 9 
of \citet{2010A&A...523A..89H}.  The difference can be explained by the
predominance of slower novae in the M31 sample.  The mean t$_2$ time 
of the nine M31 novae with reported decline times in the 
\citet{2010A&A...523A..89H} SSS sample is 31 days whereas the peak 
for our sample is significantly faster at 8 days (Figure \ref{t2vsfwhm}). 
The discrepancy in speed class between the two samples is due to selection 
effects.  By design the Galactic/Magellanic sample consists primarily 
of bright and hence faster novae.  M31 surveys sample the entire 
galaxy but with fewer \cxo, \xmm\ and \swift\ observations that are 
randomly scattered in time.  The M31 strategy finds many novae since
the observed M31 nova rate is greater, $\sim$ 30 novae yr$^{-1}$ 
\citep{1989AJ.....97.1622C} than that of the Milky Way
\citep[$\sim$ 5 novae yr$^{-1}$][]{1997ApJ...487..226S}, however, with
limited time sampling slower novae with longer SSS phases are easier to
detect than fast novae with rapid turn-offs.

\rosat\ detected 2 SSS novae out of 21 Galactic novae for a Milky Way 
detection frequency of 9.5\% \citep{2001A&A...373..542O}. 
If the 4 RNe in the \rosat\ list
are discarded because their observations were taken $\gtrsim$ 1 year
after outburst, the detection frequency increases to 11.8\%.  The M31
survey has a similar low SSS detection frequency of 6.5\% 
\citep{2010AN....331..187P}.  These two results show that it is difficult
to catch novae during their SSS phase via random time sampling.  
However, a more systematic approach that 1) targets only 
bright and low extinction novae and 2) obtains multiple
observations early in the outburst may have a greater detection
frequency.  Indeed, \swift\ has a significantly greater SSS detection rate of 
$\sim$ 45\% during its five years of operation with this more systematic
approach.  

\subsection{SSS emission in the hard X-ray spectrum of 
V407 Cyg\label{v407cygSSS}}

\begin{figure*}[htbp]
\plotone{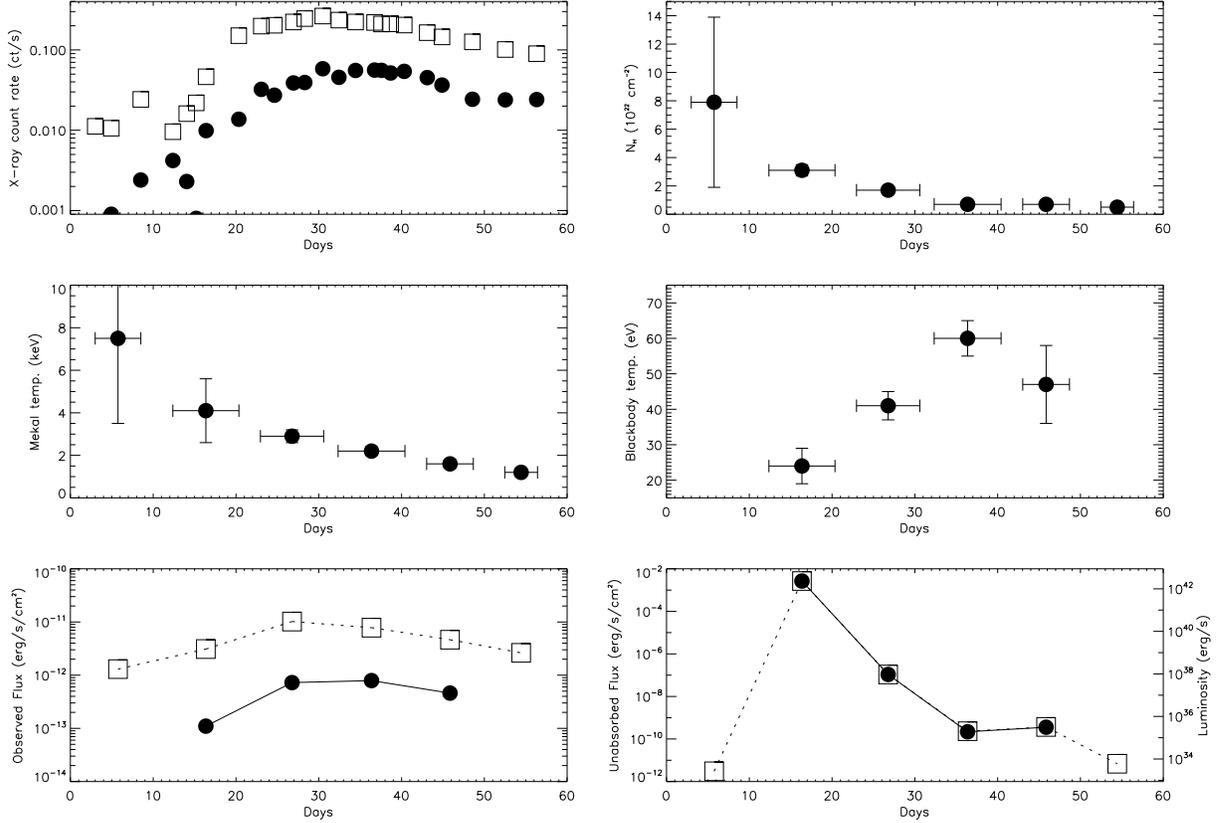}
\caption{Results of model fits to the \swift\ V407 Cyg data set.  The top
left panel shows the total 0.3-10 keV (squares) and soft 0.3-1 keV (circles)
light curves.  The derived N$_H$ column for the 6 date bins is shown in 
the top right panel.  The Mekal temperature of the hotter, optical thin
plasma model is shown in the middle left panel while the right middle panel
shows the temperature of the blackbody fit to the softer component. A 
second, soft component is not needed in the first and last date bins.
The bottom panels show the observed (left) and unabsorbed (right) fluxes.
Squares give the total from all components while the circles show just
the blackbody contributions.  The right axis of the last panel also shows
the corresponding 0.3-10 keV luminosity assuming a 2.7 kpc distance.
\label{v407cygmodel}}
\end{figure*}

In the initial analysis of the \swift\ data in \citet{2011A&A...527A..98S} 
a second soft component was required to fit some of the \swift\ X-ray spectra.
However, there were insufficient counts to distinguish between a blackbody 
and an optically-thin plasma model. Assuming a distance of 2.7 kpc
\citep{1990MNRAS.242..653M}, the unabsorbed flux of the soft component
in the day $<$30 model of Table 3 in \citet{2011A&A...527A..98S} gives a 
blackbody luminosity of 2$\times$10$^{37}$ erg s$^{-1}$ which is reasonable
for nuclear burning on a WD.  To investigate whether the soft emission
in V407 Cyg can be attributed to nuclear burning, we reanalyze the \swift\
X-ray data with twice as many time bins as previously used.  Figure
\ref{v407cygmodel} shows the results.  As in \citet{2011A&A...527A..98S}
the model abundances are allowed to vary but the temperatures are not 
significantly different if the abundances are constrained to be solar.
The data prior to day 10 and after day 50 can be fit with a single optically 
thin plasma model.  The remaining 4 time bins all require a soft component
which in this analysis is assumed to be a blackbody.  Both the derived
N$_H$ and the optically thin component temperature decline with time
in the models.  The blackbody effective temperature increases until the
day 36 bin and declines in the day 45 bin, although the error bars are
large enough that it could be constant over the last two dates.  The 
derived luminosities (over the 0.3--10~keV X-ray band) 
for the four dates with blackbody components are 
2.3$\times$10$^{42}$ erg s$^{-1}$, 9.3$\times$10$^{37}$ erg s$^{-1}$,
1.9$\times$10$^{35}$ erg s$^{-1}$, and 3.1$\times$10$^{35}$ erg s$^{-1}$,
respectively, assuming a distance of 2.7 kpc.  The extreme luminosity 
for the day 16 bin cannot be considered reliable, given the very low 
fitted temperature of $\sim$ 25~eV below the XRT 0.3~keV low-energy cut off.
Nevertheless, the results of fitting blackbodies to the \swift\ V407 Cyg 
data are consistent with a scenario where the nuclear burning proceeded 
on the WD surface near Eddington limits until about 30 days after 
visual maximum.  The fuel was consumed after that point leading to a 
rapid drop in the luminosity.  Thus, although V407 Cyg was not a 
true SSS, its soft photon light curve was consistent 
with expected evolution as seen in other novae.

\section{DISCUSSION} 

\subsection{Orbital period and turn-off time}

\citet{2003A&A...405..703G} found a correlation between the orbital 
period and X-ray turn-off time.  However, at that time only four novae 
had both well determined periods and  X-ray turn-off times, GQ Mus, 
V1974 Cyg, V1494 Aql, and V382 Vel, and limits on CI Aql and U Sco.  
The observed trend implied 
that novae with short orbital periods had the longest duration SSS phases.  
\citet{2003A&A...405..703G} attributed this relationship to 
a feedback loop between the WD and its secondary.  The luminous X-rays 
produced during the SSS phase excessively heat the facing side of the
secondary in short period systems.  The energy added to the outer
layers of the secondary causes it to expand, producing 
higher mass loss leading to enhanced accretion of material onto 
the WD.  

\begin{figure*}[htbp]
\plotone{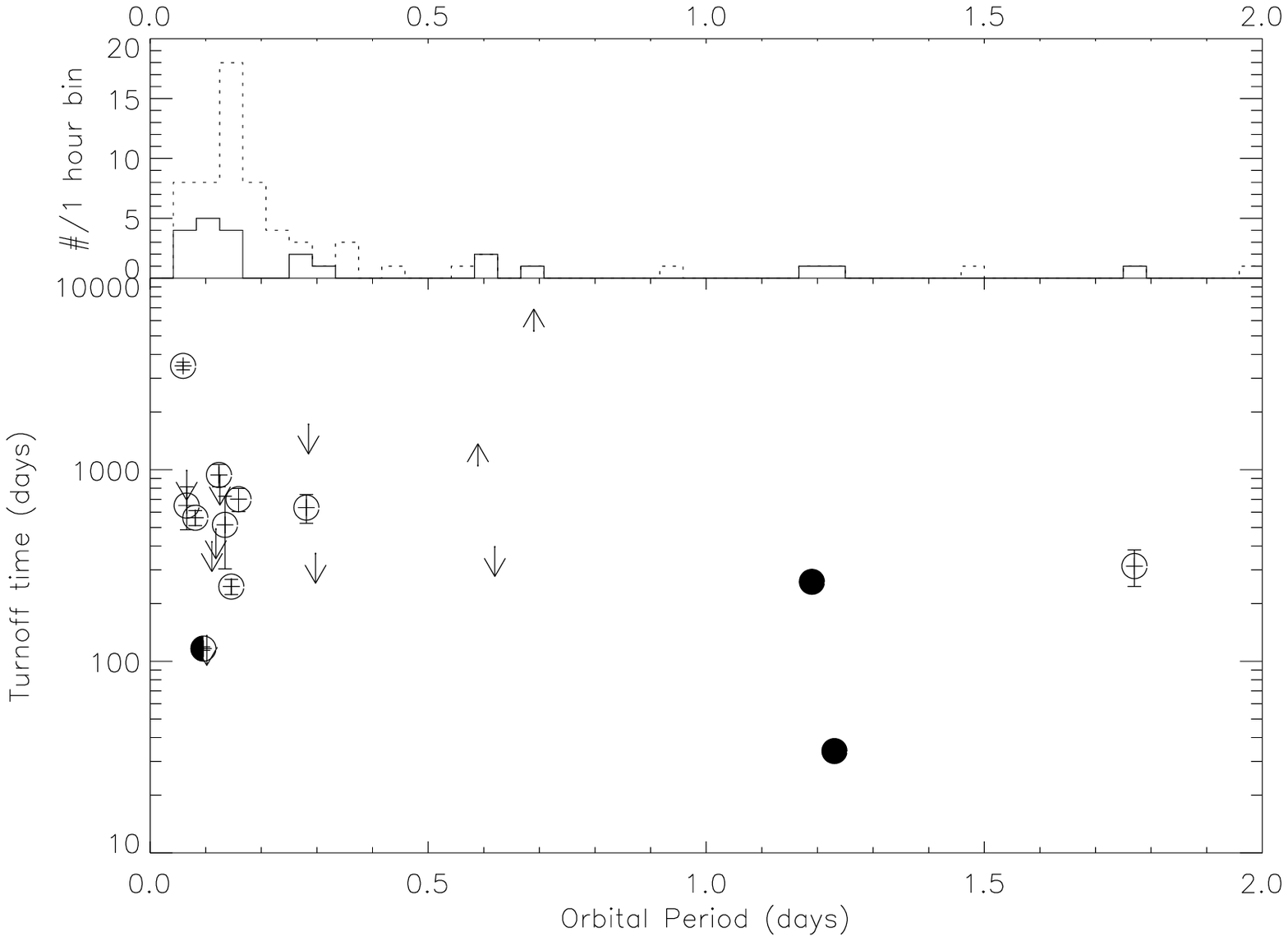}
\caption{SSS turn-off time as a function of orbital period for novae with
well established turn-off times and novae with good upper (i.e. still in
the SSS phase) and lower limits. 
Filled circles are known RNe.  Half filled circles
are suspected RNe based on their characteristics.
The top plot shows the distribution
histogram of our sample (solid line) and of all the known novae (dotted
line) from Table 2.5 in \citet{Warner2008}.
\label{periodvstoff}}
\end{figure*}

Since 2003, the turn-off times of \newturnoffperiod\ additional 
novae with known periods have been determined.  There are also strong 
limits on the turn-off times of \newturnoffperiodlimit\
other novae with known orbital periods. 
Inclusion of this expanded sample, shown in Figure \ref{periodvstoff}, 
causes the trend between orbital period and duration of the 
SSS phase noted by \citet{2003A&A...405..703G} to disappear. 
The new distribution, with an increased sample size, shows 
no discernible correlation.  Orbital separation apparently
has no effect on the duration of nuclear burning. 

To see if the lack of a trend could be explained by having a 
non-representative sample of novae, the top panel of Figure 
\ref{periodvstoff} shows the distribution in 1 hour orbital period 
bins of the updated \citet{Warner2008} sample as the solid line.  
The distribution of all novae with orbital periods is shown as the 
dotted line and shows that the SSS sample is a consistent sub-sample
of the known nova period distribution.

\citet{2010AJ....139.1831S} claim a similar relationship between turn-off 
time and orbital period, albeit in highly magnetized systems.  They find that
of the eight novae with quiescent luminosities $>$10$\times$ brighter 
than pre-eruption, all have long SSS phases, short orbital periods, highly 
magnetized WDs, and very slow declines during quiescence.  
Similar to \citet{2003A&A...405..703G}, \citet{2010AJ....139.1831S}
propose that nuclear burning on the WD is prolonged by increased 
accretion from the close secondary but in this case efficiently funneled 
on the WD by the strong magnetic fields.  The 8 novae 
\citet{2010AJ....139.1831S} cite are CP Pup, RW UMi, T Pyx, V1500 Cyg, 
GQ Mus, V1974 Cyg, V723 Cas, and V4633 Sgr.  

The hypothesis that these specific characteristics enhance the SSS duration 
can be directly evaluated using V4633 Sgr, GQ Mus, V1974 Cyg, and V723 Cas, 
since they all have X-ray observations within the first 3 years of 
outburst.  For CP Pup, RW UMi, T Pyx, and V1500 Cyg the assertion of a long 
lasting SSS emission phase depends on secondary evidence as none had any 
direct X-ray observations during outburst.  Lacking direct X-ray observations
we will ignore these 4 sources for the test.

The first X-ray observation of V4633 Sgr was obtained 934 
days after visual maximum but it and subsequent observations were of 
a hard source implying that any SSS emission was missed.
With an upper limit of 2.5 years for its SSS emission, V4633 Sgr can not
be considered a long-lived SSS nova based on the distribution shown 
in Figure \ref{histogram}a.  The SSS duration in V1974 Cyg was
even shorter and much better constrained at 1.53$\pm$0.14 years.  In
addition, V1974 Cyg was not "excessively" luminous in outburst as alleged in 
\citet{2010AJ....139.1831S}.  Its early UV plus optical fluxes were
consistent with the Eddington luminosity of a WD with a mass range of 
0.9-1.4 M$_{\sun}$ \citep{1994ApJ...421..344S}.  The later "excessive" X-ray 
luminosities of \citet{1998ApJ...499..395B} were derived from blackbody fits 
which are known to predict higher luminosities than model atmospheres 
fit to the same data.
While V723 Cas has the longest SSS duration known among novae ($\gtrsim 15$ 
yrs), its orbital period is very long at 16.62 hrs and significantly longer
than that of GQ Mus, 1.43 hrs.  The claim of magnetic activity in V723 Cas 
is based on the different periodicities observed in the early light curve 
indicating an intermediate polar (IP).  However, the multiple periodicities 
used as evidence by \citet{2010AJ....139.1831S} were from data 
obtained early in the outburst while the nova ejecta were still clearing 
\citep{1998CoSka..28..121C}.  Photometry obtained at this early stage of
development frequently results in noisy periodograms.  Data obtained
later in the outburst by \citet{2007AstBu..62..125G} and over the
the last 4 years from our own photometric monitoring \citep[Hamilton C.,
private communication,][]{2007AAS...210.0404S} reveal a well defined
16.7 hr period with a large $\sim$ 1.5 magnitude amplitude in the UV, 
optical and NIR bands.  There is no other evidence in the literature 
to support that V723 Cas is magnetic. 
Of the 4 novae with supporting X-ray observations, only GQ Mus fully
matches the criteria of a long lasting SSS on a magnetic WD in 
a short period system.  

With our expanded X-ray sample there are 3 additional novae with well 
constrained SSS durations that can potentially be used to test the hypothesis.
V4743 Sgr \citep{2006AJ....132..608K}, V597 Pup \citep{2009MNRAS.397..979W}, 
and V2467 Cyg \citep{2008ATel.1723....1S} are IP candidates and thus 
believed to have strong magnetic fields.  The orbital periods for 
V597 Pup and V2467 Cyg are relatively short at 2.66 and 3.8 hrs,
respectively, but the period in V4743 Sgr is much longer at 6.74
hrs, see Table \ref{chartable}.  While the turn-off times for
these novae are all longer than one year they are not exceptionally
long, with durations of 1.74$\pm$0.29, 1.25$\pm$0.04, and 1.85$\pm$0.33
years for V4743 Sgr, V597 Pup, and V2467 Cyg, respectively.  Thus
the data available do not imply that short orbital period or
strong magnetic fields produce significantly longer SSS 
than the average novae from our sample.

An interesting question is why there is no trend between orbital
period and SSS duration as the underlying assumption of enhanced
accretion due to heating of the secondary is sound.  One reason 
would be that there is no significant enhancement in the 
mass transfer rate from the illuminated secondary, perhaps from shielding
due to a thick disk.  Another possibility is that there is an
effect but it is subtle and affected by other variables such as the
strength of the magnetic field, composition of the accreted material,
WD mass, etc.  Another possibility is that an accretion disk can not
form under the harsh conditions during the SSS phase which inhibits 
additional mass transfer.  More observations of novae with different 
characteristics are required in order to understand the underlying
physics.

\subsection{Dusty novae}

The creation, evolution, and eventual destruction of dust occurs on 
relatively rapid time-scales in novae making them excellent objects 
for understanding dust grain formation.  One curious aspect of dust
in novae is how grains can grow within the harsh photoionizing 
environment.  A correlation of the recent \textit{Spitzer}
spectroscopic observations of dusty novae 
\citep[see][for examples]{WoodStar11,Heltonthesis}
with this large X-ray sample can bring insights to why most novae do 
not form dust and the reasons for the large differences in composition 
and amounts in the novae that do form dust.

In general, it is believed that grain growth occurs within dense clumps 
in the ejecta where they are shielded from hard radiation.  Spectroscopic 
and direct imaging show that nova shells are inherently clumpy 
\citep{1997AJ....114..258S,OBB08}.  Grain formation inside dense clumps 
also explains the higher frequency of dust in slow novae 
\citep[see Table 13.1 in][]{ER08} as they eject more material at lower 
velocities and suffer greater remnant shaping than fast novae and thus 
provide more protection for grain formation. However, even fast novae 
with small ejected masses have shown some dust formation, such as V838 Her
\citep{2007ApJ...657..453S}.  A contrary view was proposed that ionization 
actually promotes dust formation via the accretion of grain clusters through
induced dipole interactions \citep{2004A&A...417..695S}.

Known and likely dusty novae represent 31\% of the X-ray sample but only 
two, V2467 Cyg and V574 Pup, were also SSSs.  While there were no 
characteristic dips in either visual light curve indicating significant 
dust formation \citep{2009AAS...21349125L,2005IBVS.5638....1S}, both novae 
showed evidence of some dust formation from the presence of weak silicate 
emission features in the late \textit{Spitzer} mid-IR spectra 
\citep{WoodStar11}.  In V2467 Cyg the first \swift\ X-ray detection was 458 
days after maximum. It was weak but dominated by soft photons.  The following 
\swift\ observation on day 558 revealed the nova was still soft but
was also almost 3 times brighter.  The {\it Spitzer} spectra showing 
weak dust features were taken between these \swift\ observations, 
around day 480.  V574 Pup was detected as a SSS by 
\xmm\ and \swift\ 872 and 1116 days after visual maximum, respectively.
{\it Spitzer} observations taken around the same time as the \swift\ data 
showed the same weak silicate emission features seen in V2467 Cyg.  
The X-ray observations confirm that dust, albeit weak, can exist in the 
ejecta when the amount of photoionizing radiation is at its peak.  Detailed
photoionization modeling of these novae is required to determine if clumps
existed in the ejecta during this time and if the conditions were sufficient
to protect the dust grains.

There are also two strong dust formers in the sample with hard X-ray emission.
V2362 Cyg was detected numerous times by \swift\ \citep{2008AJ....136.1815L} 
and twice with \xmm\ \citep{2007ATel.1226....1H} but none of the observations
were consistent with a SSS.  However, V2362 Cyg had significant dust 
emission at the times of the \swift\ and first \xmm\ observations.  
The dust likely formed in the later extraordinary mass ejection event
that produced the large secondary peak in the light curve and increased
ejection velocities.  The additional material would have absorbed the soft
X-ray emission and delayed the onset of any SSS phase.
In the last \xmm\ observation it was extremely faint indicating that 
if there was a SSS phase it was over by 990 days after maximum.  
V1280 Sco was detected as an X-ray source late in its outburst but the X-rays 
were relatively hard and faint \citep{2009ATel.2063....1N}.  V1280 Sco 
has yet to be observed as an SSS and its internal extinction is still large.
In both V2362 Cyg and V1280 Sco, grain growth was likely enhanced to
produce the large dust events due to the effective shielding of the 
large mass ejections.

\subsection{Variability during SSS phase \label{var}}

At the maximum effective temperature, (2-8)$\times$10$^5$ K, the bulk of
the emission in a nova outburst comes from X-rays which are primarily soft. 
Assuming the external column is low enough and the effective temperature
is suitably high enough, this X-ray emission can be detected.  The theory
of constant bolometric luminosity predicts that at X-ray maximum the 
light curve should be relatively constant since one is observing the 
majority of the emitted flux.  Constant bolometric luminosity has been 
observationally verified in the early phase of the outburst from the
combined UV, optical and near-IR light data, {\it e.g.} FH Ser
\citep{1974ApJ...189..303G}, V1668 Cyg \citep{1981MNRAS.197..107S} and 
LMC 1988\#1 \citep{1998MNRAS.300..931S}.  However, the expected X-ray 
plateau in all well studied \swift\ novae has been far from constant.  
In addition, the rise to X-ray maximum also shows large amplitude 
oscillations.  What is the source of the variability during both phases?

One important caveat when discussing the \swift\ data is that the XRT 
count rate is not a direct measure of the bolometric flux, only the 
portion that is emitted between 0.3 and 10 keV.  During the SSS phase the 
vast majority of photons are emitted within this range but if the effective 
temperature varies due to photospheric expansion or contraction, the XRT 
count rate will change even if the source has a constant bolometric 
luminosity \citep[see also][]{2011ApJ...727..124O}.
Figure \ref{xrtevol} illustrates how the estimated XRT count 
rate varies as a function of effective temperature for simple blackbody 
models (WebPIMMS\footnote{http://heasarc.nasa.gov/Tools/w3pimms.html})
assuming a constant luminosity and column density
see Section \ref{Xrayvar}.  A decline from 500,000 K
to 400,000 K drops the total \swift\ XRT count rate by a factor of 6.
The change in HR1 is almost a factor of 10 while there is
essentially no change in the HR2 hardness ratio.  Thus changes
in temperature might in principle account for the observed X-ray 
oscillations, see Section \ref{teffvar}.  Why the temperature or radius 
of the WD photosphere would change on the observed 
time scales remains an open question however.  The next sections outline 
possible explanations for the variations seen during the SSS phase.

\subsubsection{Variable visibility of the WD\label{Xrayvar}}

Figures 2 and 5 in \citet{2011ApJ...727..124O} show in exquisite detail
the rapid and extreme variability in the X-ray light curve and hardness
ratio evolution in RS Oph.  In general the trend was for RS Oph to be softer
at high X-ray flux but counter examples were also observed.
\citet{2011ApJ...727..124O} cite variable visibility of the hot WD as a 
possible explanation of the observed phenomena.  Changes in the 
extinction can come from either variable ionization of the ejecta leading 
to changing extinction at higher ionization states or neutral absorption 
from high density clumps passing through the line of sight.  Changes in 
the ionization structure of the ejecta are unlikely given the rapid 
hour to day time-scales but are consistent with the crossing times of 
small, dense clumps traveling across the line of sight assuming transverse 
velocities of a few percent of the radial velocity.  There is evidence 
for this at other wavelengths.  For example, a sudden absorption component 
that appeared in the Balmer lines of V2214 Oph in July 1988 was interpreted 
by \citet{1991ApJ...376..721W} as the passage of an absorbing clump in 
front of the the emitting region.  Both types of absorption changes 
should be manifest as a hardening of the X-ray spectrum or increase in 
the hardness ratio with increasing soft flux emission consistent with 
the counter examples of \citet{2011ApJ...727..124O}.

\begin{figure*}[htbp]
\plotone{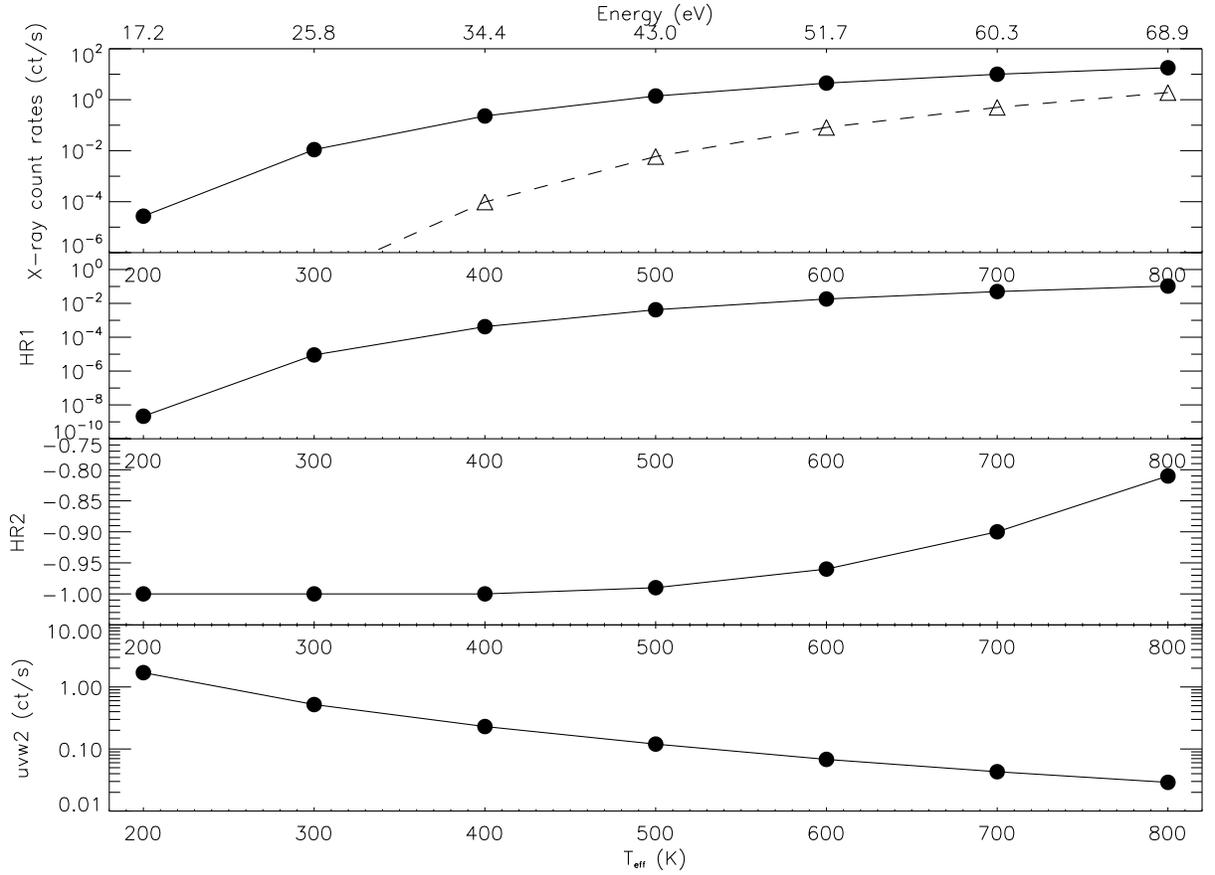}
\caption{The logarithmic X-ray count rates, hardness ratios and logarithmic
$uvw2$ count rates
as a function of blackbody temperature as calculated by webPIMMS.
An unabsorbed, bolometric flux of 3.3$\times$10$^{-8}$ erg/s/cm$^{-2}$ 
(1$\times$10$^{38}$ erg s$^{-1}$ at 5 kpc) and N$_H$ of 
3$\times$10$^{21}$ cm$^{-2}$ was used in all models.
The top panel shows the soft (0.3-1 keV, solid line and filled circles) 
and hard (0.1-10 keV, dashed line and triangles) count rates.  The soft
contribution dominates at all effective temperatures.  The middle panels 
show the hardness ratios HR1(=H/S) and HR2(=(H-S)/(H+S)).  The bottom panel 
shows how the $uvw2$ count rate increases as the blackbody temperature 
declines.  \label{xrtevol}}
\end{figure*}

As a test of the neutral absorption theory we use the model results 
from a recent photoionization analyses in WebPIMMS to determine the
count rates and hardness ratios for different column densities and 
simulate the effect of clumps.  The photoionization models require 
two components, high density clumps embedded within a larger diffuse medium
\citep[see][for details]{2007ApJ...657..453S,2010AJ....140.1347H}
to fit the emission lines of the ejected shell.  For convenience we
use the May 24th, 1991
model parameters for V838 Her in Table 2 of \citet{2007ApJ...657..453S}.
The model uses a blackbody with an effective temperature of 200,000 K to
photoionize a  two component spherical shell.  The model shell has a
clump-to-diffuse density ratio of 3 with a radius equal to the expansion
velocity multiplied by the time since outburst.  To facilitate comparisons 
with the results in Figure \ref{xrtevol}, the same unabsorbed bolometric 
flux is assumed.  WebPIMMS predicts a \swift\ soft band count rate 
of 5.3$\times$10$^{-3}$ ct s$^{-1}$ through the 
lower density diffuse gas (N$_H$ = 1.2$\times$10$^{21}$ cm$^{-2}$)
and 8.5$\times$10$^{-6}$ ct s$^{-1}$ from the higher density clumps
(N$_H$ = 3.7$\times$10$^{21}$ cm$^{-2}$).  While the total count rate
declines by over 100$\times$, the HR2 hardness ratio does not change with
this particular model. The HR2 can vary significantly when using different 
model parameters such as lower initial densities or higher clump to diffuse
density ratios.  Care is required when using hardness ratios of 
low resolution data.  In a SSS source with a hard X-ray component, such as 
RS Oph, the hardness ratio will increase if the soft component decreases 
for any reason, not just due to absorption.  

Another problem with variable visibility in RNe and very fast CNe is that
the amount of mass ejected is very low thus minimizing any effect the ejecta
have on the obscuration of the WD.  The effect should be greater
in slower novae with more ejected mass such as V458 Vul.  In addition,
\citet{2011ApJ...727..124O} find that in RS Oph the ratio of high flux
states to low flux states as a function of energy is not consistent with
either type of variable visibility of the WD.  Rather the best fit comes from
an increase in the effective temperature and declining radius,
see Section \ref{teffvar}.


\subsubsection{Periodic oscillations\label{sssperiods}}

There are several proposed explanations of the periodic X-ray variations.
In the X-ray light curve of V1494 Aql, \citet{2003ApJ...584..448D} found 
periodicities that they attributed to non-radial g$^+$-mode pulsations.  
Similar oscillations have been observed in V4743 Sgr 
\citep{2003ApJ...594L.127N,2010MNRAS.405.2668D}.

The factor of almost ten decline in the \xmm\ X-ray light curve of 
V5116 Sgr was interpreted by \citet{2008ApJ...675L..93S} as a partial 
eclipse of the WD since its duration was consistent with the orbital period.  
Finer binning of the day 762, 764, and 810 \swift\ observations of 
V5116 Sgr reveals the presence of a 500-800 second oscillation.  This
X-ray periodicity is significantly shorter than the 2.97 h orbital
period found by \citet{2008A&A...478..815D}.  In addition,
the day 810 data show a strong flare that increases the count rate by
a factor of three with no significant change in the hardness ratio.
This was similar to the flare seen in V1494 Aql \citep{2003ApJ...584..448D}.
No other flares were seen in the V5116 Sgr data set.

Other orbital periods have been detected with \swift.
U Sco is a high inclination system with deep eclipses and an orbital
period of 1.23 days \citep{2001A&A...378..132E}.  Deep eclipses were
observed in the 2010 outburst in the \swift\ UVOT light curves while the
XRT light curves showed generally lower flux levels during the UV eclipses,
but did not otherwise exhibit clear eclipse signatures.
\citep{2010ATel.2442....1O}.  A 1.19 day orbital period was deduced from 
the \swift\ UVOT light curves in the RN nova LMC 2009a 
\citep{2009ATel.2001....1B}.  This orbital period was also observed in the 
XRT light curve during the SSS phase, but with a lag with respect to the 
UV/optical of 0.24 days \citep{Bode2011}.

\begin{figure*}[htbp]
\plotone{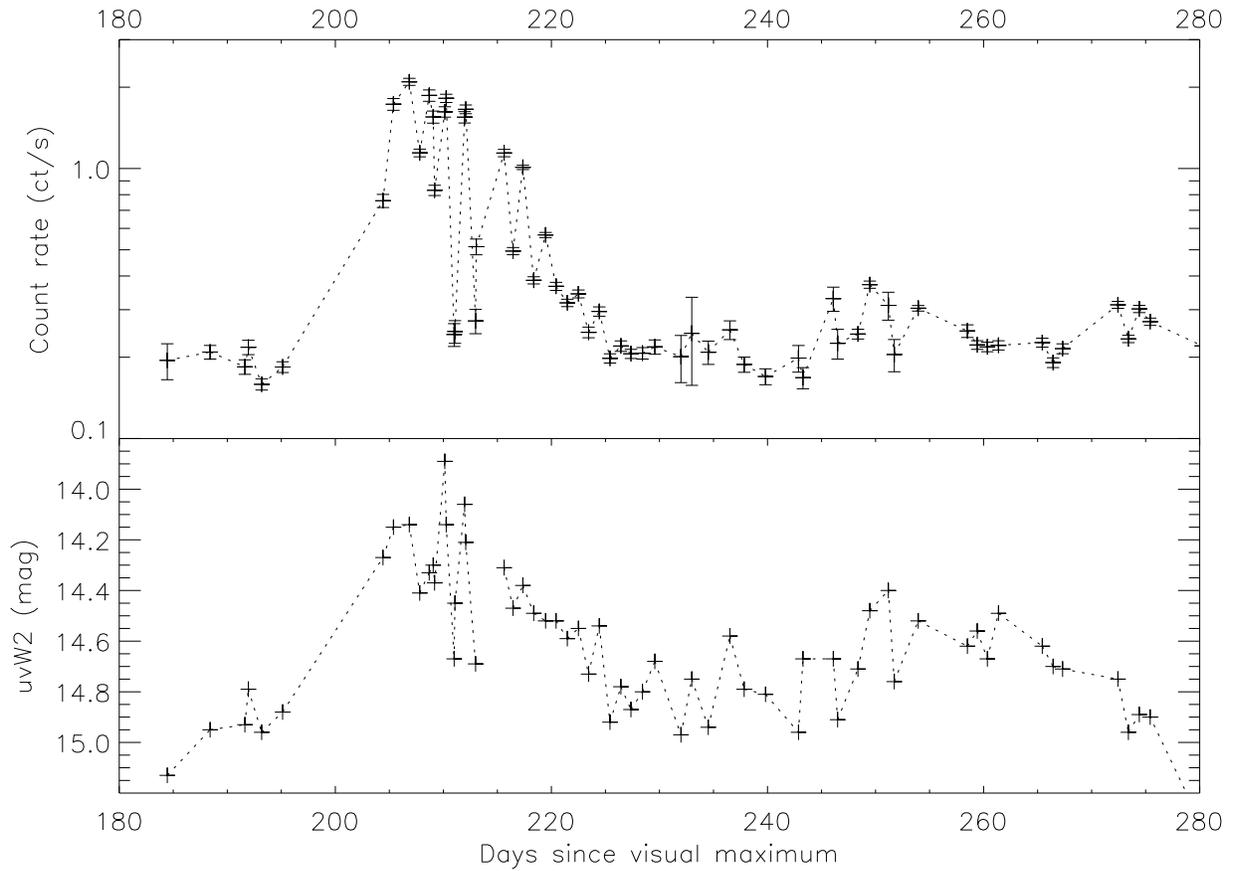}
\caption{The X-ray and uvw2 light curves of the particular nova
CSS 081007:030559+054715.  The X-ray and UV evolution are in phase.
\label{csslc}}
\end{figure*}

The X-ray behavior in CSS 081007:030559+054715 was extremely unusual.  This
odd source was discovered well after optical maximum by the Catalina Real-time
Transient Survey \citep{2008ATel.1835....1P}.  Its X-ray spectra were
extremely soft, consistent with the low extinction along its position
high above the Galactic plane ($b$ = -43.7$\arcdeg$) 
which places it well outside the plane of the Galaxy where novae
are generally not located.  Figure \ref{csslc} shows the \swift\ XRT/UVOT
light curves compiled from the data in Table 2.  
To first order both light curves are in phase with 
significant variability superimposed over three major maxima.  
\citet{2010AN....331..156B} report that the 
\swift\ light curves are unique with a 1.77 day periodicity.  They speculate 
that the period is due to obscuration of the X-ray source in a high 
inclination system with a 1.77 day orbital period.

Oscillations significantly shorter than the hours-to-days of typical 
novae orbital periods have also been detected with \swift.  Oscillations 
of order 35 s have been observed in RS Oph 
\citep{2006ATel..770....1O,2011ApJ...727..124O}
and KT Eri \citep{2010ATel.2423....1B}.  Some WDs have rotation 
periods in this range \citep[{\it e.g.} 33 s in AE Aqr;][]{2008PASJ...60..387T}.
It seems unlikely that RS Oph and KT Eri should both have nearly identical 
rotation periods unless the pulsations are tied to the mass of the WD
which for both novae are predicted near the Chandrashkar limit.  
Another reason the observed
variability might not be associated with the rotating WD is that 
the $\sim$ 35 second periodicity is not always detected in the \swift\
and \cxo\ X-ray light curves.  The 35 second pulsations could be due to 
a nuclear burning instability on the WD surface 
\citep[see ][]{2011ApJ...727..124O}.  If so, then the period then is a 
function of WD mass, and perhaps indicates that the WDs in RS Oph and KT Eri
are near the Chandrasekhar mass.

\subsubsection{Temperature variations\label{teffvar}}

\begin{figure*}[htbp]
\plotone{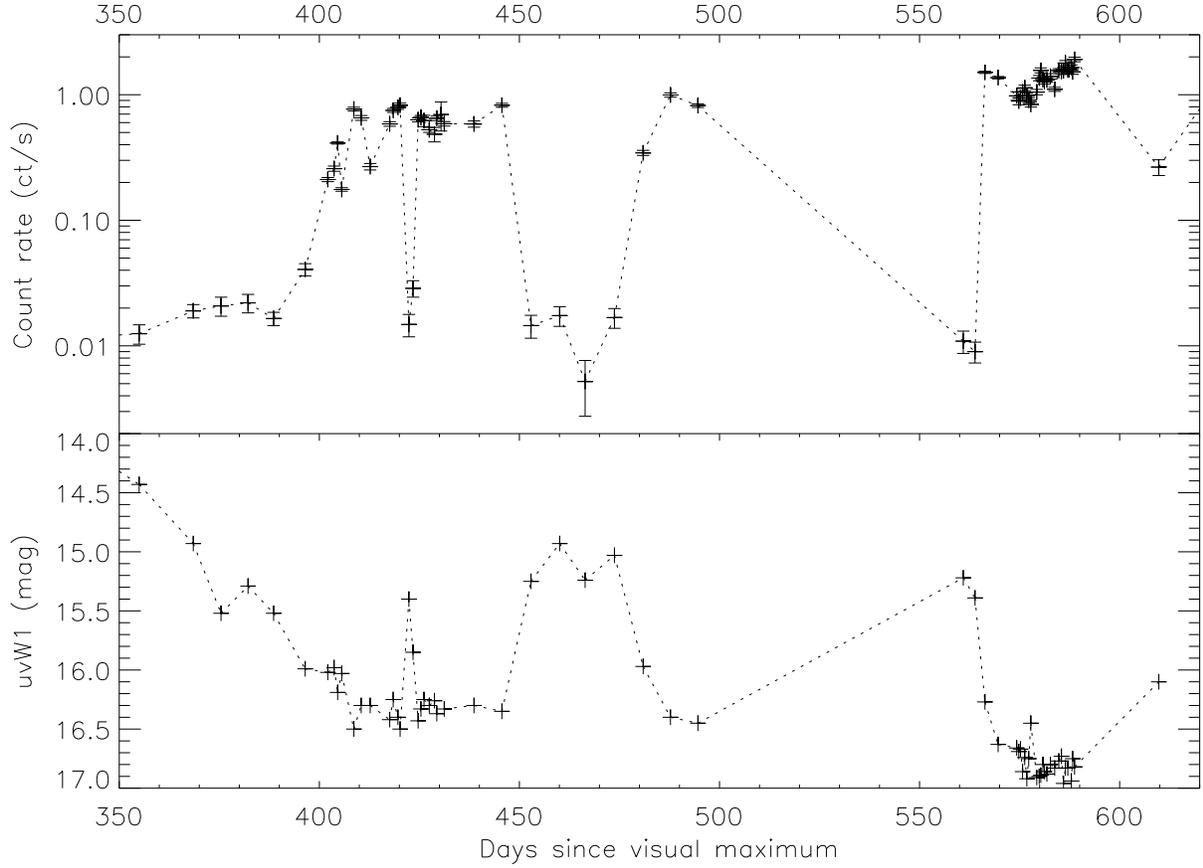}
\caption{\swift\ X-ray (top panel) and uvw1 (bottom panel) light curve 
for V458 Vul.  The X axis is the number of days after visual maximum.  
Prior to day 400, V458 Vul was in transition to X-ray maximum.  
After day 400 the majority of the X-ray observations had a count rate 
of $\sim$ 1 ct s$^{-1}$.  However, during the later phase there 
are three periods where the X-ray counts declined by about a factor 
of 100.  During these times the uvw1 ($\lambda_c$ = 2600\AA)
photometric brightness increased by a magnitude.
\label{v458vullc}}
\end{figure*}

Long lived SSS, such as Cal 83, have non-periodic X-ray on/off states.  
\citet{2000A&A...354L..37R} speculate that the 
decline in X-ray flux is due to accretion disk interactions such
as an increase in the mass accretion rate causing the WD photosphere 
to expand and shifting the SED into the EUV.  These sources then become 
optically brighter from the irradiation of the accretion disk and 
secondary by the larger WD photosphere.  The source remains X-ray faint 
until the WD photosphere shrinks back to its
original size.  Figure \ref{v458vullc} shows similar behavior in the 
\swift\ X-ray and UV light curves of V458 Vul
compiled from the data in Table 2.  The 100$\times$ decline
in the X-ray light curve is matched by a 1.5 magnitude increase
in the UV light curve.  Figure \ref{xrtevol} shows that similar X-ray
and UV variations can be achieved by large declines in the effective 
temperature. For example, a decline from 700,000 K to 500,000 K 
produces a factor of 85 decline in the X-ray count rate and a 1.1 
magnitude $uvw1$ band increase.  
If the underlying phenomenon in V458 Vul 
is the same as proposed for RX J0513.9-6951 \citep{2000A&A...354L..37R}
and Cal 83 \citep{2002A&A...387..944G} and the accretion disk has been
reestablished, V458 Vul should have an orbital period of order one
day to produce an accretion rate high enough to drive stable nuclear burning.
However, \citet{2010MNRAS.407L..21R} find a short orbital period of $\sim$ 
98 minutes implying that V458 Vul will not have a long term SSS phase.

\subsection{Estimating time scales in a variable environment}

The variability of novae also raises questions about how confident 
one can be in the determination of turn-off times.  A prime example can be 
seen in the X-ray light curve of V458 Vul in Figure \ref{v458vullc}.  If 
monitoring had stopped following the four observations between days 450 
and 480 the subsequent recovery would never have been found, and it
would have been noted that V458 Vul had a turn-off time of 1.2 years 
instead of $\gtrsim$ 2.9 years.  While this could be a significant 
problem with the determination of a turn-off time in most novae, it is 
likely that the phenomenon observed in V458 Vul is rare.  The X-ray 
behavior of V458 Vul, a 100$\times$ decline in flux and a subsequent 
recovery, is the \textit{only} case observed in the \totallimitswiftSSS\ 
novae studied by \textit{Swift} with SSS emission.
\citet{2001A&A...373..542O} found no similar "reborn" SSSs in their review
of the \rosat\ all sky survey although some of the novae in M31 previously
thought to be RNe with very rapid outburst time scales may actually be
normal novae but with on/off states similar to those in V458 Vul.  Since the 
sudden X-ray declines in V458 Vul also had corresponding UV rises, if these
source exist in M31, they should be easily found with X-ray and UV
capable facilities such as \swift\ and \xmm.

\subsection{SSS in RNe and the light curve plateau\label{RNplateau}}

A plateau in visible light of RNe is speculated to arise from the reradiation 
of the SSS emission from an accretion disk dominating the emission after 
the free-free emission has faded \citep{2008ASPC..401..206H}.  Once nuclear 
burning ends and the accretion disk is no longer irradiated, the light curve 
continues its decline to quiescence.  Figure 17.1 and 
17.2 show that the optical plateaus are nearly coincident with 
the SSS emission in RS Oph and U Sco.

The other well observed RNe in the \swift\ archive are novae LMC 2009a 
and V407 Cyg.  LMC 2009a was previously seen in outburst in 1971 
\citep{2009IAUC.9019....1L}.  It had a much longer SSS phase than RS Oph 
and U Sco and it ended 270 days after maximum.  Unfortunately, the V band light 
curve compiled from the AAVSO archives and our own SMARTS photometry 
does not extend beyond 110 days after visual maximum so we can not 
determine whether an optical plateau was observed later in this outburst,
see Figure 17.3.  However, the \swift\ uvw2 and SMARTS B
band light curves are relatively flat during the SSS phase (see Bode et al. 
submitted) indicating LMC 2009a did go through an optical plateau phase.
The data
are not as extensive for V407 Cyg but the rise in the soft X-ray emission
consistent with nuclear burning on the WD (see Section \ref{v407cygSSS})
is coincident with a short plateau in the optical light curve as shown
in Figure 17.4.

There are three other novae with well observed SSS phases in the \swift\
archive that are suspected to be RNe based on their outburst characteristics.
The novae are V2491 Cyg, KT Eri, and V2672 Oph.  Figure 17.5
shows that there is no indication of a plateau in V2491 Cyg while it was
a SSS.  However, the SSS phase in V2491 Cyg was extremely short, $<$10 days,
which may not be sufficient time to produce a noticeable optical plateau
or the system did not have its accretion disk reform this early in the
outburst. 

The early outburst spectra of KT Eri were indicative of the He/N class with 
high expansion velocities typical of RNe \citep{2009ATel.2327....1R}.  
KT Eri also had short X-ray light curve modulation similar 
to RS Oph, see Section \ref{sssperiods} and
\citet{2010ATel.2392....1B}, while Bode et al. (2011, in prep) draw 
attention to KT Eri's similarities with the X-ray behavior of LMC 2009a.
The X-ray and V band observations are shown in Figure 17.6.
The AAVSO V band light curve shows a flattening at 80 days after visual maximum
or about 10 days after KT Eri became a SSS implying there was an
optical plateau.

The case for V2672 Oph as a RNe is based on its extreme expansion velocities
at maximum \citep{2009IAUC.9064....2A} and early radio synchrotron emission
similar to that observed in RS Oph \citep{2009ATel.2195....1K}.  
\citet{2010MNRAS.tmp.1484M} also find many similarities between V2672 Oph
and U Sco.  Unfortunately, the X-ray and optical observations were 
hampered due to the faintness at visual maximum and the relatively 
large column density.
Based on the hardness ratio, V2672 Oph was in its SSS phase between days 15 
and 30 after visual maximum (Figure 17.7).  The AAVSO 
V band light curve is supplemented with SMARTS V band photometry which
shows a plateau between day 10 through 50 after visual maximum.

Of the 4 known RNe and 3 suspected RNe, there are sufficient optical data 
to reveal the presence of a plateau in six.  Of those six all but V2491
Cyg have evidence of an optical plateau correlated with the X-ray SSS
emission.  However, \citet{2010ApJS..187..275S} finds that not all Galactic 
RNe have optical plateaus.  It is interesting to note that if the plateau
phase is caused by reradiation off an accretion disk as suggested by
\citet{2008ASPC..401..206H} then there is no effect on the presence 
or strength of the plateau due to the inclination of the system.  One
would expect the effect in more face-on systems like RS Oph,
$i$ = 39$^{+1}_{-10}$$^{\circ}$ \citep{2009ApJ...703.1955R} than in
edge-on systems such as U Sco, $i = 82.7\pm2.9^{\circ}$ 
\citep{2001MNRAS.327.1323T}.  Regardless of the root cause of optical 
plateaus, their presence can clearly be used as a proxy signature of 
SSS emission.  
However, it should be stressed that while optical plateaus likely 
indicate soft X-ray emission, the start and ending of this phase in the 
optical light curve does not necessarily correspond to the turn-on and
turn-off times in the SSS phase.  Relationships between the optical
timescales and the X-ray are only weakly correlated, e.g. 
Fig. \ref{t2turnoff}, and the two phases do not always align in the 
RN and suspected RN in this sample (Fig. 17.1-17.7).

Optical/NIR plateaus should only be observed in RNe and other fast 
novae that eject very little mass.  In slower novae the later spectra 
({\it i.e.} several tens of weeks after maximum light) are dominated by 
hydrogen recombination and nebular line emission effectively hiding any 
irradiation effects.  The continuum from the WD or a hot accretion disk 
can only be observed after the ejecta have sufficiently cleared.  

\figsetstart
\figsetnum{17}
\figsettitle{X-ray and optical evolution}

\figsetgrpstart
\figsetgrpnum{17.1}
\figsetgrptitle{RS Oph}
\figsetplot{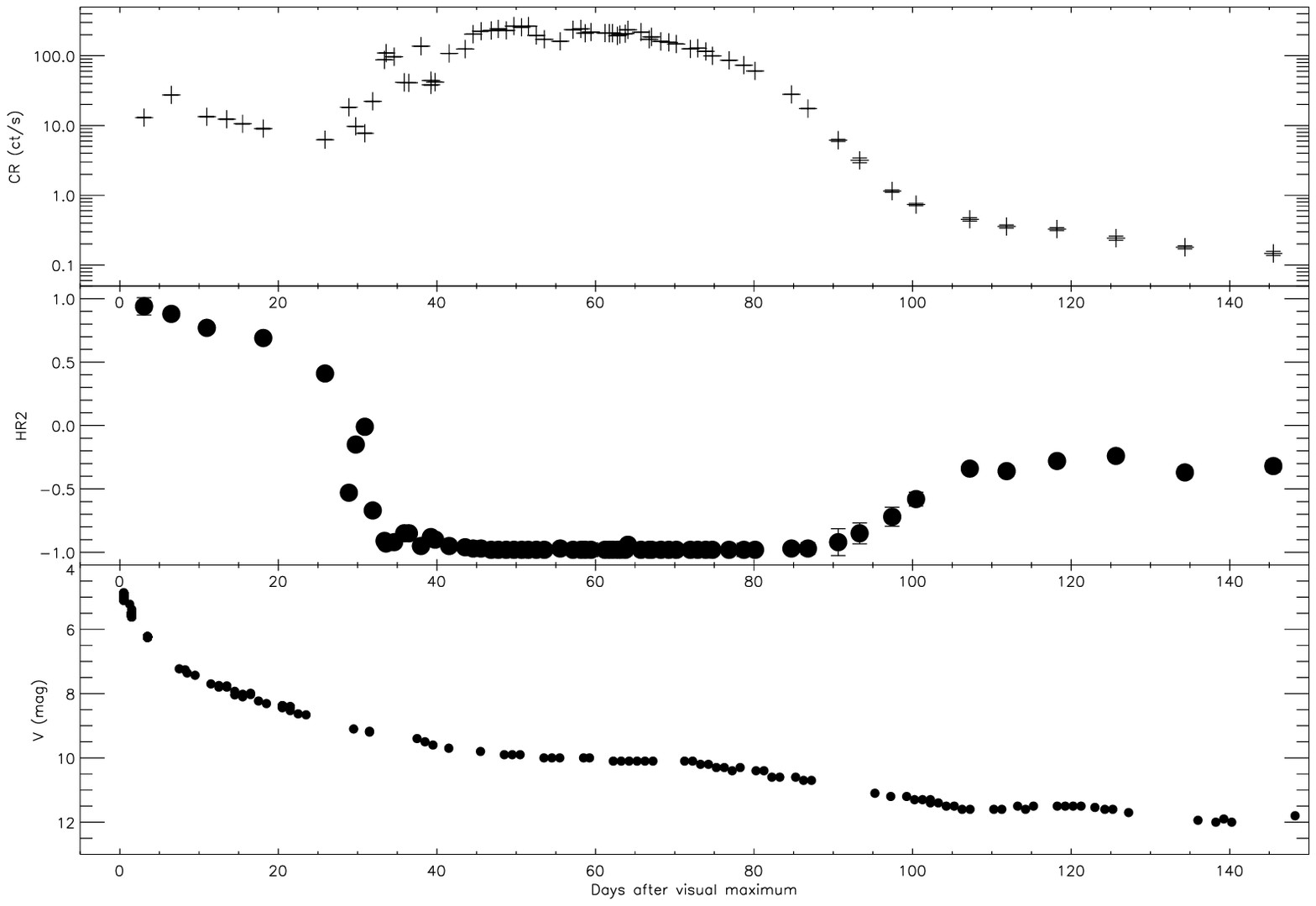}
\figsetgrpnote{X-ray and optical evolution of RS Oph.  The top panel is the \swift\ XRT (0.3-10 keV) count rate and the middle panel is the hardness ratio, (H-S)/(H+S) where H = 1-10 keV and S = 0.3-1 keV.  The bottom panel shows the AAVSO V band light curve.  \label{rsophplat}}
\figsetgrpend

\figsetend

\begin{figure*}[htbp]
\plotone{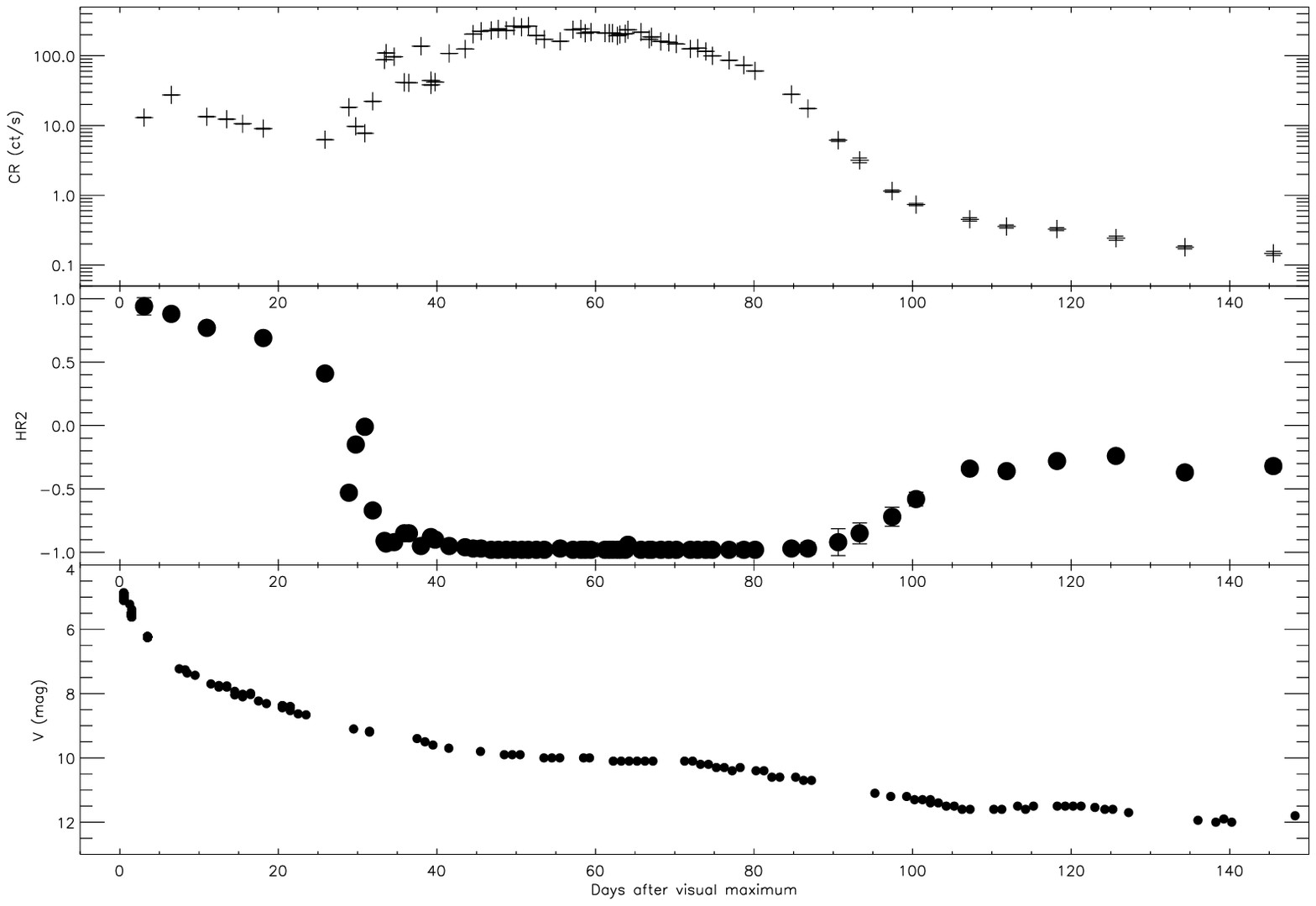}
\caption{X-ray and optical evolution of RS Oph.  The top panel is the \swift\
XRT (0.3-10 keV) count rate and the middle panel is the hardness ratio, 
(H-S)/(H+S) where H = 1-10 keV and S = 0.3-1 keV.  The bottom panel
shows the AAVSO V band light curve.  Similar figures for U Sco, Nova LMC 2009
A, V407 Cyg, V2491 Cyg, KT Eri, and V2672 Oph are available in the electronic
edition.  \label{rsophplat}}
\end{figure*}

\subsection{SSS proxies at other wavelengths: The \FeX\ line\label{fex}}

\btxt{\citet{2001AJ....121.1126V} used the evolution of UV emission
line light curves developed in \citet{1996ApJ...463L..21S} for V1974 Cyg
to estimate turn-off times.  This allowed \citet{2001AJ....121.1126V} to
determine the nuclear burning timescales of five novae with no pointed
X-ray observations but significant amounts of {\it IUE} data.  Unfortunately,
it is currently difficult to obtain sufficient UV emission line data to 
utilize this technique while the optical plateau (\S \ref{RNplateau})
only applies to fast and recurrent novae.  Another X-ray proxy is 
needed for slower novae.}

The emergence of the coronal \FeX\ 6375\AA\ line in the nebular spectra
of novae has been long recognized as a strong indication of photoionization 
of the ejecta from a hot source \citep[e.g.,][]{1989ApJ...341..968K}.  
With an ionization potential of 235 eV, an ejected shell must be highly 
ionized by a hot WD to produce \FeX.  While shocks can produce 
high temperatures, they only contribute very early in the outburst
and are insignificant during the later nebular phase when \FeX\ is typically
observed, in all but the RS Oph-type RNe. For example, strong \FeX\ 
and [\ion{Fe}{14}] 5303\AA\ emission has been observed in RS Oph in all 
outbursts with adequate spectroscopic coverage \citep{2009ApJ...703.1955R}. 
However, these lines appear well before the SSS phase begins.  A 
relationship between \FeX\ and soft X-ray emission has not been previously 
demonstrated but can be strengthened with our larger nova sample.

\begin{figure*}[htbp]
\plottwo{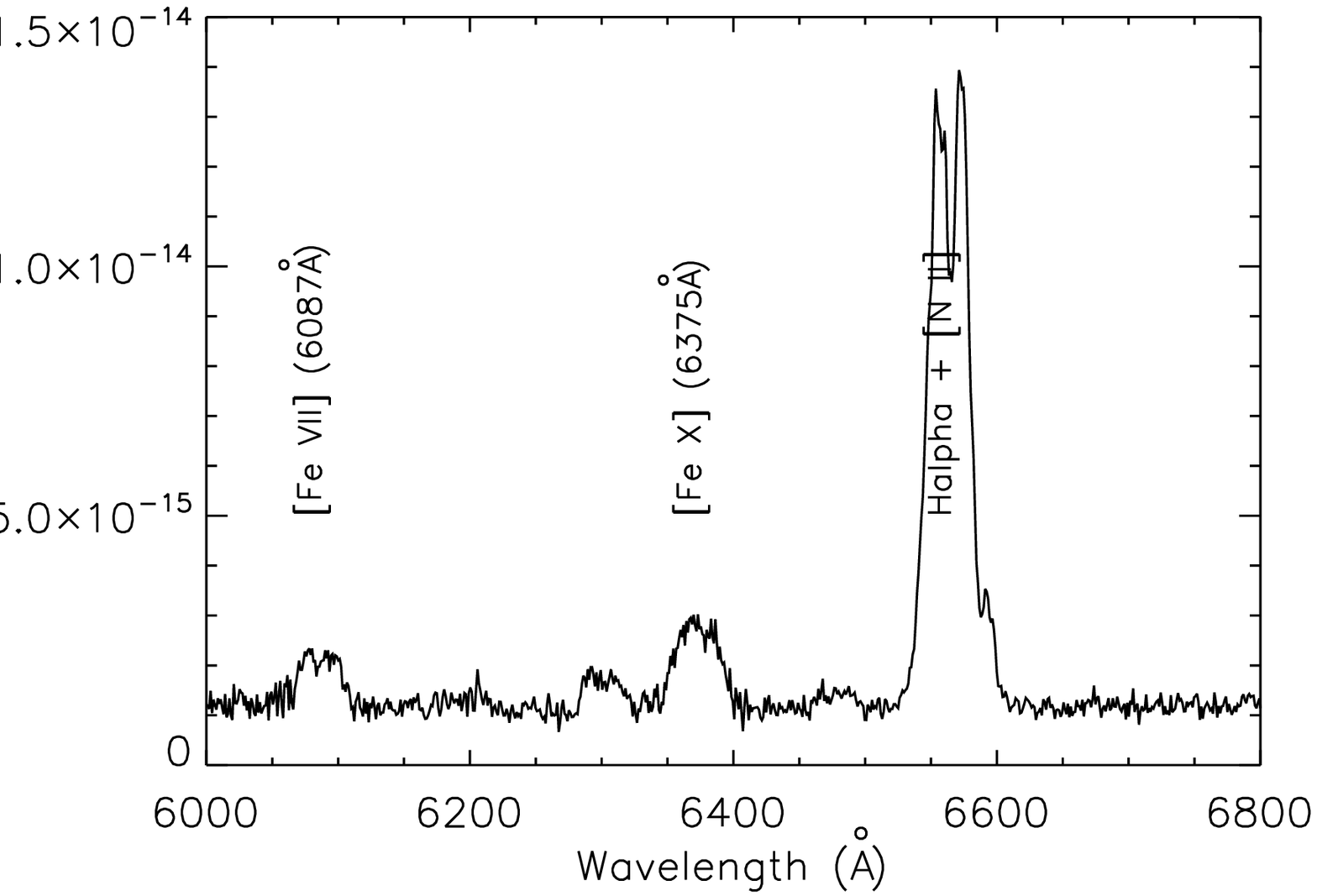}{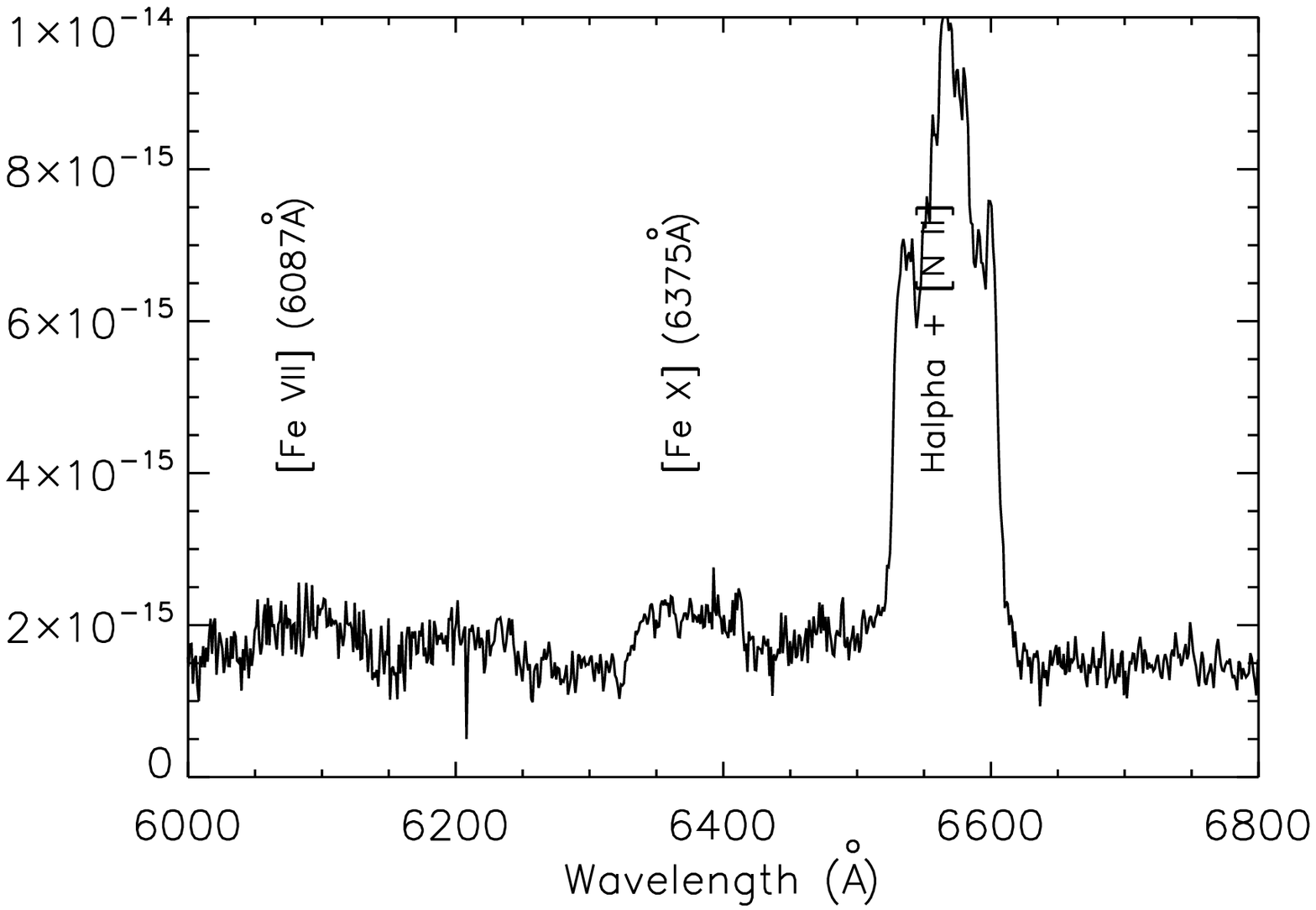}
\caption{\FeX\ 6375\AA\ emission in V1213 Cen (left) and V597 Pup (right)
obtained on June 27th, 2010 (415 days from visual maximum) 
and March 26th, 2008 (133 days from visual maximum), respectively.
The $[$\ion{Fe}{7}$]$ 6087\AA\ line is also visible in both spectra. Strong
\FeX\ emission relative to $[$\ion{Fe}{7}$]$ is a hallmark of novae in
their SSS phase.  
\label{fexplots}}
\end{figure*}

Seven novae with confirmed SSS emission, GQ Mus \citep{1989ApJ...341..968K},
V1974 Cyg \citep{1995A&A...294..488R}, V1494 Aql \citep{2003A&A...404..997I},
V723 Cas \citep{2008AJ....135.1328N}, V574 Pup \citep{Heltonthesis}, V597 Pup, 
and V1213 Cen \citep{2010ATel.2904....1S} all had strong \FeX\ lines in 
their late nebular spectra.  Example spectra of V597 Pup and V1213 Cen from 
our SMARTS archive are shown in Figure \ref{fexplots}.  In addition, 
extensive optical spectra from our Steward Observatory northern 
hemisphere nova monitoring 
campaign shows that V2467 Cyg may also have had weak \FeX\ emission at the 
same time it was a SSS but this can not be confirmed due to nearby 
\ion{O}{1} lines.  These novae clearly show that the presences of strong
\FeX\ in the optical spectrum is indicitive of underlying SSS emission.
To our knowledge there has never been a nova with strong \FeX\ emission 
that was not also a SSS during contemporaneous X-ray observations.  
While additional optically and X-ray observations are needed to fully test
this hypothesis, ground based spectroscopic monitoring is a powerful tool 
for detecting SSS novae from \FeX\ emission in novae with significant
ejected mass.  The RNe and very fast CNe 
with rapid turn-on/off times are not strong photoionization sources 
long enough to produce \FeX\ in their meager ejected shells.

\section{SUMMARY}

Over the last decade our knowledge of the X-ray behavior of 
novae has increased dramatically with the launch of 
the latest generation
X-ray facilities.  Observations of novae when they are radiating the majority 
of their flux in the soft X-ray band provide critical insight into 
the behavior of the WD and TNR processes.  Currently \totalSSS\ 
Galactic/Magellanic novae have been observed as SSSs of which \totalswiftSSS\ 
such classifications have come from 
over 2 Ms of \swift\ observations during the last five years.  

This large sample shows that individual novae can differ significantly 
from fits to smaller ensemble data sets such as the t$_2$ relationship of 
\citet{2010ApJ...709..680H} and the expansion velocity relationship of 
\citet{2003A&A...405..703G}.  Surprisingly, there is also no relationship 
between orbital period and the duration of nuclear burning.  This large
data set confirms that many factors are in play in the evolution of the 
SSS phase.

The duration of nuclear burning on the WD is short, with 89\% of the 
novae have turned off within 3 years in this expanded sample.  The median
duration of the sample is 1.4 years. This contrasts with the same distribution
in M31 which is peaked at longer burning novae.  The difference is likely
a selection effect between the two surveys.

The new \swift\ data are also challenging our understanding of novae 
with highly variable X-ray light curves both during the rise to and at
X-ray maximum.  Various mechanisms are likely at work to produce the 
variability.  Additional observations are warranted not only to help
decipher the current peculiar observations but also to be sure that we
have captured the full range of variability behaviors both periodic and
non-periodic that novae may yet produce.  
Long \xmm\ and \cxo\ grating observations can explore the short 
term oscillations more effectively than \swift\ whereas \swift\ can
easily track the long term behavior such as turn-on and turn-off times.
In addition, simultaneous X-ray/UV 
observations only available through \xmm\ and \swift\ will continue to
be a powerful tool to test the evolution of the emission from the WD 
during the outburst.

To date no strong dust-forming novae have been detected as a SSS.  
V2362 Cyg did have detectable soft X-ray photons but it was not similar 
to any of the other SSS novae.  While V574 Pup and V2467 Cyg were in
the SSS phase they had IR features indicating weak silicate dust emission.
V1280 Sco had a large DQ Her-like dust event but also ejected so much 
material and at a low velocity that it is still optically thick several
years after visual maximum.  Any SSS phase will not be detected until
this material clears.

There are optical behaviors that track SSS emission in novae.  
For the RNe with well defined plateaus 
in their optical light curves, RS Oph and U Sco, the X-ray light curves
reach maximum around the same time.  However, not all RNe and suspected
RNe in the sample had optical plateaus even though they had well 
documented observations during X-ray maximum.  An optical spectroscopic 
signature indicative of an SSS phase is the presence
of strong \FeX\ 6375\AA\ emission.  In the sample, all novae with \FeX\ that
were subsequently observed in the X-ray were SSSs.  These were slower
novae that ejected significantly more material than the RNe.  The inverse 
of the \FeX\ relationship does not hold since the source may turn off before 
\FeX\ can be created in the ejecta.  While the presences of neither optical
plateaus or \FeX\ has yet been shown to be simultaneous with SSS emission,
these relationships offer excellent oppertunities to use ground-based 
monitoring to coordinate X-ray observations during the important 
SSS phase.  

Additional X-ray data need to be collected since the sample is statistically 
meager with only \totalSSS\ known SSS novae and is smaller still for novae
with early, hard X-ray detections.  Trends can be difficult to confirm given
the wide range of behavior observed during the different X-ray phases.
With the sample heavily biased toward fast and recurrent novae, efforts 
should be expended on novae that are not 
currently well represented in the X-ray sample such as slow and 
dust forming novae. The monitoring of the two slow novae that 
have been detected as X-ray sources, V5558 Sgr and V1280 Sco, but 
have not yet evolved to a SSS state, will help in 
understanding slow systems.  Likewise, \swift\ monitoring of the 
two long lasting SSSs, V723 Cas and V458 Vul, are also of interest 
since they are rare, and thus, important to our understanding
of why they persist.  

Finally, it is important to continue to collect X-ray observations of 
novae and build on this sample.  This analysis shows that each nova is in
some ways unique and that attempts to predict their behavior based on a 
relationship to a single observational value, {\it e.g.} t$_2$ versus the 
nuclear burning timescale, is fraught with difficulties.  Some of these
problems can be addressed by expanding the sample to include regions of
the parameter space that are not well represented.  This X-ray sample
includes few slow novae which likely explains the differences between
the nuclear burning timescale of the Milky Way and M31 surveys.
It is also equally important to obtain numerous, high 
quality data for all bright novae through their evolution and at different
wavelengths from X-ray to radio.  Multiwavelength observations are critical
to properly interpret nova phenomena such as the apparent early turn-off in
V458 Vul and to verify periodicities seen in the X-ray, particularly
potential orbital periods. With the understanding that comes from a few 
well observed novae 
like RS Oph and U Sco, the entire nova data set can be anchored to nova 
theory.  These large data sets also reveal new phenomena such as the 
strong X-ray variability that is not appreciated in novae with 
sparser observations or detected at other wavelengths.

\acknowledgments

This research has made use of data obtained from NASA's \swift\ satellite.  
We thank Neil Gehrels and the \swift\ team for generous allotments of ToO 
and fill in time.  Funding support from NASA NNH08ZDA001N1. 
Stony Brook University's initial participation in the SMARTS consortium was 
made possible by generous contributions from the Dean of Arts and Sciences, 
the Provost, and the Vice President for Research of Stony Brook University.  
We acknowledge with thanks the variable star observations from the AAVSO 
International Database contributed by observers worldwide and used in this 
research.  
JPO, KP, PE \& AB acknowledge the support of the STFC.  
SS acknowledges partial support from NASA and NSF grants to ASU.  
JJD was supported by NASA contract NAS8-39073 to the \cxo\ X-ray Center.

{\it Facilities:} \facility{Swift(UVOT/XRT)}, \facility{AAVSO}, 
\facility{CTIO:1.3m}, \facility{CTIO:1.5m}, \facility{Bok(B\&C spectrograph)},
\facility{Spitzer(IRS)}


\end{document}